\documentclass[11pt,nofootinbib]{article}
\pdfoutput=1

\usepackage{cite}
\usepackage{multirow}
\usepackage{footmisc}
\usepackage{jheppub,fancyhdr,graphicx,epstopdf,color,amsmath,cases,slashed,hyperref}
\usepackage{tabularx}
\usepackage{comment}
\usepackage{caption, subcaption} 
\usepackage{dcolumn}  
\usepackage{amsmath, amssymb}       
\usepackage{color,latexsym, array,multirow,  verbatim, enumerate,cancel}
\usepackage{slashed, soul}
\usepackage[none]{hyphenat}
\usepackage{footnote}
\makesavenoteenv{tabular}
\makesavenoteenv{table}
\usepackage[titletoc]{appendix}
\usepackage{float}
\usepackage{microtype}
\usepackage[compat=1.1.0]{tikz-feynman}
\usepackage{tikz-feynman,contour}
\usepackage{xcolor}

\hypersetup{
   colorlinks=true,       
   linkcolor=blue,       
   citecolor=blue,         
   filecolor=magenta,    
   }

\title{Boosted top tagging via lepton-in-jet topology for vector-like quark searches in FCC-hh}

\author[a]{Sanchari Bhattacharyya,}
\author[a]{Biplob Bhattacherjee,}
\author[a]{Kirana K. K.,}
\author[b]{Swagata Mukherjee}

\affiliation[a]{Centre for High Energy Physics, Indian Institute of Science, Bangalore 560012, India}
\affiliation[b]{Department of Physics, Indian Institute of Technology, Kanpur 208016, India}

\abstract{In this paper, we present a machine-learning-based framework for identifying leptonically decaying, highly boosted top quarks, $t\to b\ell\nu$, at the Future Circular Collider: hadron-hadron mode (FCC-hh) operating at a center-of-mass energy of $\sqrt{s}=100$~TeV. At these energies, the large Lorentz boost causes the top-quark decay products to become highly collimated, with the lepton embedded within the resulting top-quark jet. In this regime, it is not a priori clear whether constituent-based jet taggers, which rely on information from individual jet constituents, retain the high discrimination power achieved at LHC energies. To answer this question, we employ two complementary approaches: a variable-based tagger trained with a deep neural network (DNN) using physics-motivated observables and a constituent-based Particle Transformer (ParT) acting on the jet constituents. Both approaches are optimised for the characteristic non-isolated lepton signature of boosted leptonic top decays and achieve comparable performance in discriminating leptonic top jets from hadronic top jets and heavy-flavor $b$- and $c$-jet backgrounds. We then incorporate the tagging framework into an event-level search for single production of a vector-like $B^\prime$ quark decaying through $B^\prime\to tW$, and use it to evaluate sensitivity to $B^\prime$ in the range of $3$ to $10$~TeV. For the reference $B^\prime$ benchmark, the proposed search strategy achieves a $5\sigma$ discovery reach of approximately $7$~TeV under the width requirement $\Gamma_{B^\prime}/m_{B^\prime}<25\%$, in both the statistics-only and $10\%$ background systematic uncertainty scenarios. Relaxing the width requirement to $\Gamma_{B^\prime}/m_{B^\prime}\lesssim 50\%$ extends the discovery reach to approximately $10$~TeV.} 

\emailAdd{sanchari1192@gmail.com}
\emailAdd{biplob@iisc.ac.in}
\emailAdd{kiranakk@iisc.ac.in}
\emailAdd{swagata@iitk.ac.in}

\begin{document}

\maketitle

\section{Introduction}

The Standard Model (SM) of particle physics describes the interactions of the known elementary particles to very high accuracy, and the discovery of the Higgs boson by the ATLAS~\cite{higgsdiscatlas} and CMS~\cite{higgsdisccms} collaborations at the Large Hadron Collider (LHC) completed the experimentally established particle content of the SM. Despite its remarkable success, the SM leaves several important phenomena unexplained, most notably the nature of dark matter and the origin of non-zero neutrino masses. These gaps provide strong motivation for physics beyond the Standard Model (BSM). No direct BSM signatures have been observed so far. Experimental searches are continually tightening exclusion limits and pushing the new-physics scale into very high-energy regimes. The masses of BSM particles may be beyond the direct energy reach of the LHC. Such TeV-scale BSM particles are predicted by several BSM frameworks, including simple $U(1)$ extended gauge models~\cite{Zpmodel, Zpmodel1, Zptopphil1, Zptopphil2}, Left-Right Symmetric Models (LRSM)~\cite{lrsmhiggs, Zplrsm1, Zplrsm2,  zplrsm3, ATLAS:2019isd}, multi-Higgs models~\cite{2hdm2, 2hdm3, 2hdm4, 2hdm5, 2hdm6, 2hdm7, 3hdm, multihiggs1, multihiggs2, multihiggs3, multihiggs4, multihiggs5}, Supersymmetric extensions~\cite{susy, susy1, susy2, susy3, susy4, susy5, susy6}, theories with extra-dimensions~\cite{ed1, ed2}, models with vector-like fermion (VLF)~\cite{vlq1, vlq2, vlq3, vlq4, vlq5, vlqavik, vlqavik2, vlqavik3, vlqref1, vlqref5, vlqref6, vlqref7, vlqref8, VLQmodel, VLQmodel2, VLQmodel3}, and Grand Unified Theories (GUTs)~\cite{e6gut, so10gut}. 
If the mass scale of new physics lies beyond the kinematic reach of the LHC, higher-energy hadron colliders will be required to directly probe such heavy BSM states. This motivates the development of future high-energy facilities such as the Future Circular Collider in its hadron--hadron mode (FCC-hh).

Heavy BSM resonances often decay to top quarks, making boosted top quarks an important signature at the FCC-hh. A top quark produced with very large transverse momentum ($p_T$) has strongly collimated decay products. Hadronic top decays in this regime are usually tagged using jet-substructure techniques~\cite{toptag1, hadtoptag, hadtoptag1}, as investigated in the earlier work~\cite{Bhattacharyya:2025kvl} and in dedicated HL-LHC VLQ studies~\cite{hllhc}; here we instead focus on the leptonic channel, $t \to bW \to b\ell\nu$.
When a top quark is highly boosted, the angular separation $\Delta R(b,\ell)$ between the charged lepton and the $b$ quark becomes small. The lepton can therefore fail standard isolation
requirements while remaining inside the top jet, making its
identification challenging in the boosted regime. In addition,
the neutrino produced in the leptonic top decay escapes detection,
which limits the reconstruction of the top-quark kinematics using
only the visible jet constituents.
Earlier leptonic top-tagging studies~\cite{leptoptag1,leptoptag2,leptotag3, toptag, toptag2, toptag3} have shown that this does not rule out effective tagging, and that the non-isolated lepton, together with appropriate kinematic, isolation, and jet-substructure observables, provides sufficient information to discriminate leptonic top jets from background.

The overlap between the lepton and the hadronic activity from the $b$ quark makes the internal structure of the jet an important source of information beyond conventional lepton isolation.
We investigate two complementary approaches to exploit this information. The first uses a set of physics-motivated kinematic, jet-substructure, tower- and track-based observables as inputs to a deep neural network (DNN)~\cite{Baldi:2014kfa}. The second uses a particle-level Particle Transformer (ParT)~\cite{part_transformer}, which operates directly on individual jet constituents and their correlations. Constituent-based architectures such as ParT, the Global Particle Transformer~\cite{glopart}, and the Unified Particle Transformer~\cite{UparT} have demonstrated strong performance in jet classification; on a standard hadronic top-tagging benchmark at $\sqrt{s}=14$~TeV, for instance, ParT reaches an AUC of $0.986$~\cite{part_transformer}. Comparing these two approaches allows us to quantify the additional discriminating power provided by particle-level information beyond that captured by engineered observables.

Beyond top jets, semileptonic decays of $B$ and $D$ hadrons can produce non-isolated leptons within heavy-flavour jets, particularly $b$- and $c$-jets. A similar
signature can arise in boosted $H\to b\bar{b}$ and $Z\to b\bar{b}$
jets, where the heavy-flavour hadrons decay semileptonically.
Boosted $W$ jets may contribute as well, depending on their decay
products and the surrounding hadronic activity. These backgrounds
can resemble leptonic top jets in certain respects, making them
important for evaluating whether the proposed tagger effectively distinguishes between different jet categories.

This tagging framework was then applied to a benchmark scenario involving a vector-like quark (VLQ). The phenomenology and experimental signatures of VLQs have been extensively explored in the literature~\cite{VLQmodel3, VLQpheno, VLQpheno1, VLQpheno2, VLQpheno3}. Recent LHC searches have placed lower mass limits on VLQs in the range of approximately $1$--$2$~TeV~\cite{vlqref1,vlqref2,vlqref3,vlqref5,vlqref6,vlqref7,vlqref8,vlqref9}. Depending on their electroweak representation and couplings, these TeV-scale fermions can decay into a third-generation SM quark ($b$ or $t$) in association with a $W$, $Z$, or Higgs boson. We use these scenarios to investigate the usefulness of the proposed tagging framework and compare the performance of a variable-based deep neural network and a particle-level Particle Transformer for identifying leptonic top jets. The framework may also be adaptable for tagging other boosted multi-prong objects, such as boosted $W$ jets, and applied to a broader range of BSM scenarios.

The structure of this paper is as follows. Section~\ref{model} describes the benchmark VLQ models. Section~\ref{collider} builds the ML-based leptonic top taggers from a DNN and a particle-level Particle Transformer. Section~\ref{results} presents a second DNN for event-level signal-background discrimination and determines the corresponding sensitivity to the VLQ mass range. Section~\ref{discussion} compares our results with other ML architectures, including a Graph Neural Network (GNN)~\cite{GNNegdeconv} and ParT combined with DeepSVDD~\cite{ruff2018deepsvdd}. Finally, Section~\ref{conclusion} summarises our main findings and presents our concluding remarks.

\section{The theoretical model framework}
\label{model}

This study explores a Beyond the Standard Model (BSM) scenario featuring a vector-like quark (VLQ) that decays into a top quark and a $W$ boson. Unlike standard quarks, VLQs are Dirac fermions with left- and right-handed chiral components that transform identically under the Standard Model gauge group. Theoretical models predict various kinds of VLQs, such as bottom-like ($B^\prime$) or top-like ($T^\prime$), or more exotic states possessing electric charges other than the standard $-1/3$ or $2/3$ \cite{vlq1, vlqavik, vlqavik2}.

Both the ATLAS and CMS collaborations at the LHC have searched for VLQs during the 13~TeV and 13.6~TeV LHC runs \cite{vlqref1, vlqref2, vlqref3, vlqref5, vlqref6, vlqref7, vlqref8} in several decay modes, including exotic decay modes involving other BSM particles. The investigation of VLQs exhibiting boosted top signatures has been the focus of numerous recent studies \cite{single_vlq_cms, vlq_atlas_1,  vlq_atlas_5}. Depending heavily on the specific parameters of the model (namely the VLQ species, width, and chosen decay channel), current LHC analyses push mass exclusion limits into the 1--2~TeV regime~\cite{cms_X53_2026, cms_Tphi_2026,cms_TbW_2026, cms_TtH_2026, cms_Bprime_2026,single_vlq_cms}.

The higher center-of-mass energy of the FCC-hh substantially extends the accessible mass range for VLQs, enabling detailed studies of multi-TeV resonances beyond the LHC's reach. In this work, we consider a $SU(2)$ singlet bottom-like VLQ, characterised by the $(3,1,-1/3)$ representation under the $SU(3)_C \otimes SU(2)_L \otimes U(1)_Y$ gauge group. Following electroweak symmetry breaking and quark mixing, the physical $B^{\prime}$ state couples to the first- and third-generation Standard Model quarks through the effective Lagrangian\cite{VLQmodel}:

\begin{eqnarray}
    \mathcal{L} &\supset& \dfrac{g \tilde{g}}{2} \left[ \sqrt{\dfrac{R_L}{1+R_L}} ~\bar{B}^{\prime}_L W_{\mu}^{-} \gamma^{\mu} u_L + \sqrt{\dfrac{1}{1+R_L}} ~\bar{B}^{\prime}_L W_{\mu}^{-} \gamma^{\mu} t_L \right]  \nonumber \\
    &+& \dfrac{g \tilde{g}}{2\sqrt{2} ~\cos \theta_W} \left[ \sqrt{\dfrac{R_L}{1+R_L}} ~\bar{B}^{\prime}_L Z_{\mu} \gamma^{\mu} d_L + \sqrt{\dfrac{1}{1+R_L}} ~\bar{B}^{\prime}_L Z_{\mu} \gamma^{\mu} b_L \right]  \\
    &-& \dfrac{g \tilde{g} ~m_{B^\prime}}{2\sqrt{2} ~m_W} \left[ \sqrt{\dfrac{R_L}{1+R_L}} ~\bar{B}^{\prime}_R H d_L + \sqrt{\dfrac{1}{1+R_L}} ~\bar{B}^{\prime}_R H b_L + \sqrt{\dfrac{1}{1+R_L}} ~\dfrac{m_b}{m_{B^\prime}} ~\bar{B}^{\prime}_L H b_R \right] + \mathrm{h.c.} \nonumber
\end{eqnarray}

where $g$ is the SM $SU(2)_L$ gauge coupling, and $\tilde{g}$ and $R_L$ are free real parameters that govern the overall strength of the interaction and the relative mixing between first- and third-generation quarks, respectively. The $B^\prime$ quark can be pair-produced by QCD interaction or single-produced through electroweak interaction at the LHC and FCC-hh.

Once the VLQ mass is fixed, the phenomenology is essentially controlled by the two free parameters $\tilde{g}$ and $R_L$, which determine the production cross section, the total decay width, and the flavour structure of the interactions. Relative couplings to the first- and third- generation quarks are given by 
\begin{equation}
  C_3=\frac{1}{1+R_L}, \qquad
  C_1=\frac{R_L}{1+R_L},
\end{equation}
where $R_L=0$ corresponds to an exclusive coupling to the third generation, and  $R_L=1$  generates equal couplings to the first and third generations.

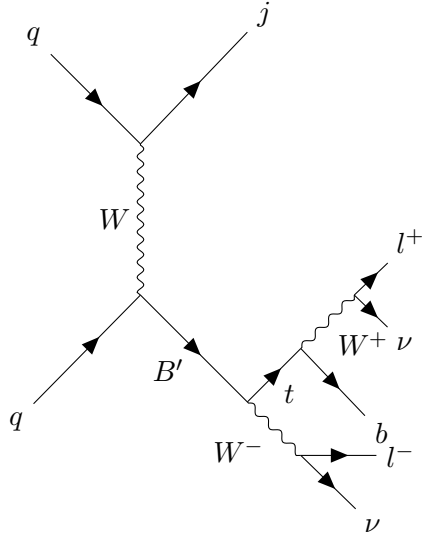
\begin{figure}
\begin{center}
\begin{tikzpicture}
		\begin{feynman}
			\vertex(a) {\(q\)};
			\vertex[below right=2 cm of a] (b);
			\vertex[above right=2 cm of b] (c){\(j\)};
			\vertex[below=2 cm of b] (d);
			\vertex[below right=2 cm of d] (e);
			\vertex[below left=2 cm of d] (f){\(q\)};
            \vertex[above right=1 cm of e] (g);
            \vertex[below right=1 cm of e] (h);
            \vertex[above right=1 cm of g] (i);
            \vertex[below right=1.2 cm of g] (j){\(b\)};
            \vertex[right=1 cm of h] (k) {\(l^-\)};
            \vertex[below right=1 cm of h] (l) {\(\nu\)};
            \vertex[above right=0.6 cm of i] (m) {\(l^+\)};
            \vertex[below right=0.6 cm of i] (n) {\(\nu\)};
            \diagram* {(a) -- [fermion] (b) -- [fermion] (c),(b) -- [boson,edge label'=\(W\)] (d),(d) -- [fermion, , edge label'=\(B^{\prime}\)] (e),(d) -- [anti fermion] (f), (e) -- [fermion, edge label'=\(t\)] (g), (e) -- [boson, edge label'=\(W^-\)] (h), (g) -- [boson, edge label'=\(W^+\)] (i), (g) -- [fermion] (j), (h) -- [fermion] (k), (h) -- [fermion] (l), (i) -- [fermion] (m), (i) -- [fermion] (n)};
		\end{feynman}
	\end{tikzpicture}
\end{center}
\caption{Feynman diagram for singlet VLQ, $B^\prime$ production along with a light jet.}
\label{feynman_bpj}
\end{figure}

In this work, we consider the single production of a $B^\prime$ quark in association with a light jet through the process $pp \to B^\prime j \to t W j$. We specifically focus on this production mode because its cross-section is significantly larger (although model-dependent) than that of pair production, particularly in the multi-TeV regime where the available phase space for pair production is limited, as shown in Table~\ref{Sig_BP}. The corresponding Feynman diagram for this process is illustrated in Figure~\ref{feynman_bpj}. To implement this scenario, we adopt the model file detailed in Ref.~\cite{VLQmodel}.

To determine representative benchmark points for the collider analysis, we perform a scan over the model parameter space, $\tilde{g}\in[0.01,\,0.10]$ and $R_L\in[0.1,\,1.0]$, for $B^\prime$ masses between 3 and 20 TeV. For each mass point, we use \textsc{MadGraph5\_aMC@NLO}~v3.5.8~\cite{madgraph} to evaluate
the total decay width and the single-production cross section throughout the $(\tilde{g}, R_L)$ parameter space. The NNPDF2.3 LO~\cite{nnpdf23} parton distribution function set is employed for this purpose.

The choice of the benchmark is driven by two requirements. We first impose a constraint $\frac{\Gamma_{B^{\prime}}}{m_{B^{\prime}}}<25\%$ on the resonance, thus restricting the analysis to the part of the parameter space where the $B^\prime$ is a reasonably well-defined resonance. This criterion is adopted as a benchmark selection requirement in the decay-chain simulation framework used in this work.
Second, among the parameter combinations that satisfy this condition, we consider benchmark points with sizeable production cross sections to maintain an appreciable signal yield while extending the accessible resonance mass range.

The parameter scan shows that, for fixed $m_{B^\prime}$ and $\tilde{g}$, the width-to-mass ratio $\Gamma_{B^{\prime}}/m_{B^\prime}$ has only a weak dependence on the mixing parameter $R_L$ in the range considered. Thus, in this work, we set $R_L$ to the maximum scanned value, $R_L=1.0$. Within the adopted parameter space, this choice enhances the single-production cross section while leaving the total width essentially unchanged. This behaviour is illustrated in Table~\ref{RL_table}.

\begin{table}[htbp]
\centering
\renewcommand{\arraystretch}{1.2}
\begin{tabular}{|c|c|c|}
\hline
\textbf{$R_L$} &
 \textbf{$\Gamma_{B^{\prime}}/m_{B^\prime}$ (\%)} & \textbf{$\sigma(pp\to B^{\prime}j)$ [fb]} \\
\hline
0.1 & 24.60 & 7.27 \\
1.0 & 24.60 & 39.84 \\
\hline
\end{tabular}
\caption{Dependence of the width-to-mass ratio and production cross section on the mixing parameter $R_L$ for a $B^\prime$ quark with $m_{B^\prime}=10$~TeV and $\tilde{g}=0.05$ at $\sqrt{s}=100$~TeV.}
\label{RL_table}
\end{table}

The choice of the coupling $\tilde{g}$ determines the maximum resonance mass that can satisfy the imposed width criterion. For small couplings ($\tilde{g}\leq0.05$), the condition $\Gamma/m_{B^\prime}<25\%$ can be fulfilled even for resonance masses above 10~TeV. The corresponding production cross sections, however, are too small to yield meaningful sensitivity at the FCC-hh. For larger couplings ($\tilde{g}\geq0.09$), in contrast, the production cross section is significantly increased, but the allowed mass range is limited to $m_{B^\prime}\leq5$~TeV, since the resonance width becomes larger than the imposed limit for larger masses. An intermediate choice of $\tilde{g}=0.07$ therefore provides a good compromise, allowing resonance masses up to $7$~TeV within the imposed width criterion while maintaining a sizeable production cross section. The resulting benchmark points are summarised in Table~\ref{tab:g_scan}.

\begin{table}[htbp]
\centering
\renewcommand{\arraystretch}{1.3}
\begin{tabular}{|c|c|c|c|}
\hline
 $\tilde{g}$ & $m_{B^\prime}$ [TeV] & $\Gamma_{B^\prime}/m_{B^\prime}$ (\%) & $\sigma(pp\to B^{\prime}j)$ [fb] \\
\hline
0.1  & 5 & 24.57 & 1009 \\
0.07  & 7 & 23.62 & 217.9 \\
0.05 & 10 & 24.60 & 39.8  \\
0.03 & 16 & 22.67 & 2.6  \\
0.02 & 20 & 15.75 & 0.41  \\
\hline
\end{tabular}
\caption{Maximum accessible $B^\prime$ resonance mass satisfying the imposed width criterion, $\Gamma_{B^\prime}/m_{B^\prime}<25\%$, for different values of $\tilde{g}$ with $R_L=1.0$. For each coupling, the quoted mass corresponds to the largest resonance mass in the scan that satisfies this criterion.}
\label{tab:g_scan}
\end{table}

Among the benchmark points listed in Table~\ref{tab:g_scan}, the
parameter choice $(\tilde g,R_L)=(0.07,1.0)$ is adopted as the reference benchmark for the remainder of this study. Within the scanned parameter space, this choice provides a favourable balance between the accessible resonance mass and the expected signal yield while satisfying the imposed width criterion, $\Gamma_{B^\prime}/m_{B^\prime}<25\%$.

\begin{table}[htbp]
    \centering
    \renewcommand{\arraystretch}{1.3}
    \begin{tabular}{|c|c|c|}
    \hline 
    \textbf{Mass of $B^\prime$ [TeV]} &
    \multicolumn{2}{c|}{\textbf{Cross section [fb]}} \\
    \cline{2-3}
    & \textbf{$\sigma(pp\to B^{\prime}j)\times
    \mathrm{BR}(B^{\prime}\to tW)$} &
    \textbf{$\sigma(pp\to B^{\prime}\bar{B}^{\prime})
    \times \mathrm{BR}(B^{\prime}\to tW)^2$} \\
    \hline
    3  & 457  & 7.06 \\
    5  & 161.7 & 0.36 \\
    10 & 20.7 & 0.002 \\
    \hline
    \end{tabular}
    \caption{Single- and pair-production cross sections at
    $\sqrt{s}=100$~TeV for $\tilde{g}=0.07$ and $R_L=1.0$, including
    the branching fractions for the decay $B^\prime\to tW$.}
    \label{Sig_BP}
\end{table}

Having established the benchmark scenario, we now examine the
characteristic kinematic properties of the final-state leptons that
motivate our tagging strategy. The signal events are generated at
leading order using \textsc{MadGraph5\_aMC@NLO}~v3.5.8 and subsequently interfaced with \textsc{PYTHIA}~8.315~\cite{pythia8315} for parton showering and hadronisation. The lepton origin is identified using the particle ancestry information, allowing leptons originating from the top-quark decay chain and from the directly produced $W$ boson in $B^\prime\to tW$ to be distinguished.

\begin{figure}[htbp]
	\begin{center}
		\includegraphics[width=7.5cm]{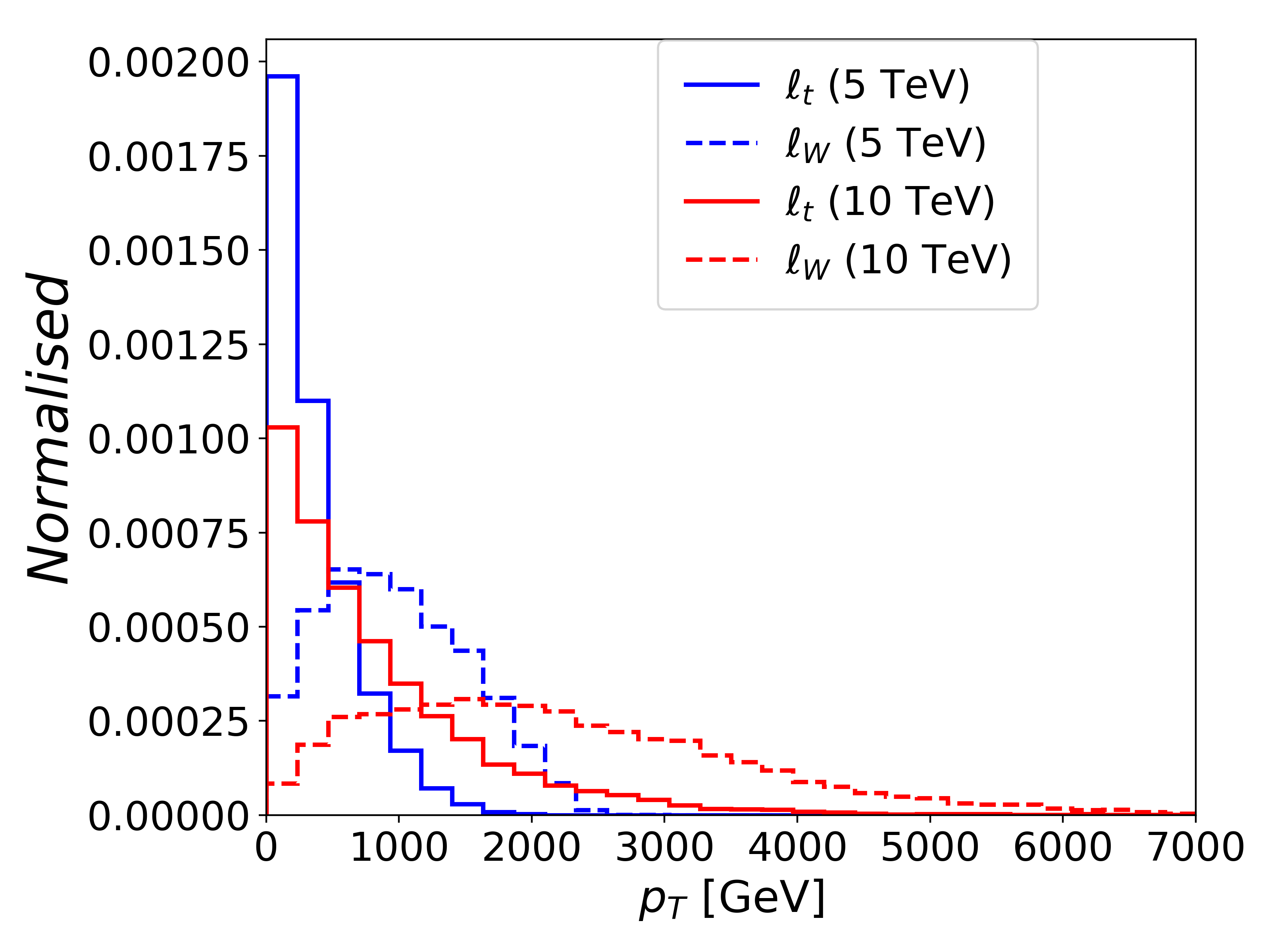}
        \includegraphics[width=7.5cm]{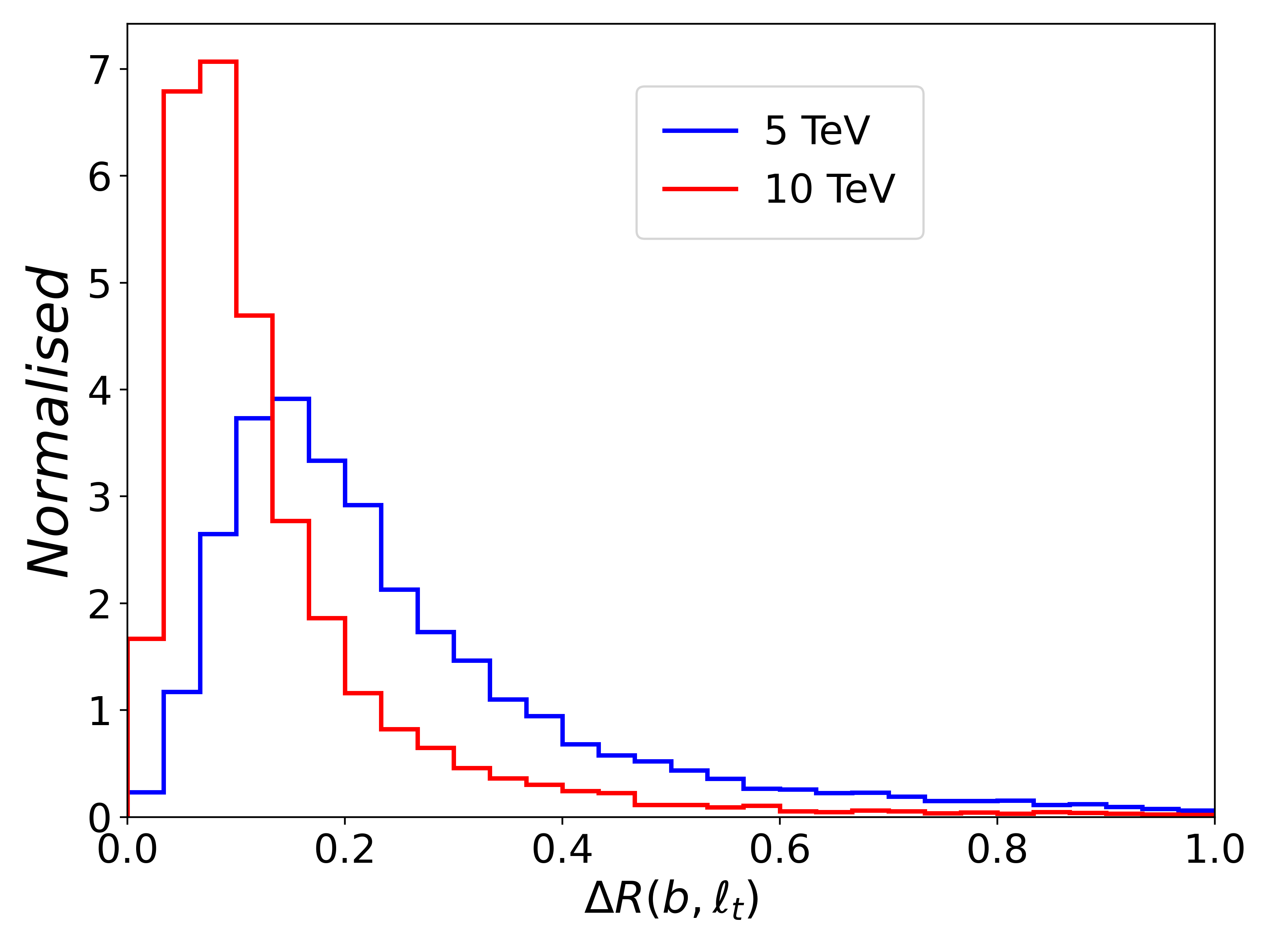}
	 \end{center}
  \caption{Transverse momentum of leptons $\ell_t$ (from top-quark decays, $t \to bW \to b\ell\nu_\ell$) and $\ell_W$ (from the $W$ boson in $B^\prime \to tW$, $W \to \ell\nu_\ell$) (left), and angular separation $\Delta R(b,\ell_t)$ between the lepton and $b$-quark from the same top-quark decay (right). Distributions are shown for $B^\prime$ masses of 5 and 10~TeV at $\sqrt{s}=100$~TeV.}

 \label{iso_noniso_pt_deltaR}
\end{figure}

The Feynman diagram in Figure~\ref{feynman_bpj} shows that there are two leptons in the final state, one originating from the direct decay of the $W$ boson, whereas the other arises from the top-quark decay chain through $t\to bW$. Figure~\ref{iso_noniso_pt_deltaR} (left) shows the transverse momentum ($p_T$) distributions of leptons from top quark decays ($\ell_{t}$) and leptons from $W$ boson decays ($\ell_{W}$) in the signal process, for benchmark $B^\prime$ masses of 5 and 10~TeV. Leptons from the top decays are generally softer than those from $W$ decays. Still, for both mass points, sizable tails are extending into the 2--3~TeV range. 
Figure~\ref{iso_noniso_pt_deltaR} (right) shows the angular separation $\Delta R(b,\ell_t)$ between the $b$-quark and the charged lepton from the same top-quark decay $(t\rightarrow bW,\, W\rightarrow \ell\nu)$ for the two $B^\prime$ benchmarks.~\footnote{The angular separation between the $b$-quark and the lepton is defined as $\Delta R(b,\ell_t)=\sqrt{(\eta_b-\eta_{\ell_t})^2+ (\phi_b-\phi_{\ell_t})^2}$, where $\eta$ and $\phi$ denote pseudorapidity and azimuthal angle, respectively.}
The signal distributions peak at small $\Delta R$, indicating that the top decay products are collimated due to the large Lorentz boost of the top quark. This effect is more pronounced in the $m_{B^\prime}=10$~TeV sample than in the $m_{B^\prime}=5$~TeV sample, as the more massive $B^\prime$ imparts a larger boost to the top.

These characteristic kinematic features motivate the dedicated boosted-top lepton-tagging strategy presented in the following sections.

\section{Leptonic-top jet tagger}
\label{collider}

In the highly boosted regime, resolved reconstruction techniques lose their effectiveness, since the top decay products are collimated into a single jet rather than reconstructed as separate objects. The identification of the top must therefore rely on the jet's internal structure. In this study, we focus on the leptonic channel, $t \to bW \to b\ell\nu$. However, the neutrino escapes detection; the non-isolated lepton inside the jet provides a distinctive handle for suppressing hadronic backgrounds. In the following subsections, we describe the substructure and kinematic features used to tag these leptonic top jets.

The samples used for the training and evaluation of the tagger are
generated using \textsc{PYTHIA}~8.315~\cite{pythia8315}, with the
NNPDF2.3 LO PDF set accessed through LHAPDF6~\cite{lhapdf,nnpdf23}.
The generated events are subsequently passed through
\textsc{Delphes}~3.5.0~\cite{delphes} for fast detector simulation
using the FCC-hh detector card. Jets are reconstructed using the
anti-$k_T$ algorithm~\cite{antikt} with a radius parameter $R=0.4$.

\begin{table}[htbp]
\centering
\renewcommand{\arraystretch}{1.2}
\begin{tabular}{|l|c|c|c|}
\hline
 & $b$-jet & $c$-jet & Light jet \\
\hline
Fraction passing $p_T(\ell)>100$ GeV (\%) & 21.0 & 8.2 & 0.9 \\
Production cross section (fb) & $2.74\times10^{5}$ & $3.01\times10^{5}$ & $6.53\times10^{7}$ \\
\hline
\end{tabular}
\caption{Fraction of simulated $b$, $c$, and light-flavour jet events
containing a reconstructed lepton satisfying
$p_T(\ell)>100$~GeV and $|\eta(\ell)|<2.5$.}
\label{tab:events}
\end{table}

For training the jet tagger, our signal is the leptonic top, where the lepton ($\ell$) can be an electron ($e$), muon ($\mu$), or a
leptonically decaying tau ($\tau_{\ell}$). The hadronic top is one of
the principal backgrounds. In addition, $b$- and $c$-jets constitute
important backgrounds, as they can contain genuine non-isolated leptons inside the jet cone. The following preselection requirements are applied to leptons: $p_T(\ell)>100$~GeV and $|\eta(\ell)|<2.5$.

As shown in Table~\ref{tab:events}, for samples generated with a parton-level cut of $\hat{p}_T > 500$~GeV, only about $0.9\%$ of light-flavour jet events contain a reconstructed lepton satisfying these requirements, compared with $21\%$ and $8.2\%$ for $b$- and $c$-jets, respectively. However, because the production cross-section for light-flavor jets is much larger than that for heavy-flavor jets, this small fraction alone does not imply that their contribution is negligible. We therefore expect the light-flavour background to be further suppressed by the full event selection, in particular by the requirement of both an isolated and a non-isolated lepton, as required by the $B^\prime \to tWj$ signal topology, together with the jet-tagger selection. In contrast to heavy-flavour jets, a lepton candidate inside a light-flavour jet is generally not expected to exhibit the displaced-track and substructure characteristics associated with the leptonic top and heavy-flavour backgrounds considered in this study. Consequently, the object-level tagger is trained using hadronic top, $b$-, and $c$-jets as the background classes. We observe from Figure~\ref{iso_noniso_pt_deltaR} that the $p_T$ of the lepton can span a wide range, and a single tagger is not sufficient to capture the full kinematic behavior across this spectrum. Therefore, the tagger is trained separately in the following lepton $p_T$ bins: $0.1$--$0.5$~TeV, $0.5$--$1$~TeV, $1$--$1.5$~TeV, $1.5$--$2$~TeV, $2$--$2.5$~TeV, $2.5$--$3$~TeV, and $p_T>3$~TeV. This binning enhances sensitivity in the high-energy regime, which is more relevant for FCC-hh.

To separate leptonic top jets from backgrounds, we use kinematic and substructure variables that exploit the characteristic topology of leptonic top decays. Both leptonic top jets and $b$-jets can have a lepton inside the jet cone, but the energy sharing and angular pattern are different. For boosted tops, the lepton comes from the $W$ boson and carries a larger fraction of the jet $p_T$ than leptons from semileptonic $B$-hadron decays in $b$-jets. In addition, the $b$-quark in the top jet gives displaced tracks, which helps separate it from light-quark and gluon jets.

After pre-selecting leptons with $p_T > 100$~GeV, we define a cone of radius $\Delta R < 0.4$ around the lepton direction in the $(\eta,\phi)$ plane. Tracks within the cone must have $p_{T}^{\text{trk}} > 5$~GeV and $|\eta^{\text{trk}}| < 2.5$, while calorimeter towers must have $E_{T}^{\text{tow}} > 5$~GeV and $|\eta^{\text{tow}}| < 2.5$. Jets are required to have $p_{T}^{\text{jet}} > 50$~GeV and $|\eta^{\text{jet}}| < 2.5$.
All these objects within the cone are used to define the variables listed in Table~\ref{object_features}, which are organised into four categories: lepton kinematics, track-based variables, calorimeter tower-based variables, and jet variables. Each variable category captures a distinct physical feature of the leptonic top quark decay topology. For example, the track-based variables capture the displaced decay products of the $B$-hadron, tower-based variables capture the calorimeter energy distribution around the lepton, and jet substructure observables characterise the internal radiation pattern of the jet.

\begin{table}
    \centering
    \scriptsize
    \renewcommand{\arraystretch}{1.2}
    \begin{tabular}{|c|l|}

\hline
\textbf{Notation} & \textbf{Definition} \\
\hline

\multicolumn{2}{|c|}{\textbf{ Kinematics}} \\
\hline

$DZ^{\text{lep}}$ & Lepton longitudinal impact parameter(mm) \\
$D0^{\text{lep}}$ & Lepton  transverse  impact parameter(mm) \\

\hline

\multicolumn{2}{|c|}{\textbf{Track-based variables} ($\Delta R_{\ell,\mathrm{trk}} < 0.4$)} \\
\hline 

$\langle \Delta R \rangle_{p_T}^{\ell}$ & $p_T$ weighted $\Delta R$ \, :\, $\dfrac{\sum \Delta R_{\ell,\mathrm{trk}} \, p_{T}^{\mathrm{trk}}}{\sum p_{T}^{\mathrm{trk}}}$ \\[2ex]

$N^{\text{trk}}$ & Number of tracks \\

$\sum p_{T}^{\text{trk}}$ & Absolute track-isolation variable based on sum of track $p_T$  \\[2ex]

$\text{Iso}_{p_T}$ & Relative track isolation  \, $\dfrac{\sum p_{T}^{\text{trk}}}{p_{T}^{\text{lep}}}$ \\[4ex]

$\langle \Delta R \rangle_{p_T}^{\ell,\, \text{disp}}$ & $\dfrac{\sum \Delta R_{\ell,\mathrm{trk}}\, \times p_{T}^{\text{trk}}}{\sum p_{T}^{\text{trk}}}$ with transverse impact parameter (IP) $>$ 1 mm  or \\
& track transverse distance (TD) = $\sqrt{\text{trk}_{X}^{2} + \text{trk}_{Y}^{2}} > 1~\text{mm}$ \\[2ex]

$N^{\text{trk}}_{IP>1}$ & Number of tracks with IP $>1$ mm or TD $>1$ mm \\[2ex]

$Q_{\kappa}$ & $\dfrac{\sum (p_{T}^{\mathrm{trk}})^{\kappa} \,Q_{\text{trk}}}{(\sum p_{T}^{\text{trk}})^\kappa}$ with $\kappa = 0.3, \,0.5, \,1, \,2$, where $Q_{\text{trk}}$ is track charge. \\[2ex]

\hline

\multicolumn{2}{|c|}{\textbf{ Tower-based variables}  ($\Delta R_{\ell,\mathrm{tow}} < 0.4$) } \\
\hline
$\sum E_{T}^{\text{tow}}$ & Absolute tower isolation based on sum of $E_T$ of all towers   \\[2ex]

$\text{Iso}_{E_{T}}$ & Relative tower isolation \, $\dfrac{\sum E_{T}^{\text{tow}}}{E_{T}^{\text{lep}}}$ \\[2ex]

$\sum E_{T,\text{had}}^{\text{tow}}$ & Absolute tower isolation based on sum of $E_T$ of hadron calorimeter towers \\

$\langle \Delta R \rangle_{E_T}^{\ell}$ & $E_T$ weighted $\Delta R$ \, $\dfrac{\sum \Delta R_{\ell,\mathrm{tow}} \, E_{T}^{\mathrm{tow}}}{\sum E_{T}^{\mathrm{tow}}}$ \\

$N^{\text{tow}}$ & Number of calorimeter towers\\

\hline

\multicolumn{2}{|c|}{\textbf{ Jet variables}  ($\Delta R_{\ell,\mathrm{jet}} < 0.4$)} \\
\hline
$m_{\text{jet}}^{\ell}$ & Jet mass  \\

$N_{b\text{-jet}}$ & Number of $b$-tagged jets  \\

$\tau_N$ & N-subjettiness, where $N=1,\,2,\,3$. \\[2ex]

$\dfrac{\tau_2}{\tau_1}$, $\dfrac{\tau_3}{\tau_2}$, $\dfrac{\tau_3}{\tau_1}$ & Ratios of subjettiness variables  \\[2ex]

\hline

\end{tabular}
\caption{Summary of variables used in training the variable-based top-tagger.
}
\label{object_features}
\end{table}

\begin{figure}[H]
	\begin{center}
		\includegraphics[width=16cm]{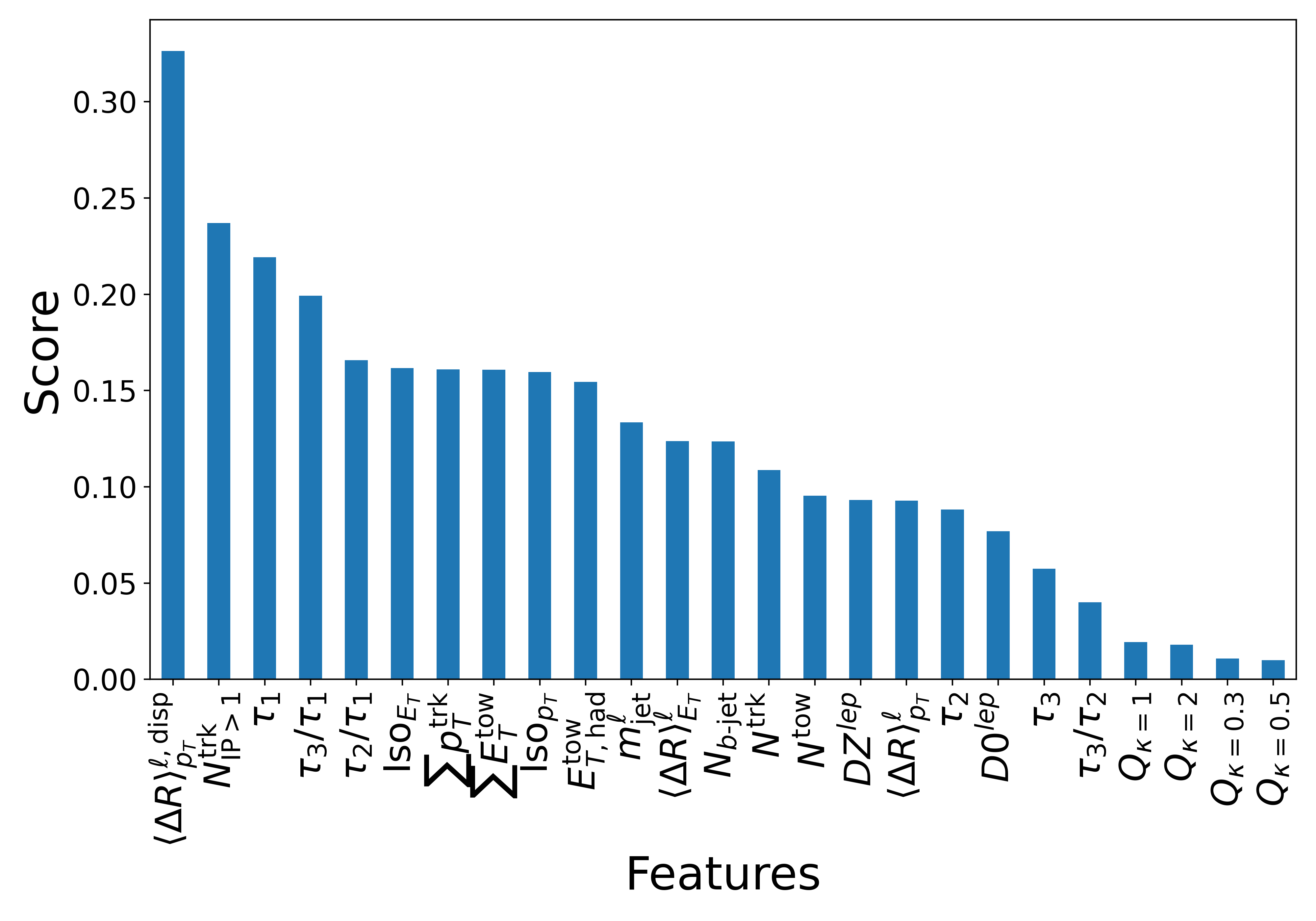}
       
	 \end{center}
 \caption{Mutual information scores of the selected kinematic and jet features employed for signal–background discrimination, computed in the lepton $p_T$ bin 1.0–1.5 TeV. }
 \label{mutual}
\end{figure}

\begin{figure}[H]
	\centering	{\includegraphics[width=7.5cm]{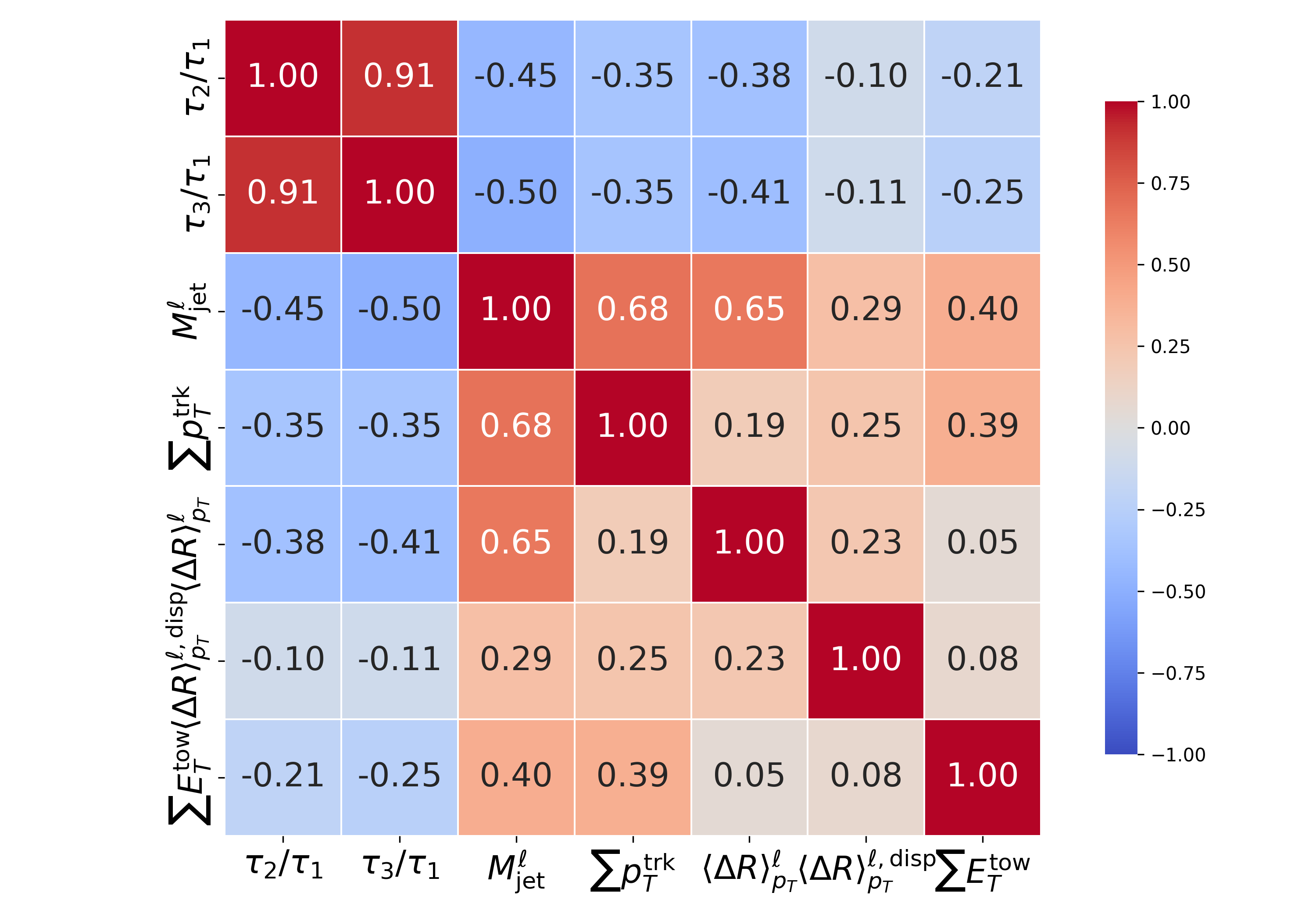}}
    {\includegraphics[width=7.5cm]{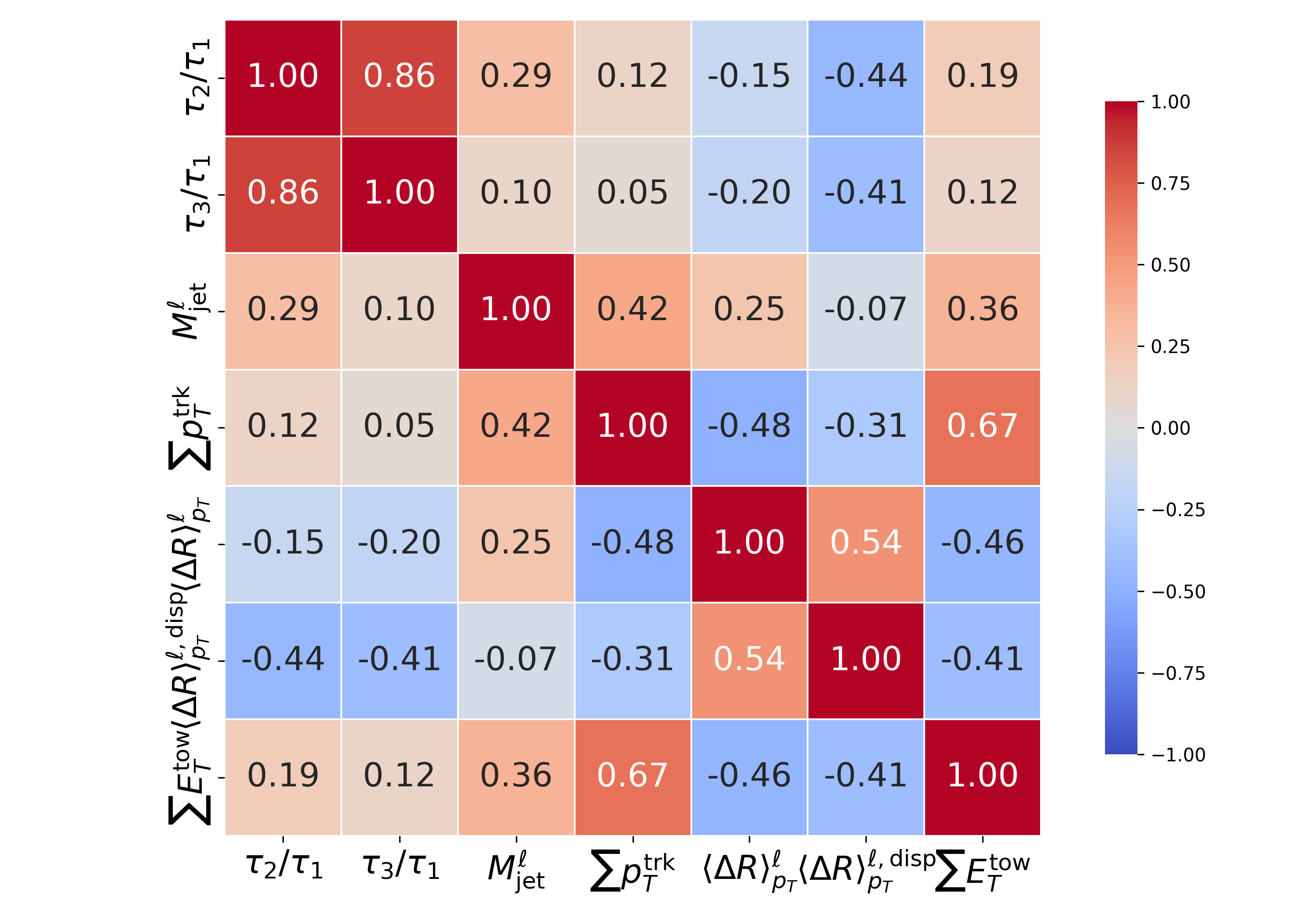}}

 \caption{The correlations among some high-performing variables are shown through the correlation matrices for $b$-jet (left) and leptonic top jet (right) in the lepton $p_T$ interval of 1.0--1.5~TeV.}
 \label{corr}
\end{figure}

Although all the variables are listed in Table~\ref{object_features}, not all of them carry equal discriminating power. To rank the input features by their discriminating power across four jet classes -- leptonic top, hadronic top, $b$-jet, and $c$-jet -- the Mutual Information (MI) score was computed for each feature $X$,  
where a higher MI score indicates greater discriminating power. MI score was estimated using the $k$-nearest-neighbor estimator~\cite{Ross2014} as implemented in \texttt{scikit-learn}~\cite{Pedregosa2011}. Figure~\ref{mutual} shows the MI scores of all the variables used for the training of the tagger, for a particular $p_T$ bin. Figure~\ref{corr} shows the correlation matrices of a set of selected high-performing variables for a particular $p_T$ bin for signal jet and one type of background jet. The plot shows that most variable pairs exhibit modest correlations, with a few expected exceptions such as the $N$-subjettiness~\cite{Thaler} ratios.

In the following subsections, we discuss some of the most discriminating variables.

\subsection{Track-based variables}

\begin{figure}[H]
	\begin{center}
		\includegraphics[width=7.5cm, height=6.5cm]{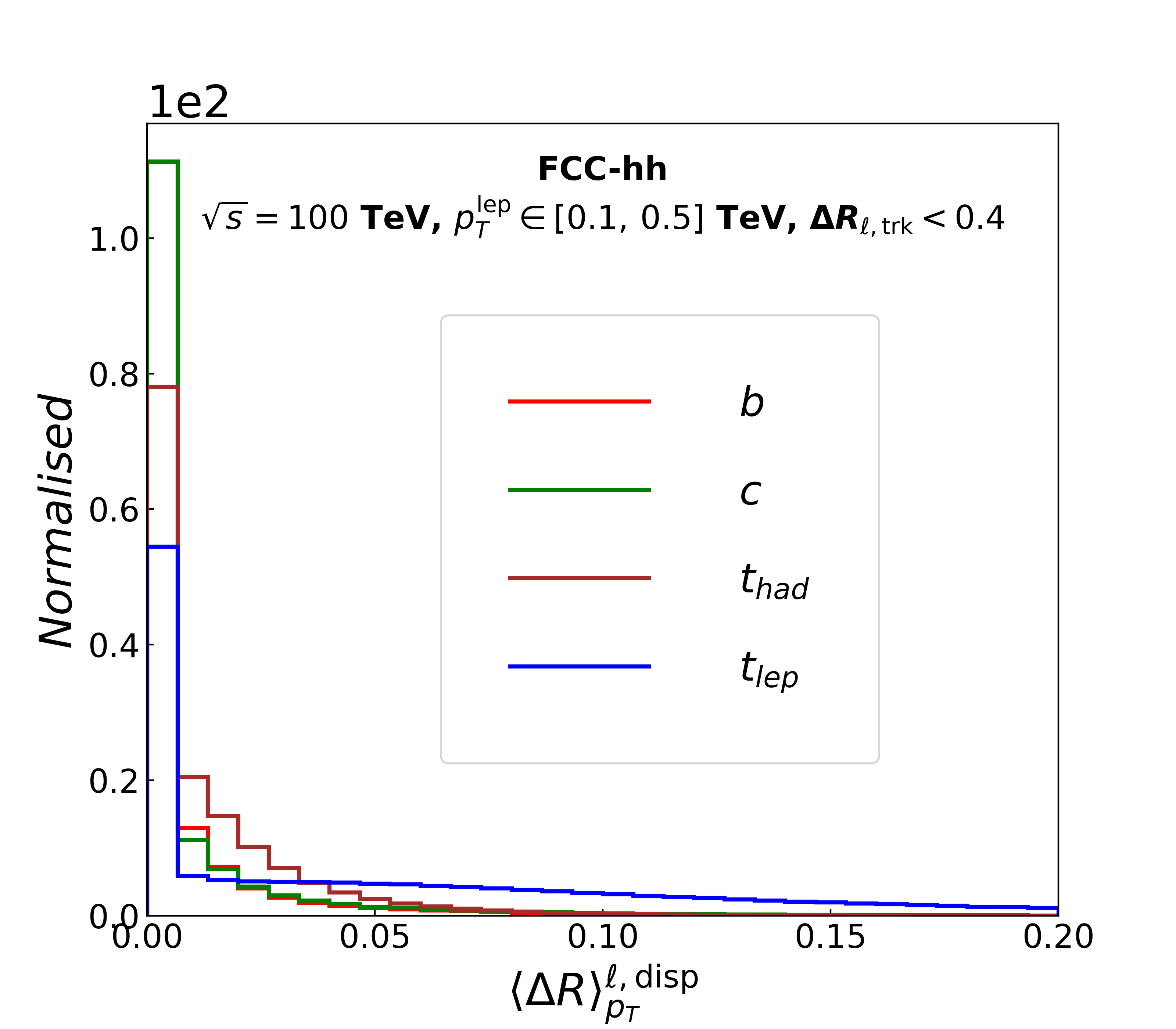}
        \includegraphics[width=7.5cm, height=6.5cm]
        {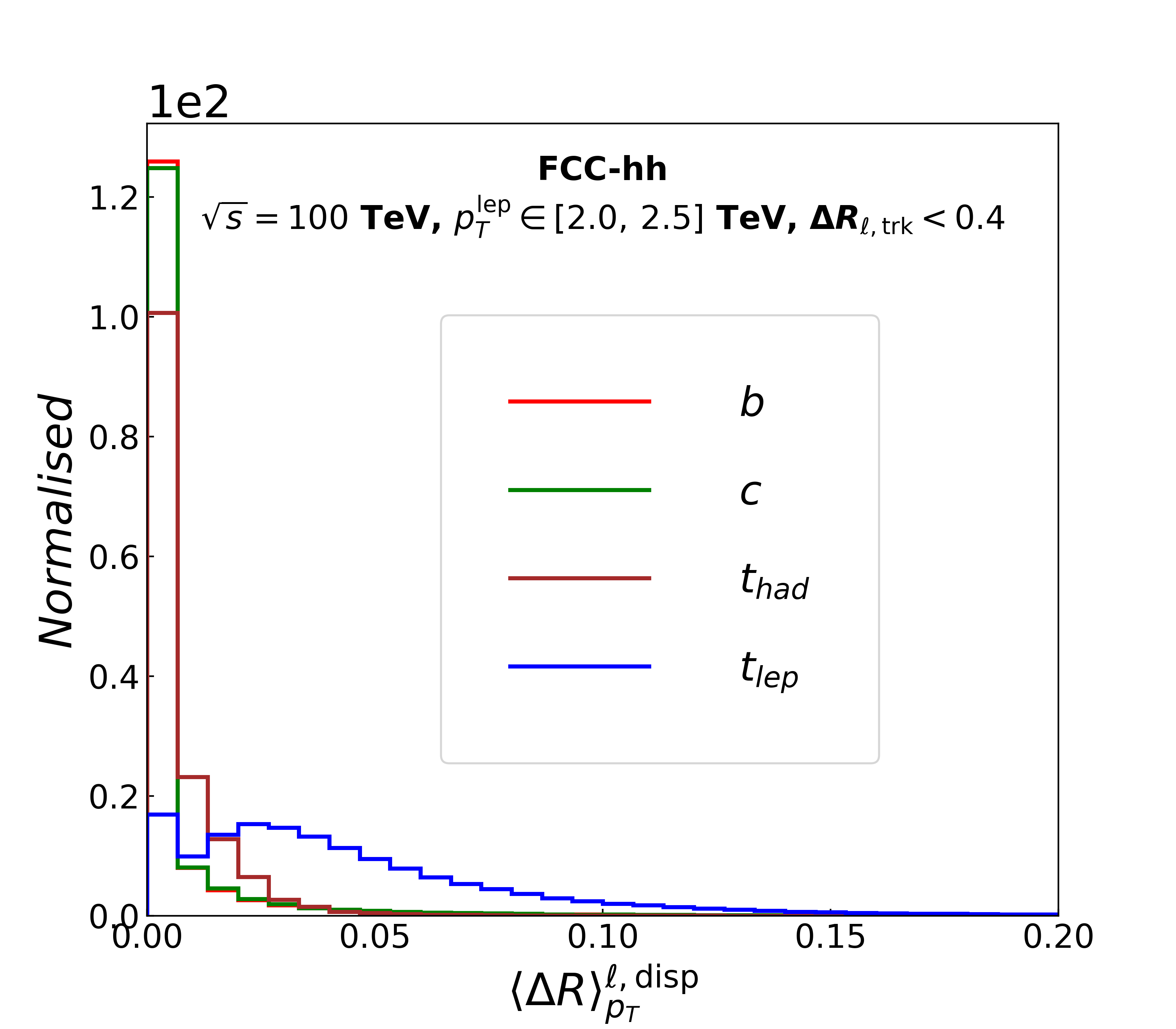}
	 \end{center}
 \caption{Distribution of the track-based variable $\langle \Delta R \rangle_{p_T}^{\ell,\, disp}$  (see Table~\ref{object_features}), computed within $\Delta R_{\ell,\,\mathrm{trk}} < 0.4 $ around the lepton for different processes in the lepton $p_T$ ranges $0.1$--$0.5$~TeV (left) and $2.0$--$2.5$~TeV (right).}
 \label{iso}
\end{figure}

Displaced-track variables take advantage of the fact that heavy-flavor hadrons, such as $B$ and $D$ hadrons, are slightly long-lived, so they do not decay right at the main interaction point. As a result, the tracks from these decays have large impact parameters and appear as displaced tracks (i.e., tracks with $IP > 1$[mm] or $TD > 1$[mm]). In leptonic top quark decays, a $b$-quark is always present, making these displaced-track variables helpful for distinguishing the signal from some backgrounds where all tracks are prompt. Figure~\ref{iso} shows the distributions of the displaced-track variable $\langle \Delta R \rangle_{p_T}^{\ell,\, \text{disp}}$ for signal and backgrounds, for two different lepton $p_{T}$ bins. 
This is among the most discriminating variables used in the training.

\begin{figure}[H]
	\begin{center}
		\includegraphics[width=7.5cm, height=6.5cm]{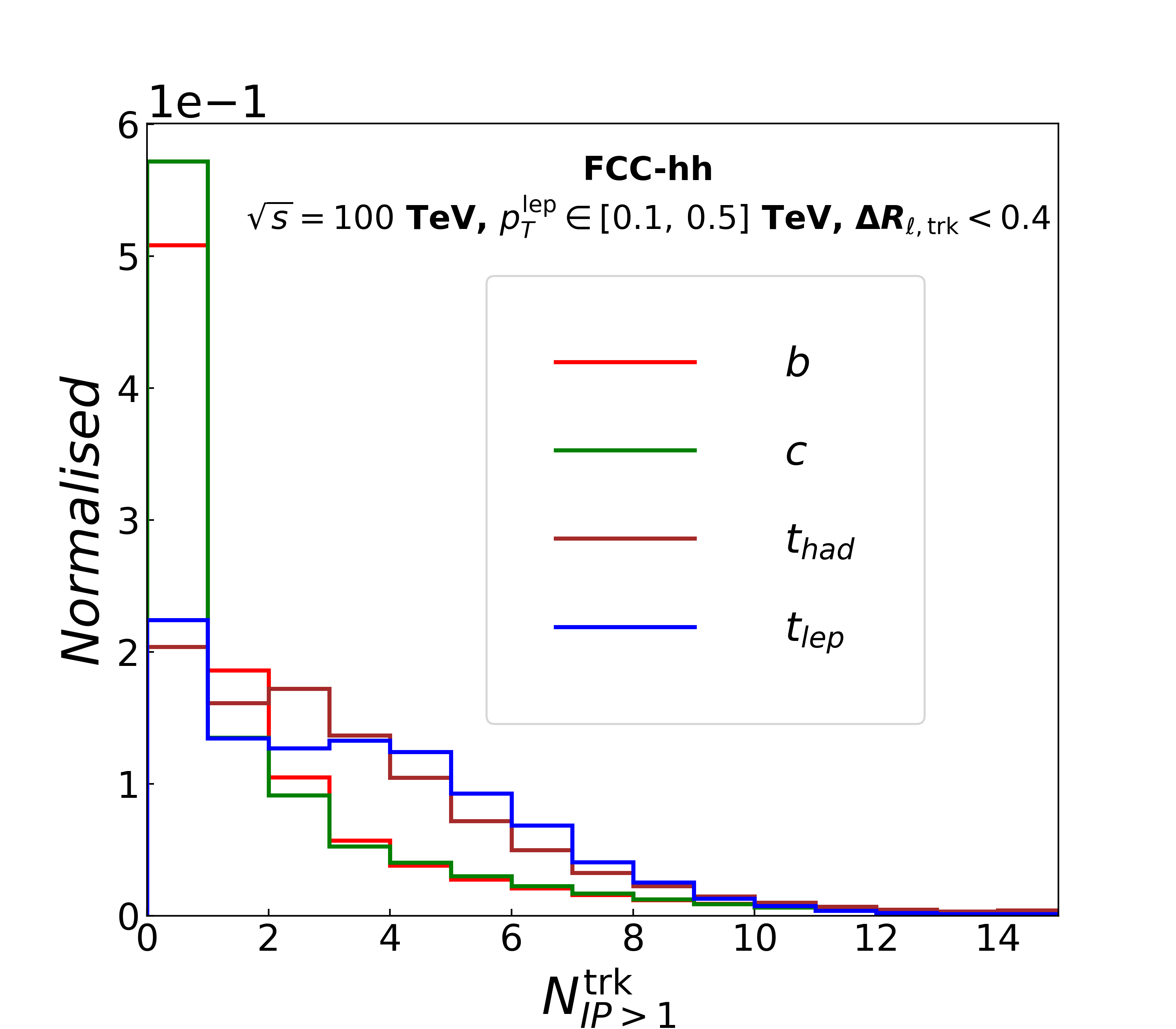}
        \includegraphics[width=7.5cm, height=6.5cm]{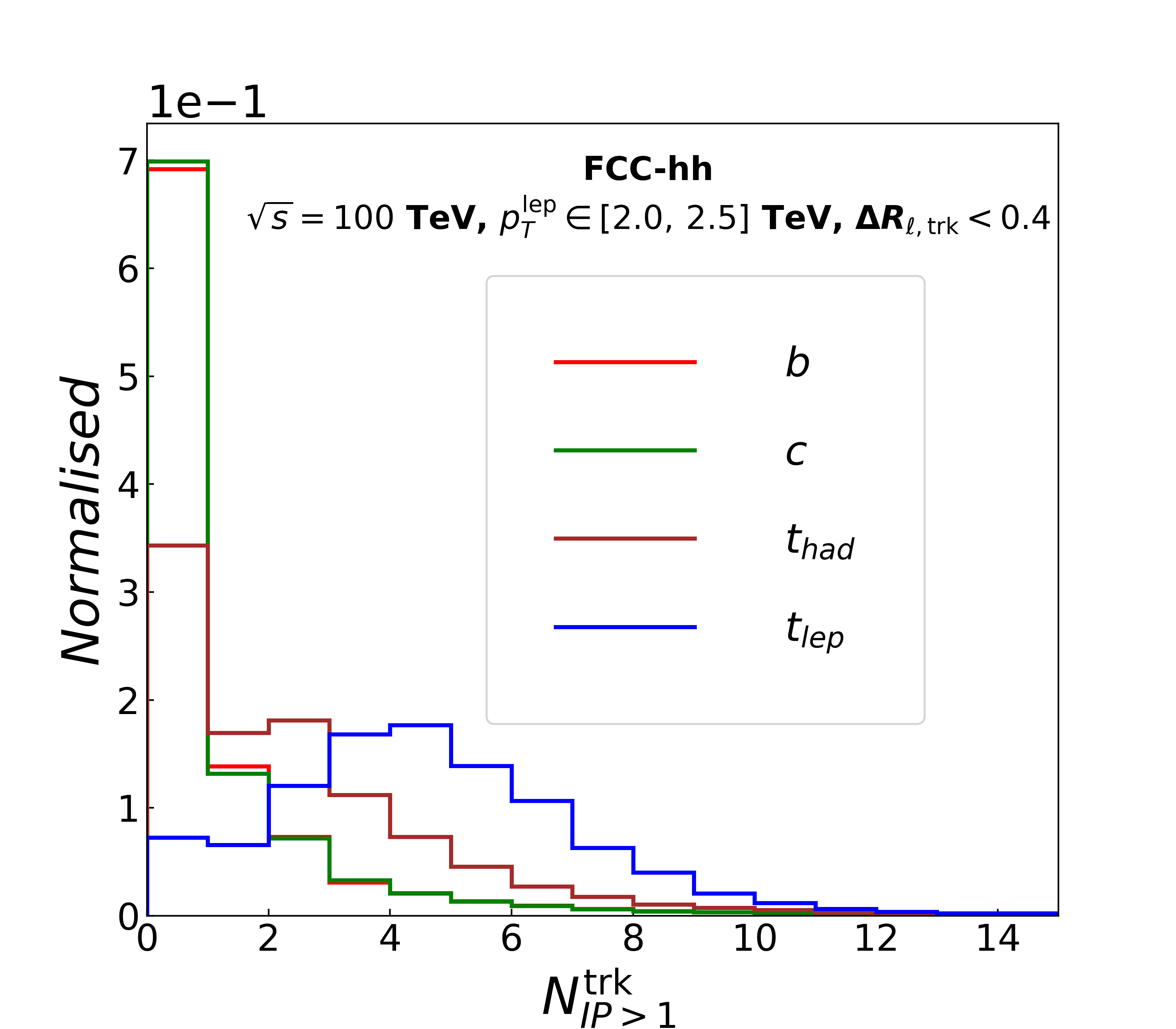}
	 \end{center}
 \caption{Displaced track multiplicity for signal and background jets in lepton $p_T$ intervals $0.1$--$0.5$~TeV (left) and $2.0$--$2.5$~TeV (right). 
}
 \label{track}
\end{figure}

Figure~\ref{track} shows how many displaced tracks are found within a cone of $\Delta R_{\ell,\,\mathrm{trk}} < 0.4$ around the lepton. 
The distributions show good discrimination between signal jets and background jets.

\subsection{Jet variables}
\begin{figure}[H] 
	\begin{center}
		\includegraphics[width=7.5cm, height=6.5cm]{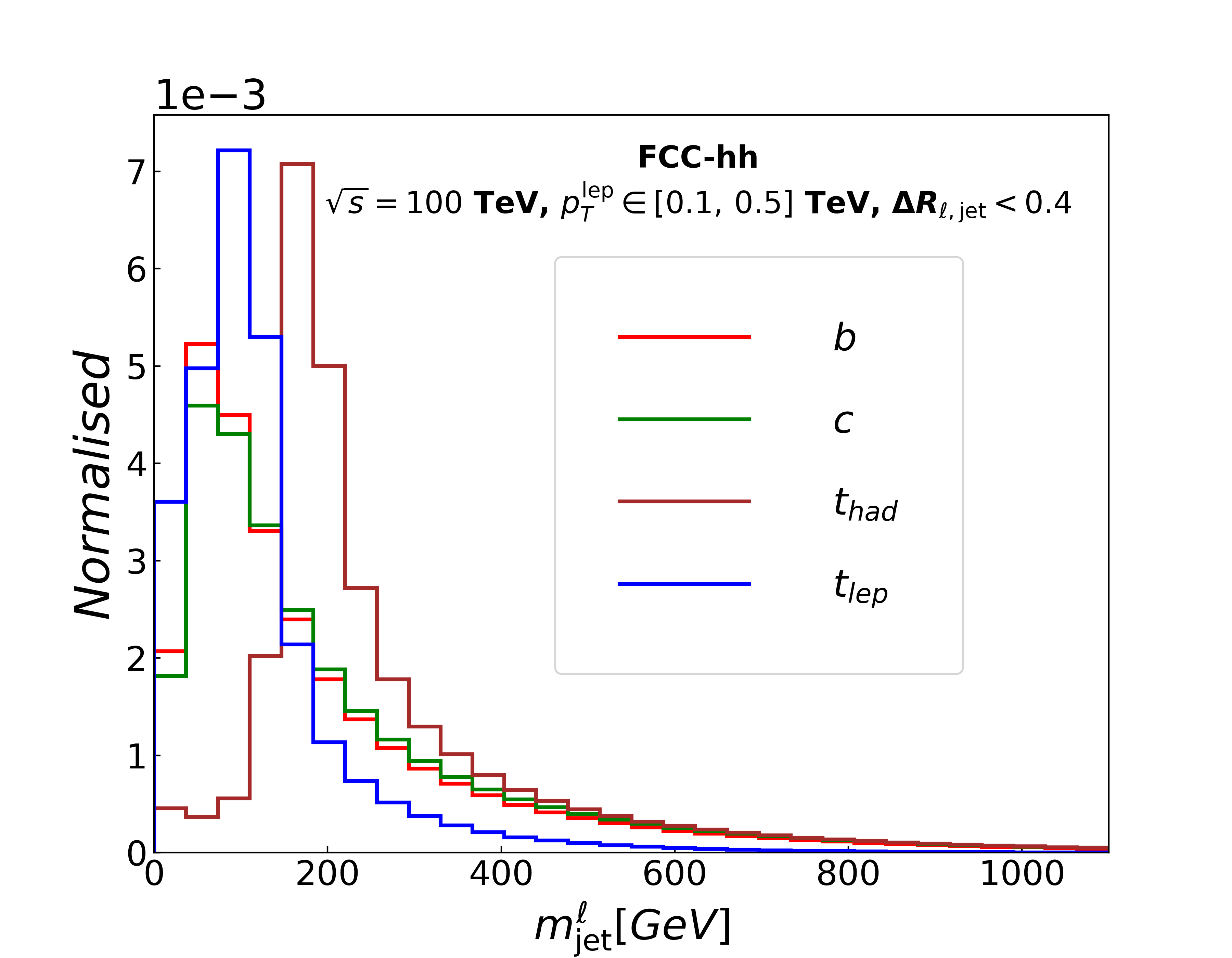}
        \includegraphics[width=7.5cm, height=6.5cm]
        {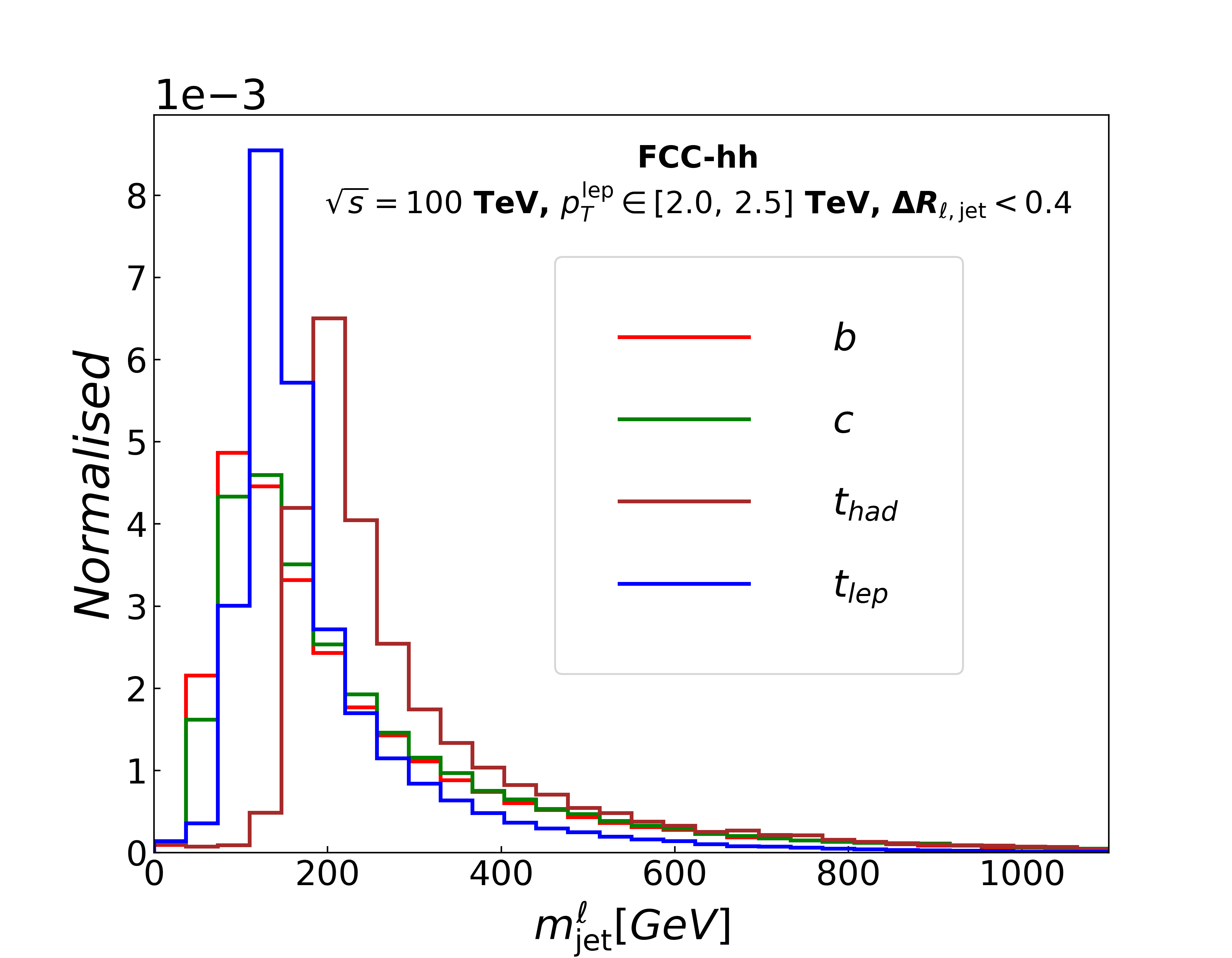}
	 \end{center}
 \caption{ Jet mass distribution within $\Delta R=0.4$ around the lepton for $t_{\mathrm{lep}}$, $t_{\mathrm{had}}$, $b$, and $c$ in the lepton $p_T$ ranges $0.1$--$0.5$~TeV (left) and $2.0$--$2.5$~TeV (right).}
 \label{jet_mass}
\end{figure}

When the top quark is highly boosted,  
we are very likely to capture the full top decay inside the cone of $\Delta R < 0.4$ around the lepton.
This effect is visible in Figure~\ref{jet_mass}, especially for the higher lepton $p_{T}$ bin shown in the plot on the right.

The jet mass distribution peaks quite close to the mass of the top quark -- this means that the $b$-jet near the lepton is actually very close to the lepton and they are fully contained within the cone.  
In contrast, when the lepton $p_T$ is smaller (the left plot), the $b$-jet is more separated from the lepton, so the cone around the lepton may not fully capture the $b$-jet, leading to a broader and softer distribution of jet mass.

\begin{figure}[H]
	\begin{center}
		\includegraphics[width=7.5cm, height=6.5cm]{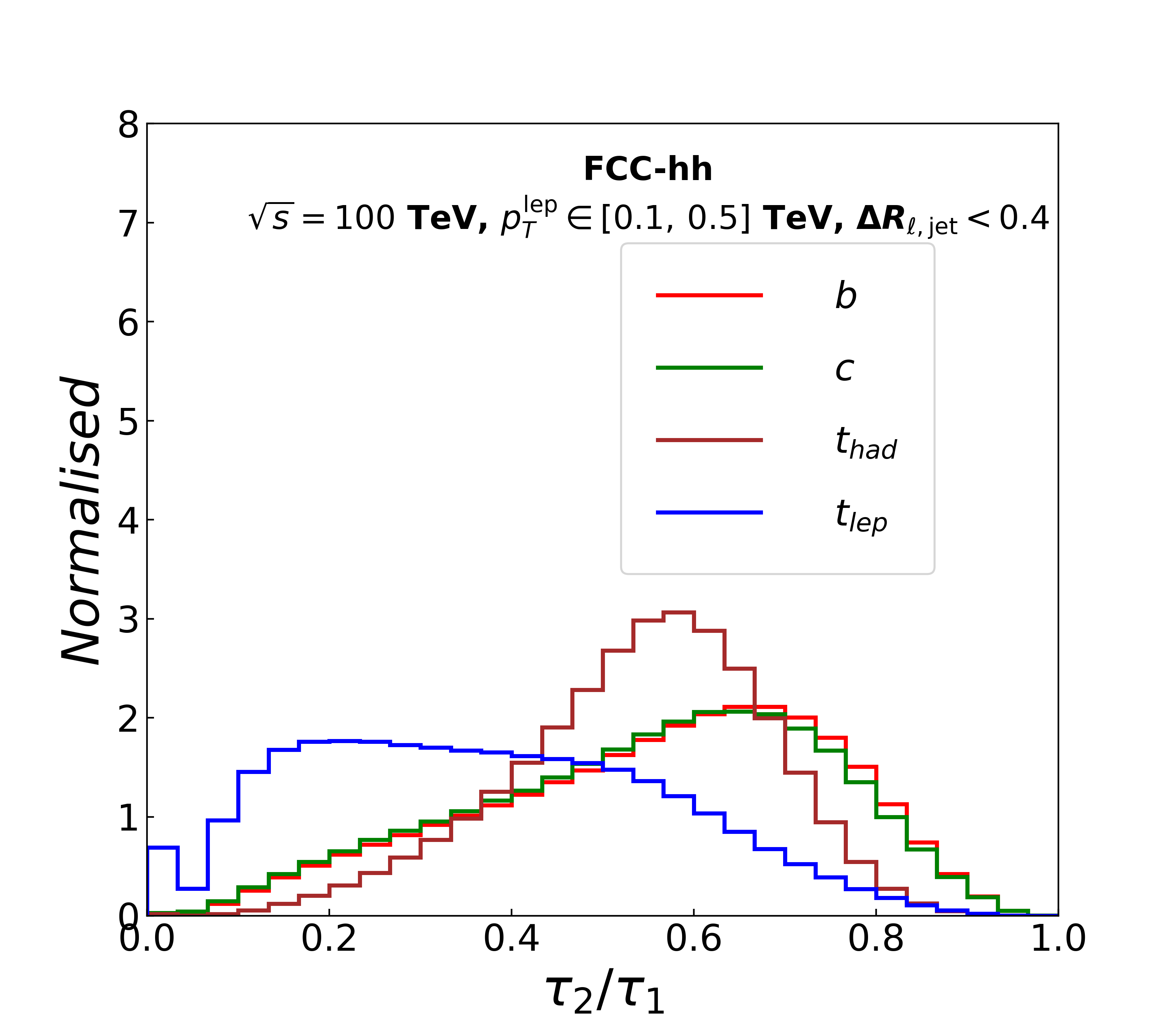}
        \includegraphics[width=7.5cm, height=6.5cm]{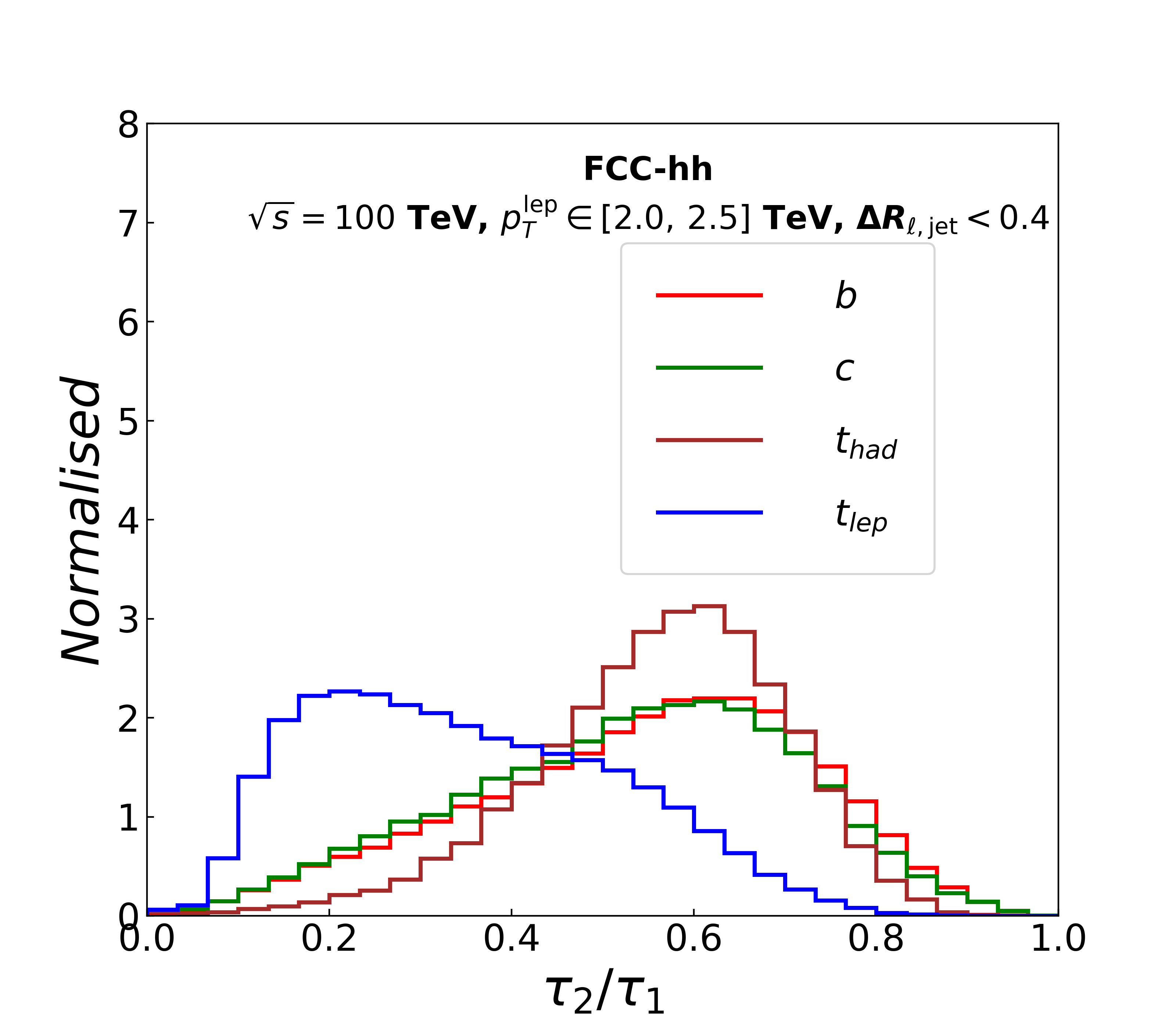}
	 \end{center}
 \caption{ Distribution of the subjettiness ratio $\tau_2/\tau_1$ for $t_{\mathrm{lep}}$ and the backgrounds $b$, $c$, and $t_{\mathrm{had}}$ in the lepton $p_T$ ranges $0.1$--$0.5$~TeV (left) and $2.0$--$2.5$~TeV (right). }
 \label{tau_ratio}
\end{figure}

\vskip 5 pt 
Another discriminating variable is N-subjettiness ($\tau_N$), 
which measures how well the radiation within a jet aligns with candidate axes. In the limit $\tau_N \to 0$, the jet radiation is well aligned with $N$ candidate axes, indicating an $N$-pronged structure. Conversely, $\tau_N \gg 0$ suggests the presence of more than $N$ hard cores. The ratio $\tau_{N+1}/\tau_N$ provides stronger signal-to-background discrimination than individual $\tau_N$ values alone. Figure~\ref{tau_ratio} shows the distribution of $\tau_2 / \tau_1$ for the leptonic top quark and other jets for two different $p_T$ bins. As we can see from the plots, it is a very efficient discriminating variable.  

\subsection{Tower-based variables }

\begin{figure}[H]
	\begin{center}
		\includegraphics[width=7.5cm, height=6.5cm]{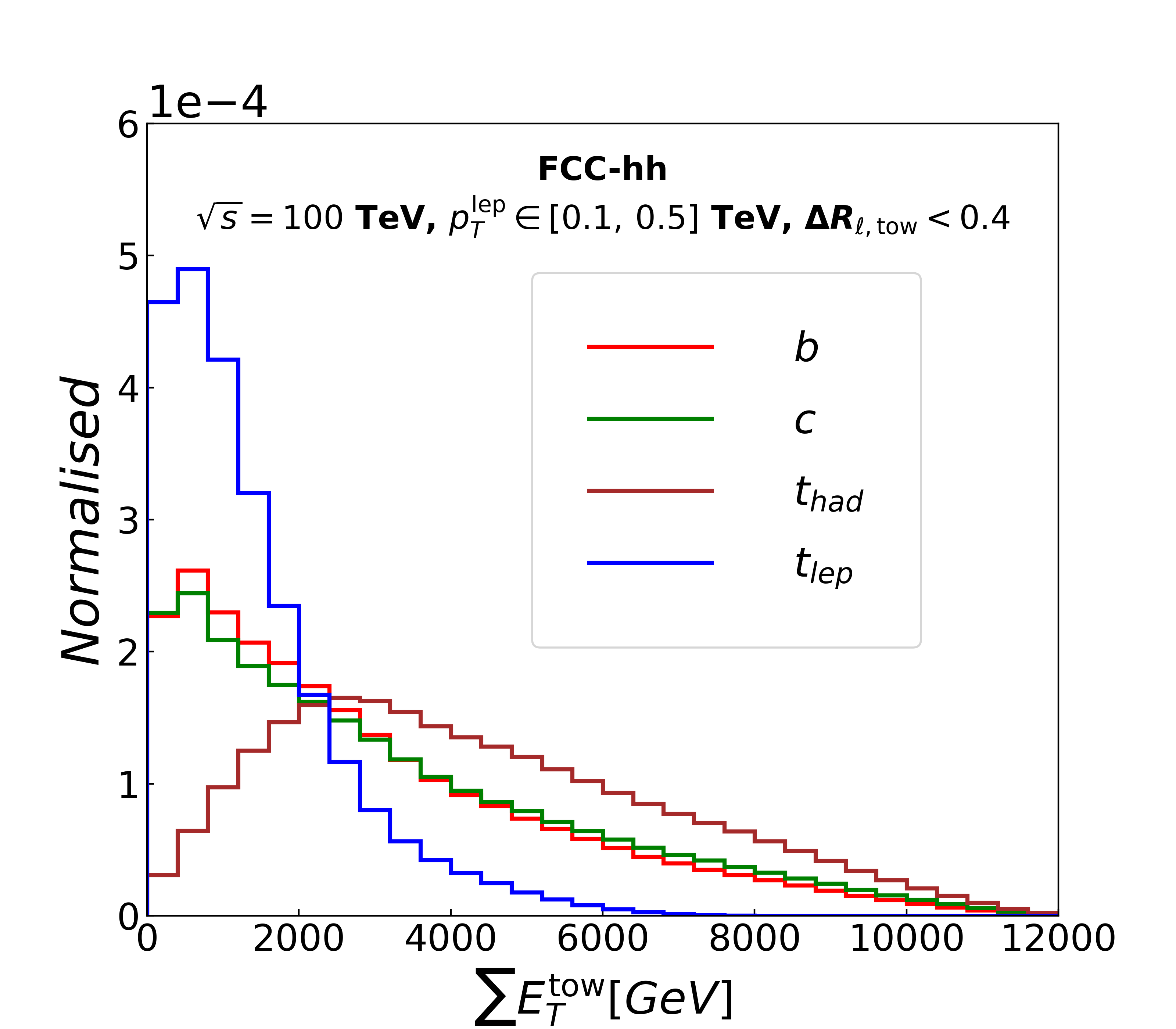}
        \includegraphics[width=7.5cm, height=6.5cm]{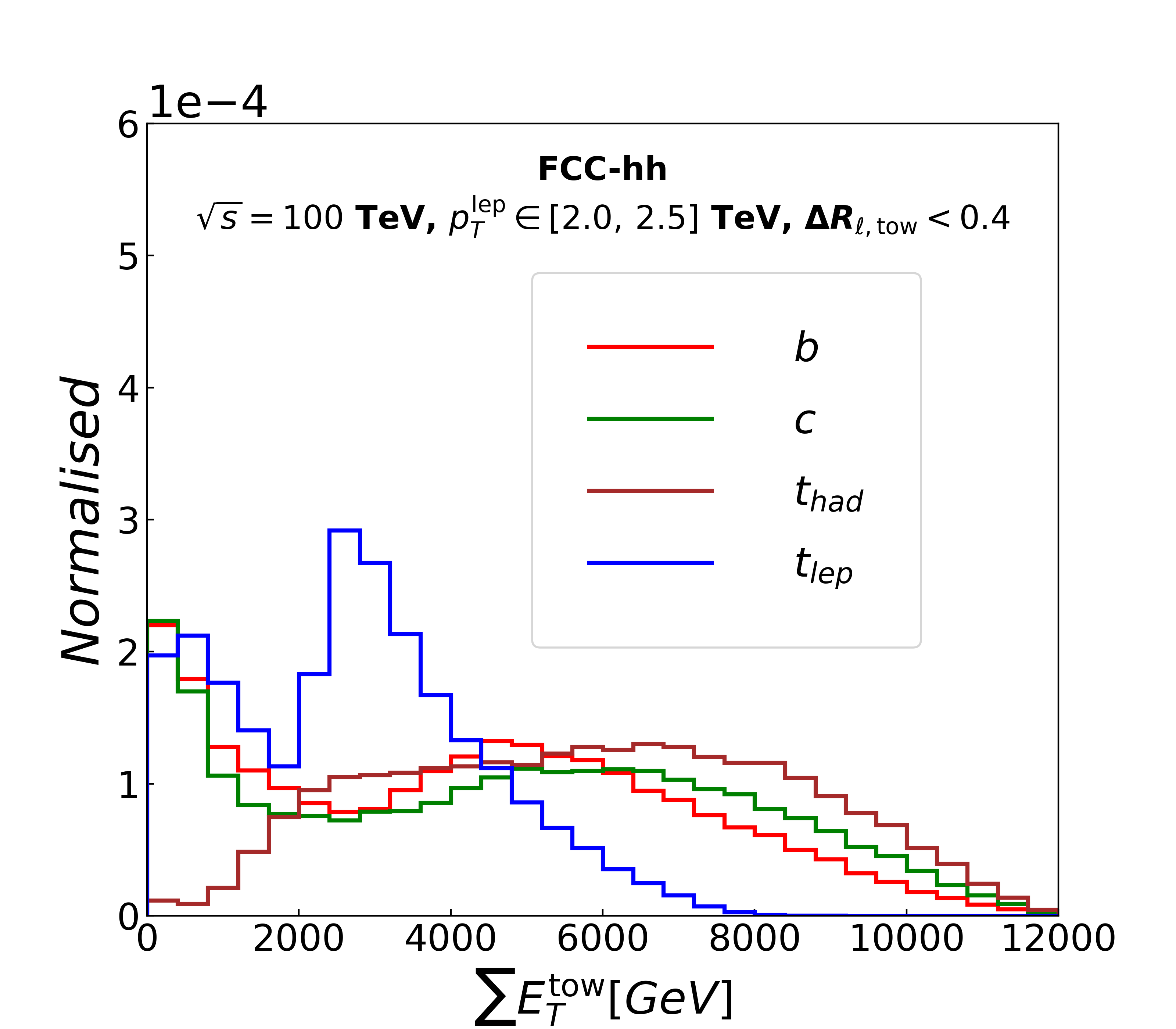}
	 \end{center}
 \caption{Distribution of absolute isolation variable constructed using $E_T$ of all calorimeter towers around the lepton within $\Delta R < 0.4$. Results are shown for lepton $p_T$ intervals $0.1-0.5$~TeV (left) and $2.0-2.5$~TeV (right).}
 \label{tow_sum}
\end{figure}

\begin{figure}[H]
	\begin{center}
		\includegraphics[width=7.5cm, height=6cm]{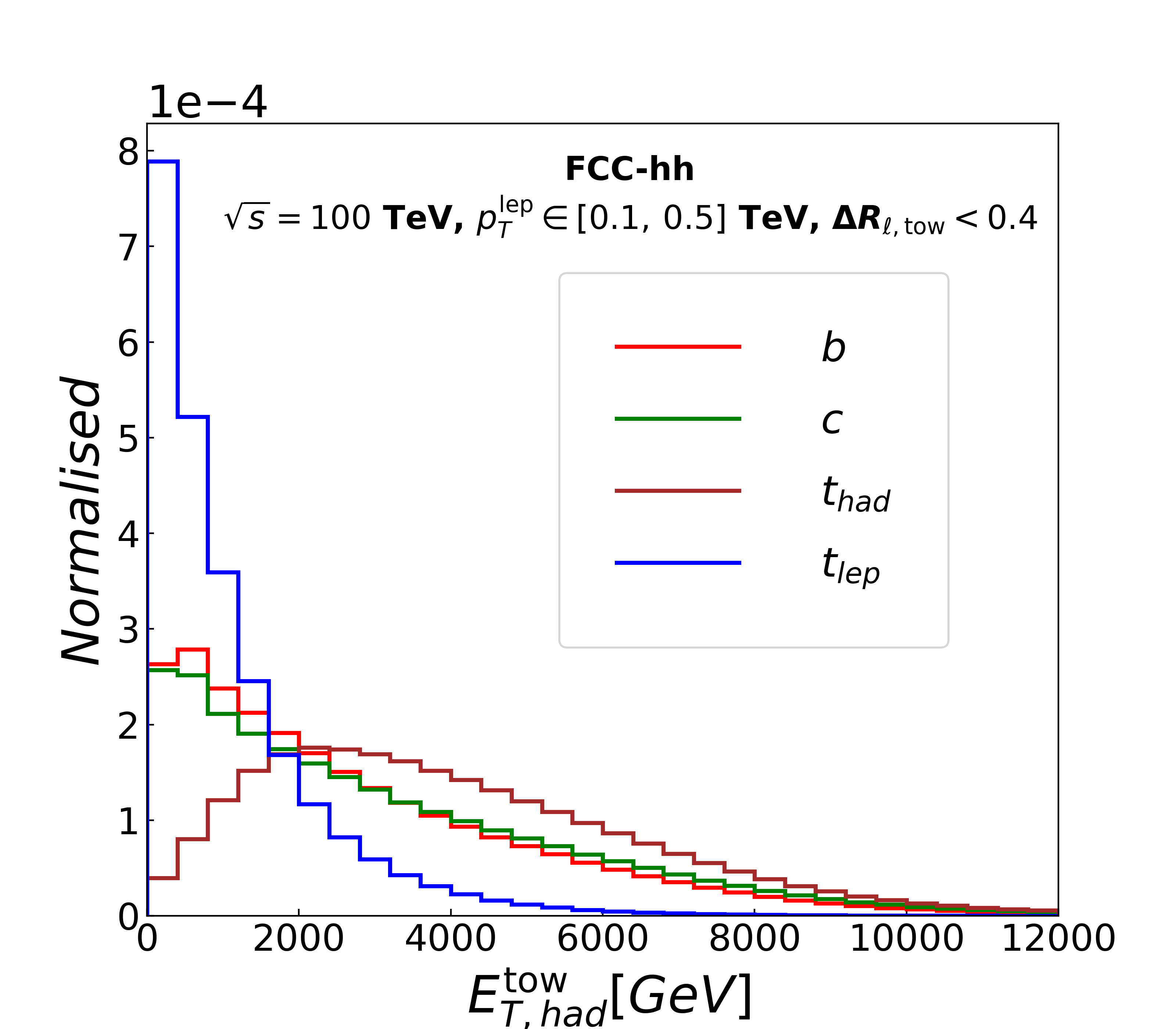}
        \includegraphics[width=7.5cm, height=6cm]{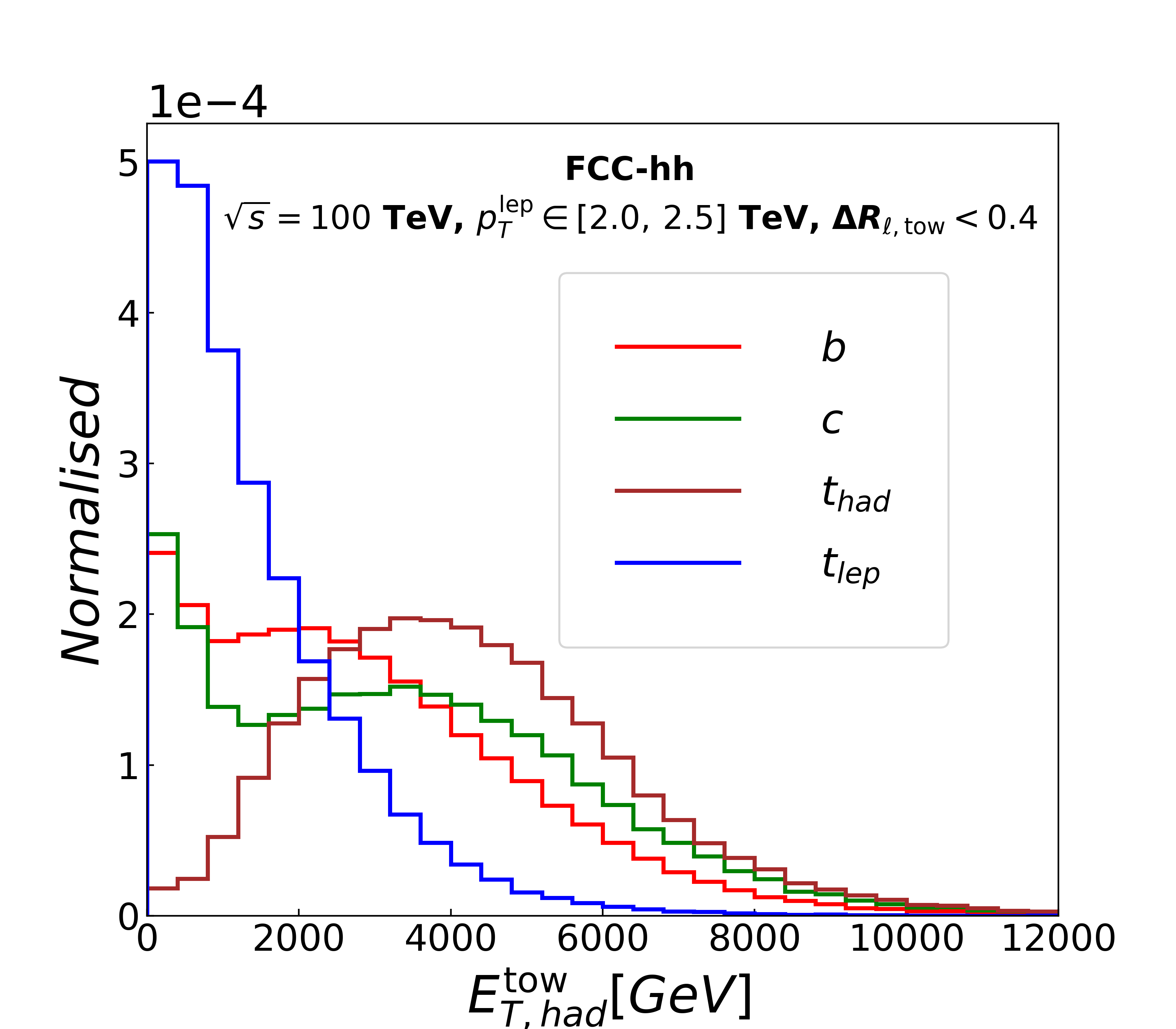}
	 \end{center}
 \caption{Hadronic activity around the lepton within $\Delta R <0.4$, shown for the lepton $p_T$ ranges $0.1-0.5$~TeV (left) and $2.0-2.5$~TeV (right).
 }
 \label{tow_had}
\end{figure}

The tower-based absolute isolation variable measures the total transverse energy within a cone of $\Delta R < 0.4$ around the lepton.  
The distribution of this variable shows good discrimination power, as we can see in Figure~\ref{tow_sum}. A similar behavior is observed for the purely hadronic tower-based absolute isolation variable, as presented in Figure~\ref{tow_had}.

\clearpage
\subsection{Performance of the jet tagger}
\label{jet tagger}

To distinguish leptonic top jets from hadronic top jets, $b$-jets, and $c$-jets, we investigate two supervised deep learning approaches that use different representations of jet information. A distinctive feature of boosted leptonic top jets is the presence of a high-$p_T$, non-isolated lepton inside the jet cone, originating from the decay $t \to b\ell\nu$. To exploit this feature, we implement a variable-based Deep Neural Network (DNN) classifier in
PyTorch~\cite{Paszke2019}, using a set of high-level, physics-motivated observables, and a Particle Transformer (ParT) that instead operates directly on the particle-level constituents of the jet. While constituent-based architectures such as ParT have demonstrated strong performance at the LHC, their performance in the ultra-boosted FCC-hh regime remains less explored.
At these energies, the top-quark decay products become collimated, providing a dense particle environment within the jet. This motivates us to investigate whether carefully designed high-level observables can achieve discrimination comparable to that of a constituent-based architecture. We address this by comparing the two tagging approaches.

A fully connected DNN with three hidden layers (512, 256, and 128 neurons) was implemented in PyTorch. The network takes as input the set of high-level physics-motivated observables summarised in Table~\ref{object_features}. GELU activation functions~\cite{Hendrycks2016}, dropout regularisation, and early stopping were employed to improve model generalisation. The network was trained using the Adam optimiser with a categorical cross-entropy loss function. Further details of the DNN architecture and training procedure are provided in Appendix~\ref{dnn_part_arct}.

In contrast to the DNN, the ParT classifier operates directly on the individual jet constituents. Each constituent (the lepton, tracks, and calorimeter towers) is represented by nine input features: transverse momentum, the angular variables $\Delta R(\ell,\mathrm{trk/tow})$, $\Delta\eta(\ell,\mathrm{trk/tow})$, and $\Delta\phi(\ell,\mathrm{trk/tow})$, the transverse momentum (or transverse energy) fraction, the transverse and longitudinal impact parameters ($d_0$ and $d_z$), and the electromagnetic and hadronic energies. These constituent features are embedded in a latent feature space and processed through four transformer blocks that employ multi-head self-attention, residual connections, layer normalisation, and feed-forward networks. A learnable classification (CLS) token is used to aggregate the global jet information. Its final representation is passed to a fully connected classifier with four output nodes corresponding to the leptonic top ($t_{\mathrm{lep}}$), hadronic top ($t_{\mathrm{had}}$), $b$-jet, and $c$-jet classes. The attention mechanism additionally incorporates learnable geometric biases constructed from the pairwise angular variables $(\Delta\eta,\Delta\phi,\Delta R)$ between jet constituents. The complete network architecture and training configuration are summarised in Appendix~\ref{dnn_part_arct}.

The classification performance of the DNN and ParT classifiers is evaluated using confusion matrices, where the diagonal elements represent the tagging efficiencies and the off-diagonal elements correspond to the mistag rates for the four jet classes.

\begin{figure}[H]
	\centering
	\includegraphics[width=7.5cm, height=6cm]{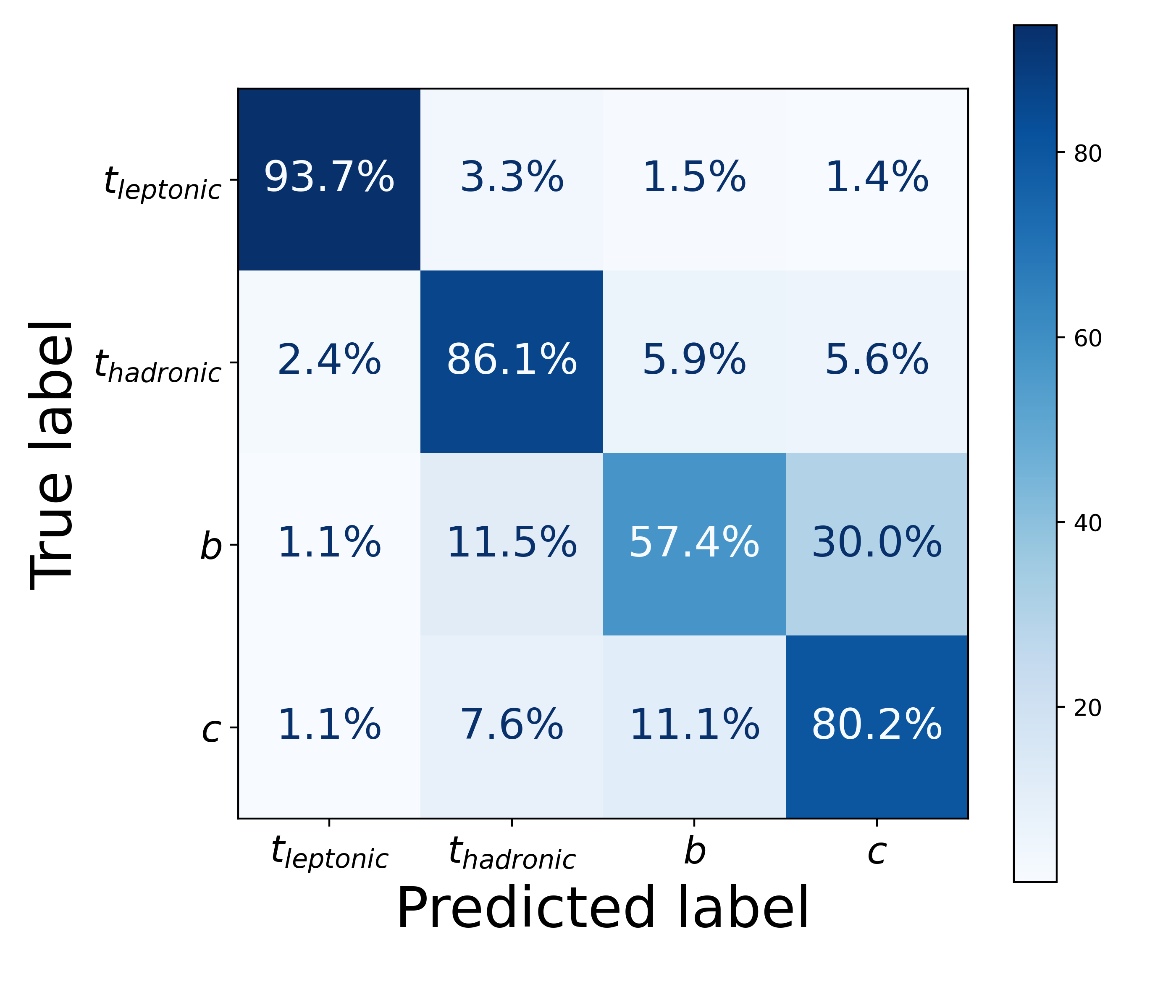}
    {\includegraphics[width=7.5cm, height=6cm]{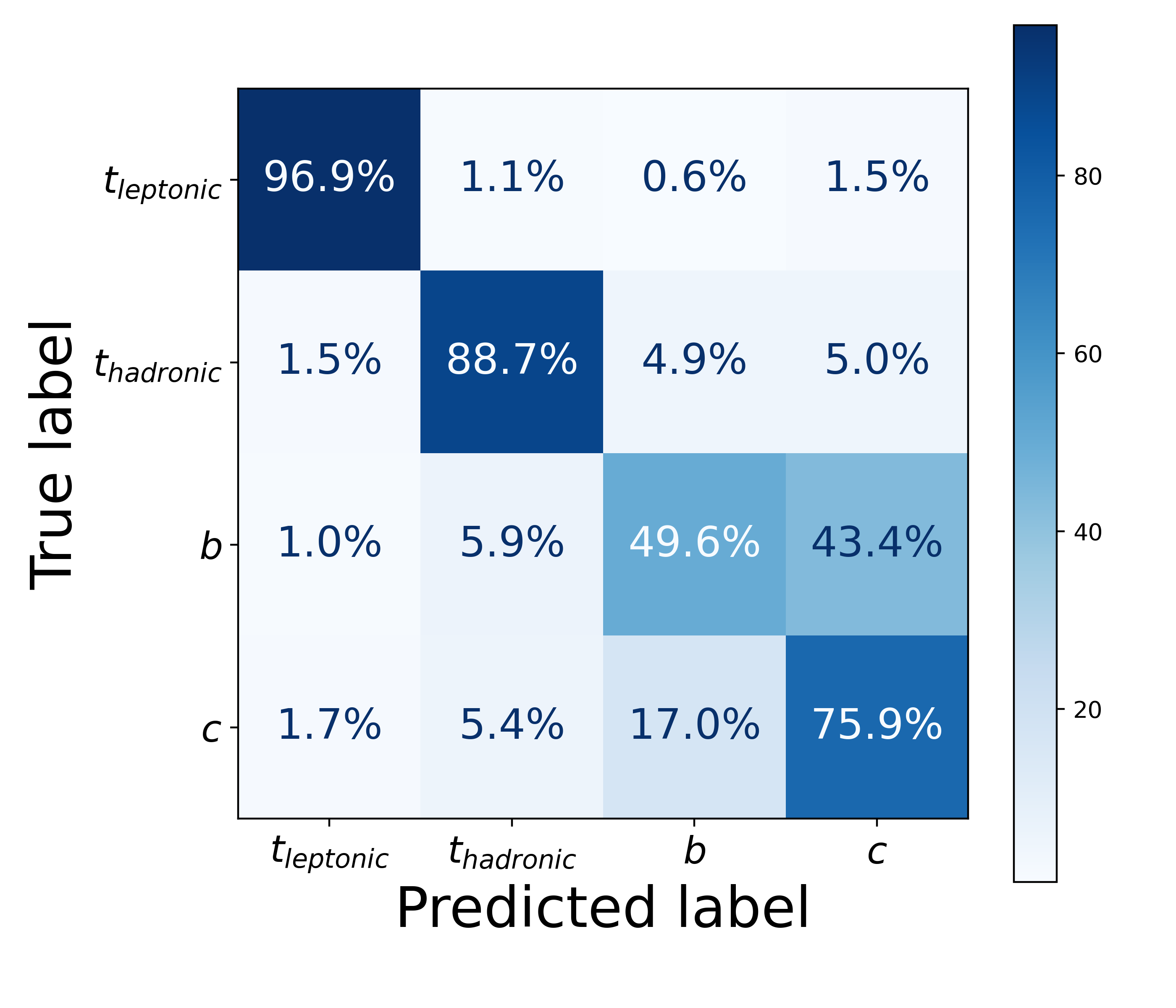}}
 \caption{
 Representative multi-class confusion matrices for the DNN (left) and ParT (right) classifiers in the $1.0$--$1.5$~TeV lepton transverse momentum interval.
}
 \label{conf_mat}
\end{figure}

Both classifiers yield excellent efficiency for the signal class ($t_{\mathrm{lep}}$) across the considered kinematic range. As a representative example, Figure~\ref{conf_mat} presents the confusion matrices for the $1.0$--$1.5$~TeV interval, where the ParT classifier achieves a leptonic-top tagging efficiency of $96.9\%$, compared with $93.7\%$ for the DNN.

To further evaluate the tagging power of the DNN and ParT frameworks, Receiver Operating Characteristic (ROC) curves were constructed for each jet category within each lepton transverse momentum interval. The Area Under the Curve (AUC) provides a threshold-independent measure of the classifier's ability to distinguish between the target jet class and the remaining categories.

\begin{figure}[H]
	\centering
	\includegraphics[width=7.5cm, height=6cm]{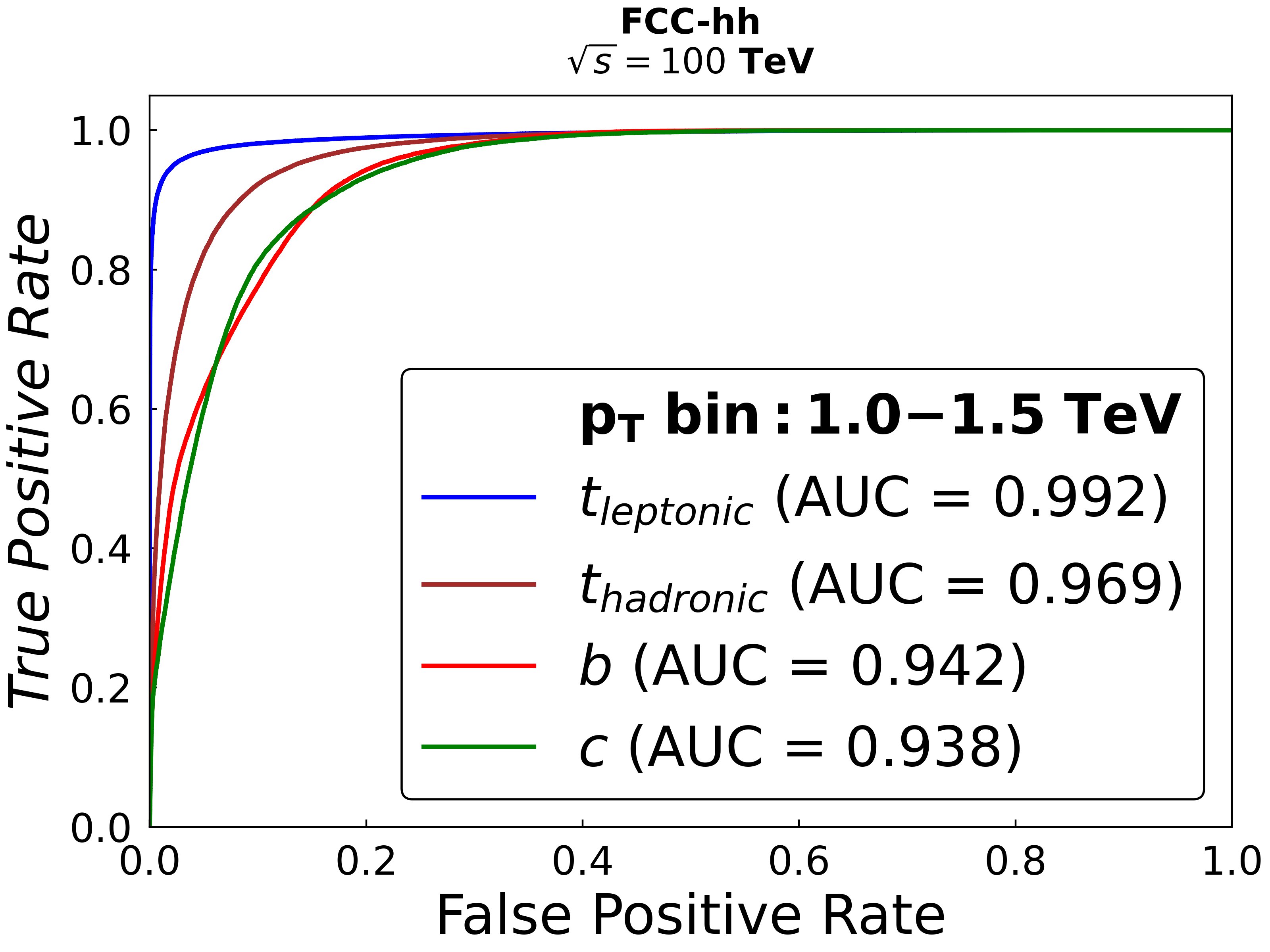}  
    \includegraphics[width=7.5cm, height=6cm]{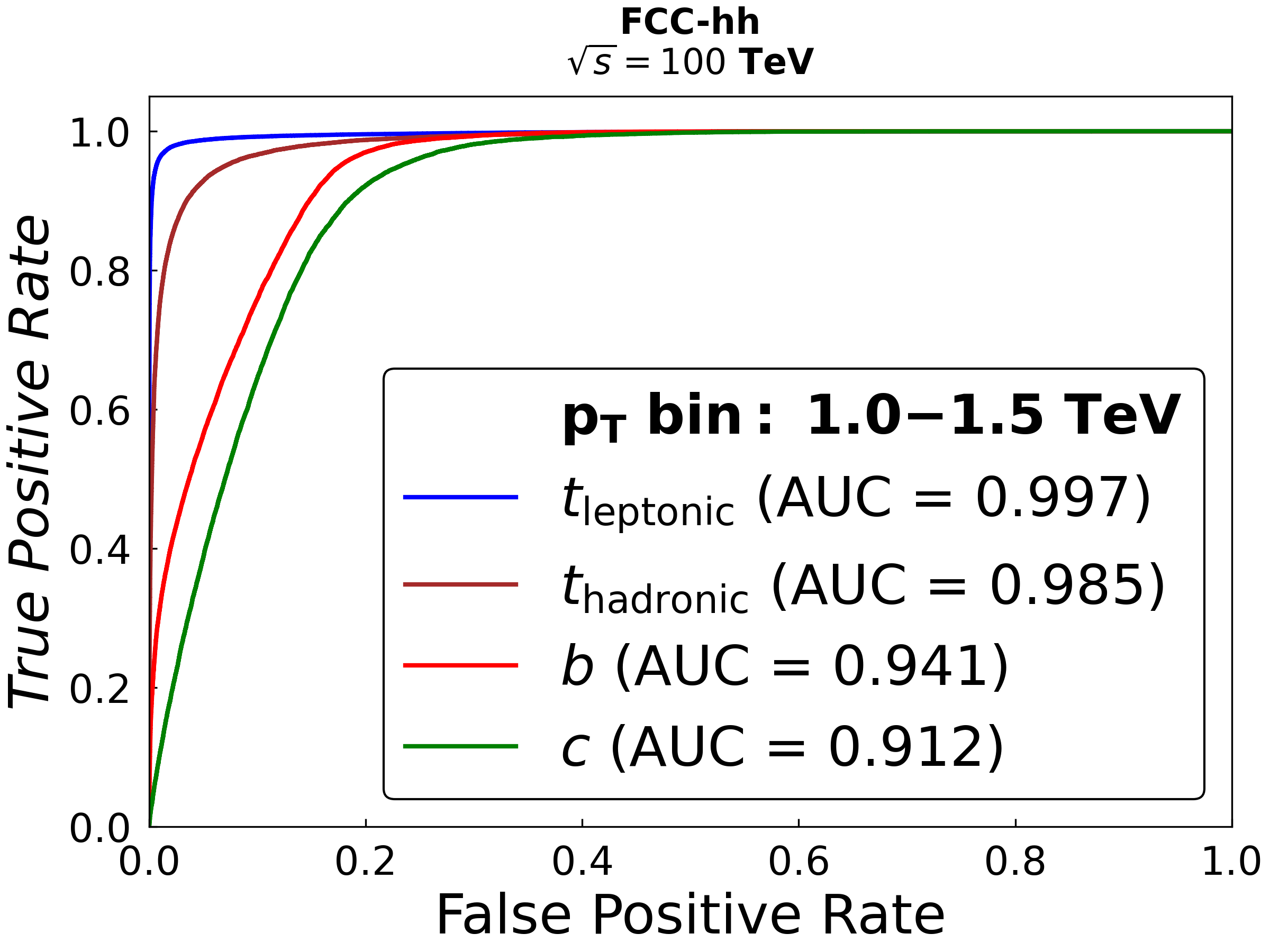}
 \caption{Receiver Operating Characteristic (ROC) curves and corresponding Area Under the Curve (AUC) values for the multi-class DNN (left) and ParT (right).  Performance is shown for the $1.0$–$1.5$~TeV $p_T$ interval. The curves illustrate the one-vs-rest identification efficiency for leptonic top jets ($t_{\text{lep}}$), hadronic top jets ($t_{\text{had}}$), $b$-jets, and $c$-jets.
}
 \label{roc} 
\end{figure}

For the DNN, the leptonic-top AUC increases from 0.981 in the $0.1$--$0.5$~TeV interval to 0.996 for $p_T > 3$~TeV, whereas the ParT classifier maintains a nearly constant AUC, ranging from 0.994 to 0.997 across all $p_T$ bins. For the other categories, ParT separates hadronic tops better in every $p_T$ bin, the two classifiers perform similarly for $b$ jets, and the DNN separates $c$ jets better. As an illustration, Figure~\ref{roc} shows the ROC curves for the $1.0$--$1.5$~TeV $p_T$ interval. \footnote{The cone radius $\Delta R<0.4$ is chosen to capture the collimated $b$+lepton system, whose angular separation peaks below $0.4$ in the boosted regime. We verified that the tagging performance is not sensitive to this choice: repeating the DNN training with a tighter cone of $\Delta R<0.2$, while keeping the same input variables and network architecture, leaves the leptonic-top discrimination essentially unchanged, with the AUC increasing from $0.992$ to $0.993$ and the tagging efficiency from $93.7\%$ to $94.8\%$ in the $1.0$--$1.5$~TeV bin. The corresponding ROC curves and confusion matrix are shown in Appendix~\ref{cone_02}.}

\begin{figure}[H]
	\centering
	\includegraphics[width=7.5cm, height=6cm]{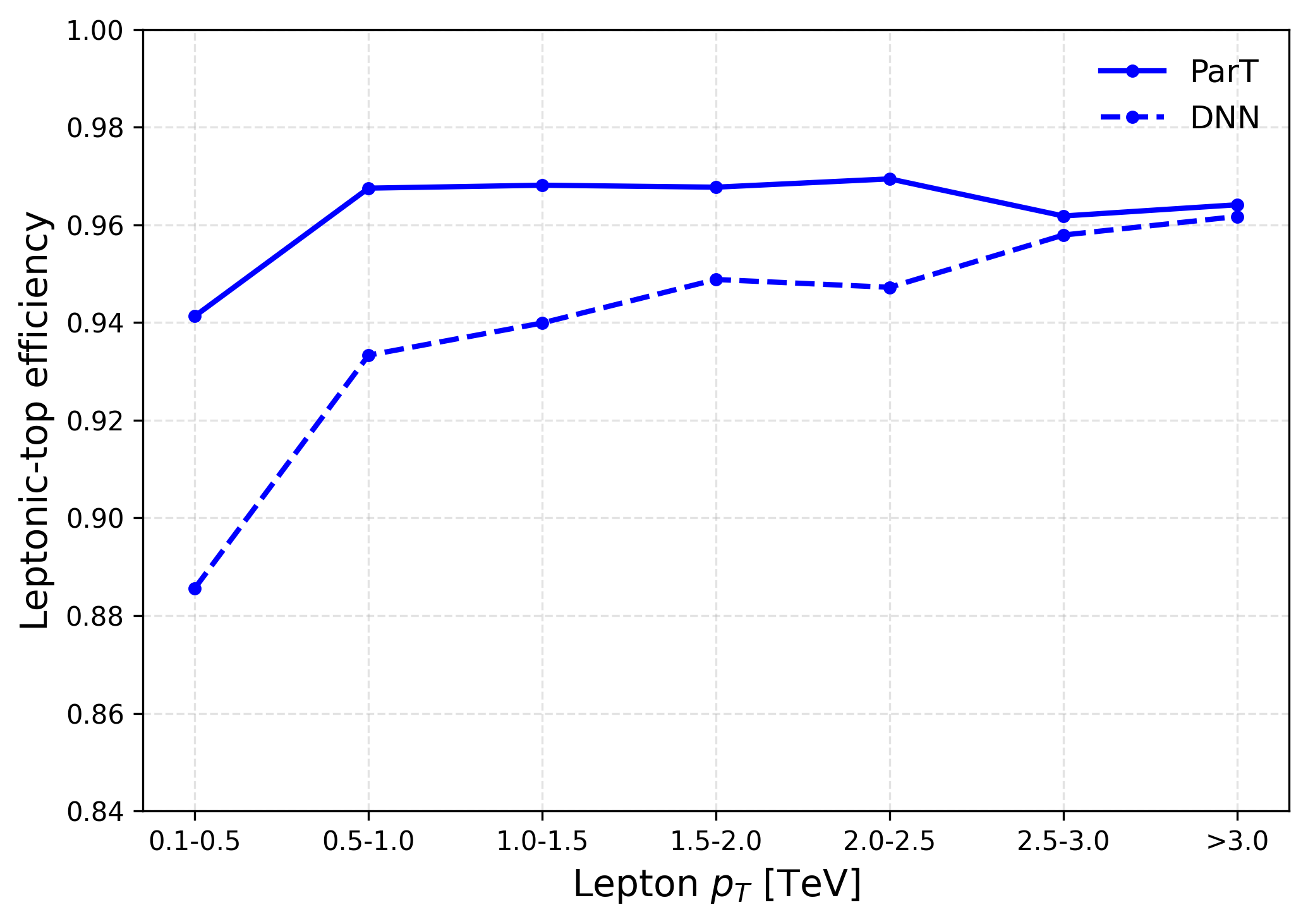}  
    \includegraphics[width=7.5cm, height=6cm]{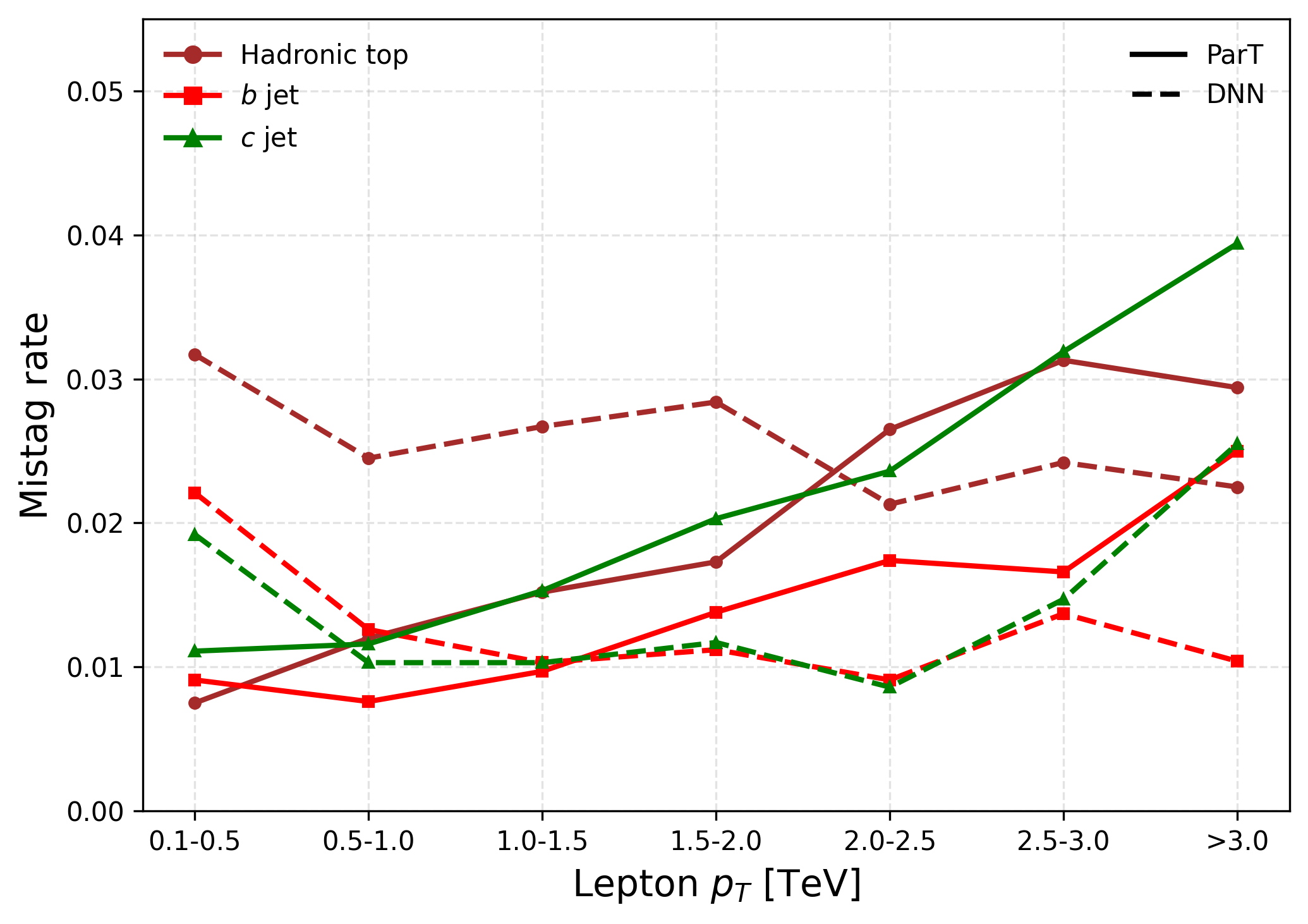}
 \caption{Leptonic-top tagging efficiency (left) and mistag rates for the hadronic-top, $b$-jet and $c$-jet categories (right) as a function of the lepton transverse momentum, for ParT (solid) and the DNN (dashed). A jet is classified as $t_{\mathrm{lep}}$ if its $t_{\mathrm{lep}}$ score exceeds $0.4$.}
 \label{eff_pt} 
\end{figure}

Figure~\ref{eff_pt} summarises the $p_T$ dependence of the two classifiers at a working point of $0.4$ on the leptonic-top score. ParT achieves a higher leptonic-top tagging efficiency than the DNN in every $p_T$ bin, reaching approximately $94\%$ in the lowest bin and $96$--$97\%$ above $0.5$~TeV, while the DNN efficiency increases from approximately $89\%$ to $96\%$. These results demonstrate that the presence of a non-isolated high-$p_T$ lepton provides a robust discriminating signature even at extreme Lorentz boosts. ParT also yields lower mistagging rates across all three background categories in the lower-$p_T$ bins. With increasing transverse momentum, however, the DNN efficiency approaches that of ParT and, above about $2$~TeV, yields lower mistag rates for all three background categories. This behaviour suggests that the constituent-level information exploited by ParT is particularly useful when the decay products of the parent quark remain sufficiently angularly resolved, while the engineered jet-level observables used by the DNN retain substantial discriminating power in the ultra-boosted regime.

\clearpage
\section{Multivariate Event-Level Analysis}
\label{results}

The variable-based jet-tagger and ParT tagger, described in the previous section, are then used in a well-motivated collider search analysis, which will be described in this section. In the boosted topology of a 100~TeV collider, the decay products of heavy, vector-like $B^\prime$ quarks will carry very high $p_T$.

Following the object-level jet-tagging analysis, we implement a second DNN for the event-level signal-versus-background classification. The signal of our interest is the single production of a heavy vector-like $B^\prime$ quark in association with a jet, $pp \to B^{\prime}\,j$, where the $B^\prime$ quark decays to $tW$ final state, $B^\prime \to tW$, with the top quark decaying leptonically as $t \to b\ell\nu$ and the $W$ boson decaying as $W \to \ell\nu$.
The final state contains one isolated lepton from the $W$ boson decay and one non-isolated lepton from the boosted top-quark decay, along with significant $E_T^{\text{miss}}$ from the two neutrinos. We select leptons with $p_T > 100$~GeV and $|\eta| < 2.5$. A lepton is classified as isolated if both $\text{Iso}_{p_T} < 0.1$ and $\text{Iso}_{E_T} < 0.1$, while non-isolated leptons are required to satisfy $\text{Iso}_{p_T} > 0.1$. We require exactly one isolated and one non-isolated lepton in each event to select the target topology and suppress backgrounds with different lepton
multiplicities.

The primary SM background events considered in this analysis are
$t\bar{t}$, $b\bar{b}$, $c\bar{c}$, diboson ($VV$), $W+$jet, and
$tW$ production. The $tW$ process constitutes an irreducible
background for this channel. The Standard Model background samples
are generated using \textsc{PYTHIA}~8.315~\cite{pythia8315}, with the
NNPDF2.3 LO PDF set accessed through LHAPDF6~\cite{lhapdf,nnpdf23}.
The events are generated with appropriate kinematic filters to ensure
sufficient statistics in the high-$p_T$ tails relevant to the
100~TeV FCC-hh environment.

We utilise a set of global kinematic observables that exploit the distinct topological differences between the $B^\prime$ signal and the SM background events. These variables serve as the input features for the event-level DNN. The distributions for some of these key observables are presented in Figure~\ref{event_var}.
\begin{itemize}

\item Total hadronic activity ($H_{T}$): It is defined as the scalar sum of the $p_T$ of all reconstructed jets in the event. It characterises the overall jet activity in the event. As we see in the $H_{T}$ plot, it has decent discriminating power between the signal and certain backgrounds.

\item $b$-jet multiplicity near isolated lepton: The number of $b$-tagged jets within $\Delta R < 0.2$ of the isolated lepton provides an additional discriminating observable. In the signal process, the isolated lepton is typically well separated from $b$-jets, whereas in dominant backgrounds, $b$-jets are more likely to be found in its vicinity.

\item $E_{T}^\text{miss}$: This variable accounts for the momentum imbalance in the transverse plane, primarily attributed to neutrinos in the final state that go undetected. The signal events involve multiple neutrinos from the leptonic decays of the $W$ boson and the top quark. In most events, this leads to a high $E_{T}^\text{miss}$ signature that provides significant separation from certain backgrounds.

\item Dilepton invariant mass: We compute the invariant mass of the isolated and non-isolated leptons, $m(\ell_{\text{iso}}, \ell_{\text{non-iso}})$. For heavier $B^\prime$ masses, the signal tends to populate the high-mass region of this distribution, providing discrimination against the backgrounds.
\end{itemize}

\begin{figure}[H]
	\centering
	\subfloat[]{\includegraphics[width=7.5cm]{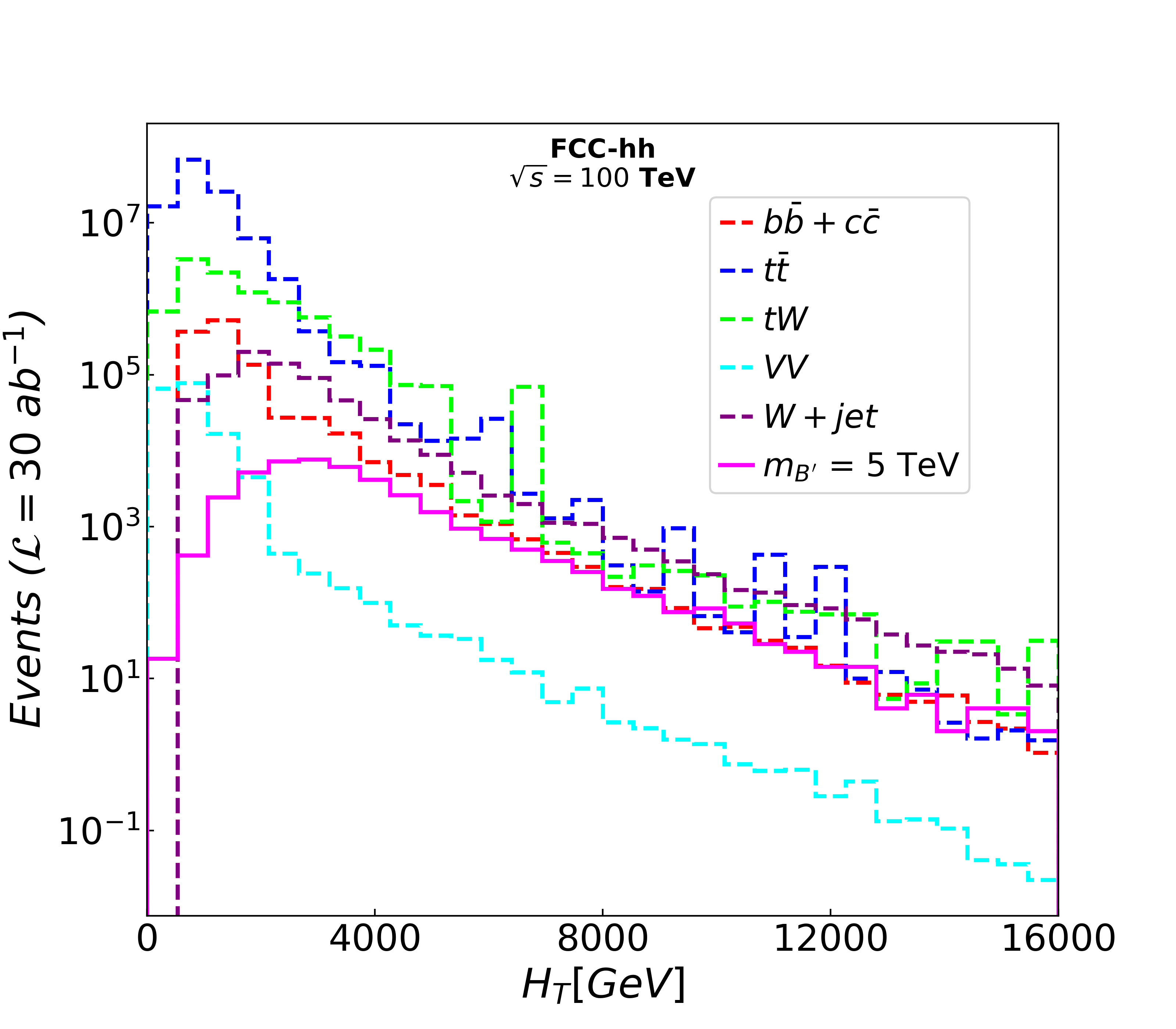}}
     \subfloat[]{\includegraphics[width=7.5cm]{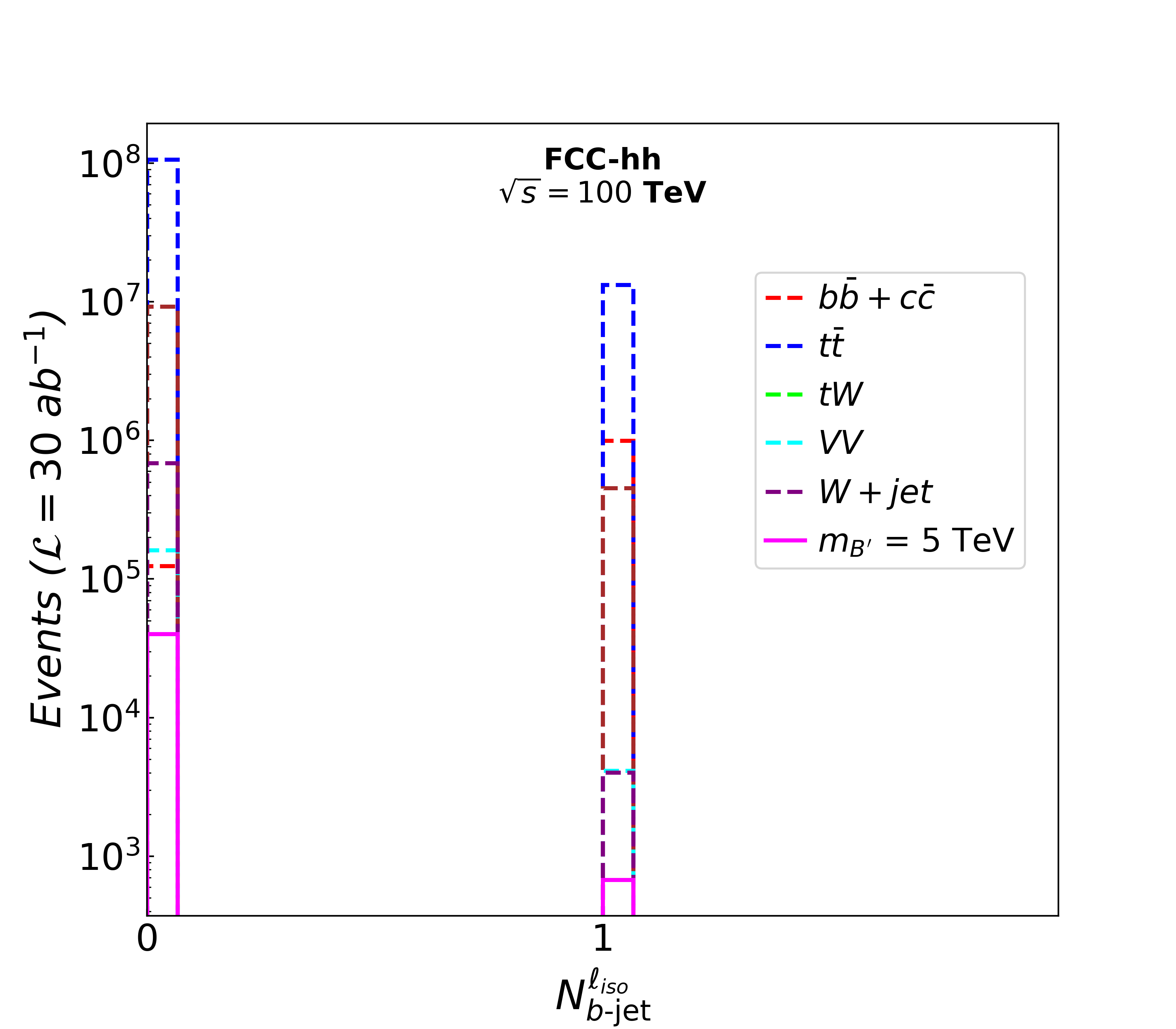}}

    \phantomcaption
\end{figure}   
\begin{figure}[H]
\ContinuedFloat
    \centering
    \subfloat[]{\includegraphics[width=7.5cm]{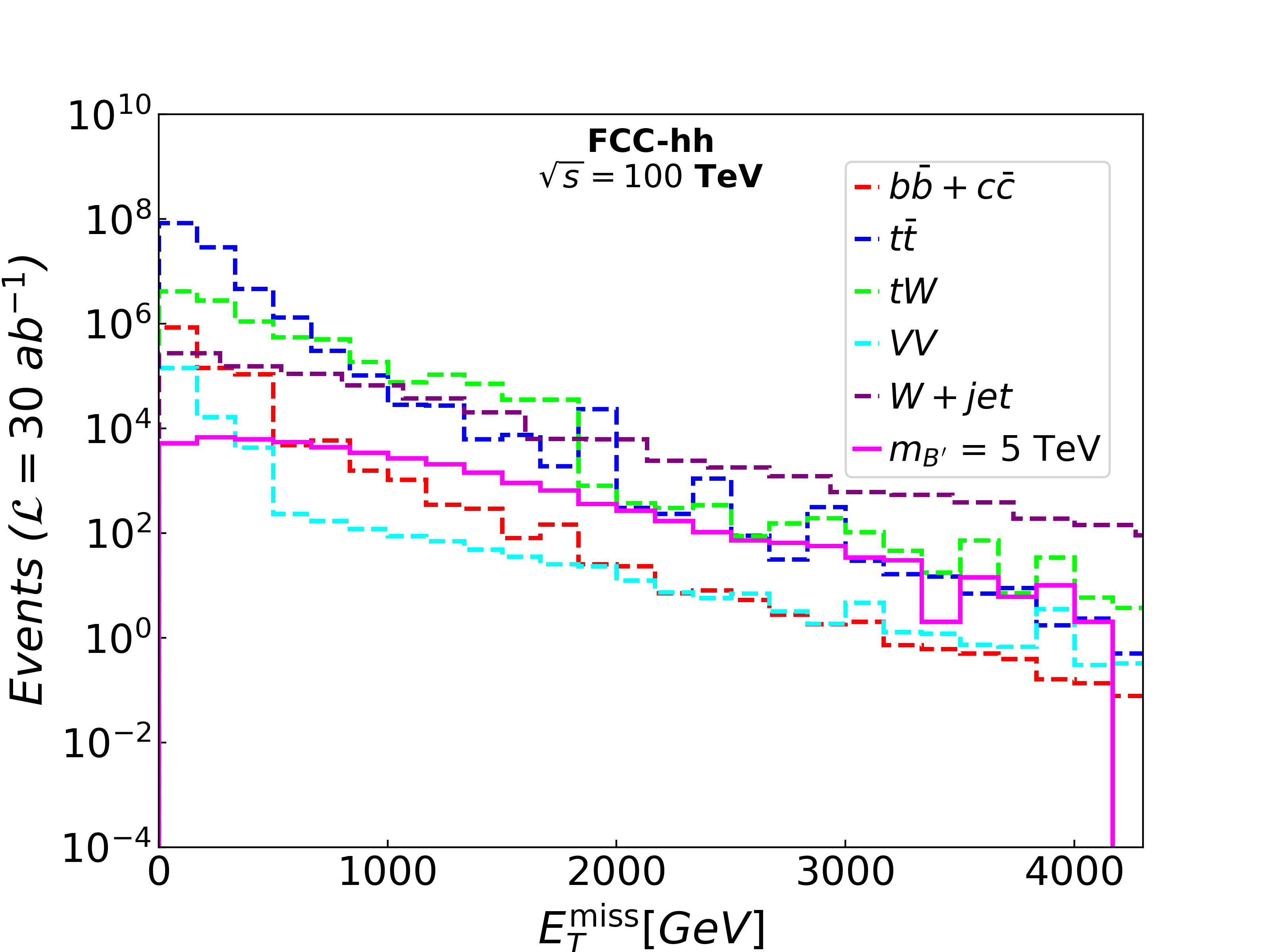}}
    \subfloat[]{\includegraphics[width=7.5cm]{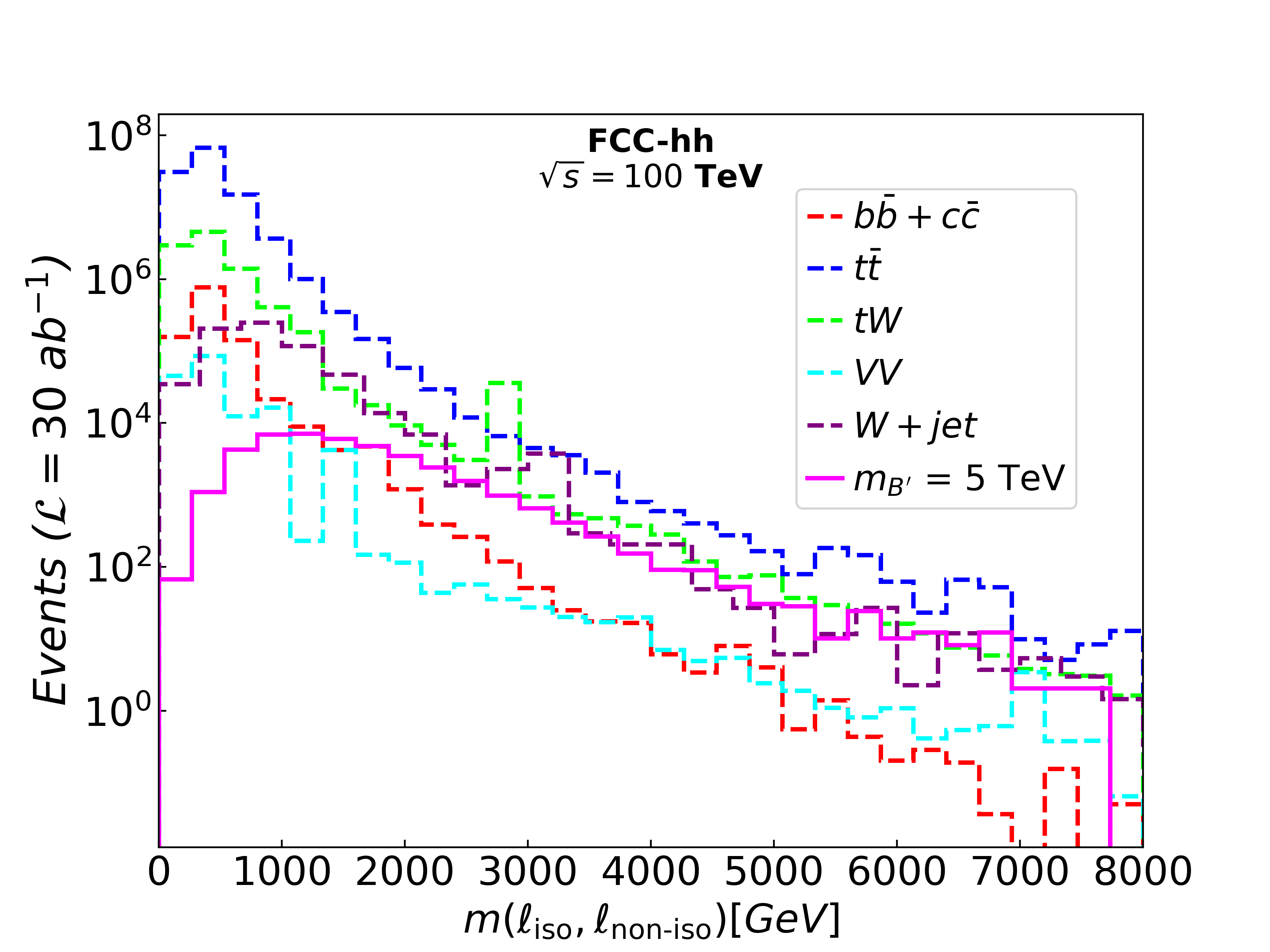}}
 \caption{ Distributions of the event-level variables for the $B^\prime$ signal ($m_{B^\prime} = 5$~TeV) and dominant SM backgrounds. The panels show: (a) the scalar sum of jet transverse momenta $H_T$, 
 (b) the multiplicity of $b$-jets within a cone of $\Delta R < 0.2$ around the isolated lepton,
 (c) the missing transverse energy $E_{T}^{\text{miss}}$, (d) the invariant mass of the isolated and non-isolated leptons $m(\ell_{\text{iso}}, \ell_{\text{non-iso}})$.
}
 \label{event_var}
\end{figure}

\begin{figure}[H]
	\begin{center}
	\includegraphics[width=14cm]{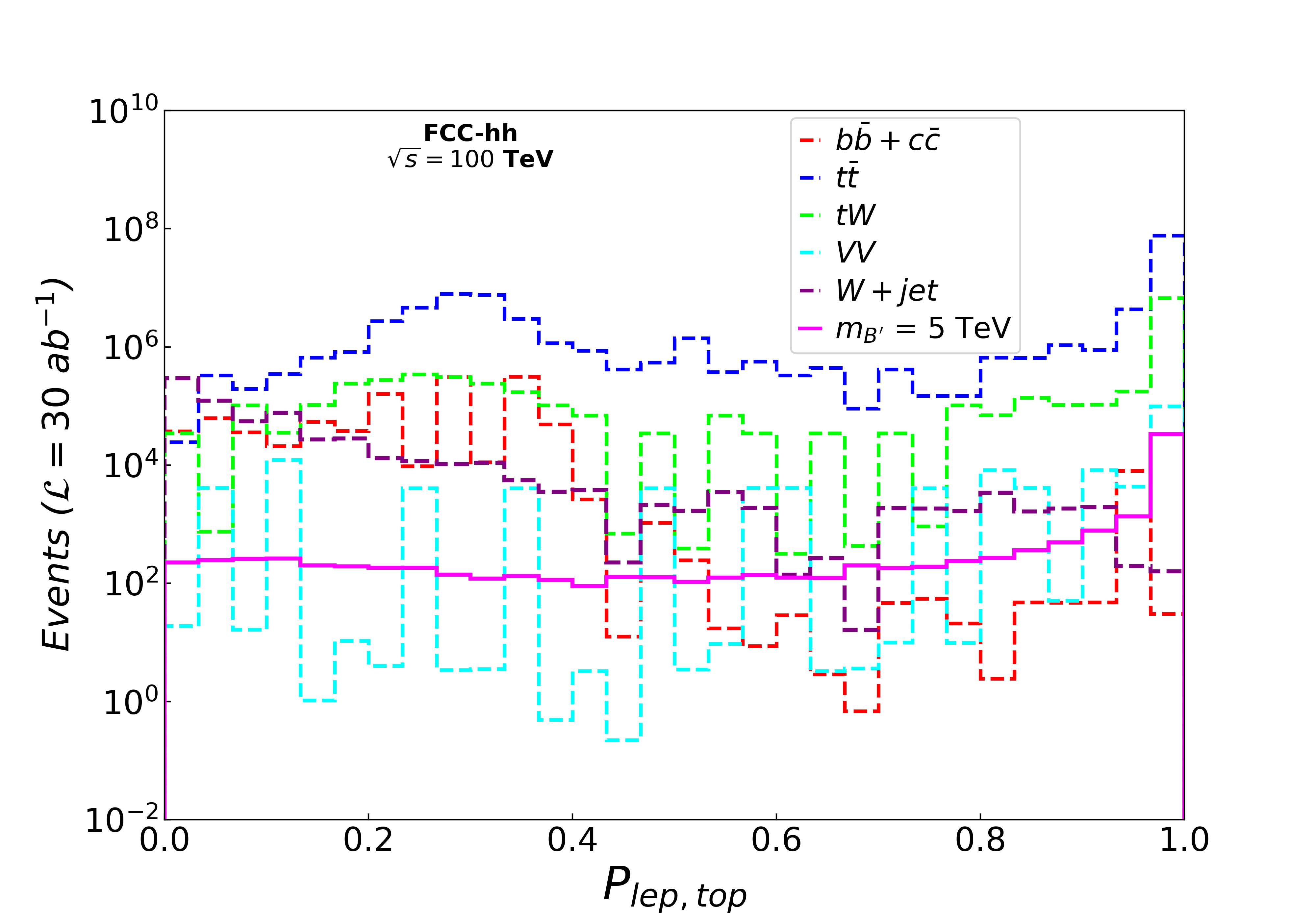}
	 \end{center}
 \caption{Output score of the top-tagger DNN evaluated on the non-isolated lepton in signal and background events.
}
 \label{event_level_prob}
\end{figure}

The score of the DNN top-tagger, $P_{\text{lep, top}}$, evaluated on the non-isolated lepton is plotted in Figure~\ref{event_level_prob} for signal and background events. The $B^\prime$ signal and the SM $tW$ and $t\bar{t}$ backgrounds exhibit high scores, with a pronounced accumulation of events near unity. 
Such behaviour is expected because these backgrounds can produce genuine boosted leptonic top quarks. Near $P_{\text{lep, top}} = 1$, the peak in the diboson ($VV$) background likely originates from leptonic decay of boosted $Z$ and boosted $W \to cs$, yielding non-isolated leptons. This can closely fake a boosted leptonic top signature. In contrast, most $b\bar{b}$, $c\bar{c}$, and $W+$jet events are pushed into the low-probability region, heavily suppressing those backgrounds.

\begin{table}[htb]
    \centering
    \renewcommand{\arraystretch}{1.4}
    \begin{tabular}{|c|l|}

\hline
\textbf{Notation} & \textbf{Definition} \\
 
\hline
$E_T^{\text{miss}}$& Missing transverse energy  \\
$H_{T}$ &  Scalar sum of jet $p_T$ \\

$p_{T,\text{lep}}^{\text{noniso}}$ & Transverse momentum of the non-isolated lepton \\

$p_{T,\text{lep}}^{\text{iso}}$ & Transverse momentum of the isolated lepton \\

$M_{T,\text{lep}}^{\text{iso}}$ & Transverse mass of the isolated lepton and $E_T^{\text{miss}}$ \\

$M_{T,\text{lep}}^{\text{noniso}}$&  Transverse mass of the non-isolated lepton and $E_T^{\text{miss}}$  \\

$f_{p_T}^{\ell,jet}$ & $p_T$ fraction of non-isolated lepton relative to its jet ($\Delta R < 0.2$)\\

$m(\ell_{\text{non-iso}}, \text{jet})$ & Invariant mass of non-isolated lepton and jet ($\Delta R < 0.2$) \\

$N_{b\text{-jet}}^{\ell_{\text{non-iso}}}$  & Number of $b$-jets near the non-isolated lepton \\

$N_{b\text{-jet}}^{\ell_{\text{iso}}}$ & Number of $b$-jets near the isolated lepton \\

$\Delta\phi (\ell_{\text{iso}},E_T^{\text{miss}}) $ & $\Delta\phi$ between isolated lepton and $E_T^{\text{miss}}$\\

$\Delta\phi (\ell_{\text{non-iso}},E_T^{\text{miss}}) $ & $\Delta\phi$ between non-isolated lepton and $E_T^{\text{miss}}$ \\

$\Delta\phi (\ell_{\text{iso}},\ell_{\text{non-iso}}) $ & $\Delta\phi$ between isolated and non-isolated leptons\\

$m (\ell_{\text{iso}}, \ell_{\text{non-iso}})$ & Invariant mass of the isolated and non-isolated lepton system \\

$N_{\text{jets}}$ & Total jet multiplicity \\

$D0^{\ell_{\text{iso}}}$ & Transverse impact parameter(mm) of isolated lepton \\

$N_{\text{b-jets}}$ & Total $b$-jet multiplicity \\

$P_{\text{lep,top}}(DNN/ ParT)$ & Probability of the non-isolated lepton to be a leptonic top quark \\

$P_{\text{had,top}}(DNN/ ParT)$ & Probability of the non-isolated lepton to be a hadronic top quark \\

$P_{b}(DNN/ ParT)$ & Probability of the non-isolated lepton to be a $b$-jet \\

$P_{c}(DNN/ ParT)$ & Probability of the non-isolated lepton to be a $c$-jet \\

\hline
\end{tabular}

\caption{Summary of the event-level variables used as input features for the DNN classifier. The set includes global event observables, lepton-specific kinematics, and output scores from the object-level top-tagger.}
\label{event_variable}
\end{table}

After constructing the event-level feature set, as outlined in Table~\ref{event_variable}, these are fed to a DNN for event-level classification. This DNN is designed to discriminate the $B^\prime$ signal ($m_{B^\prime} = 5$ TeV) from the SM backgrounds ($t\bar{t}$, $c\bar{c}$, $b\bar{b}$, $VV$, $W+jet$ and $tW$) through a 7-class classification scheme. The architecture consists of three fully connected hidden layers with 512, 256, and 128 neurons, respectively. To ensure stable training and mitigate overfitting, batch normalisation is applied after each hidden layer, complemented by a dropout rate of 0.1. We employ GELU as the activation function and utilise a batch size of 64.
The performance of the classifier is evaluated using the ROC for signal versus all backgrounds, achieving an AUC of 0.994.

The expected signal significance is evaluated assuming an integrated luminosity of
$30~\mathrm{ab}^{-1}$. 

The significance $Z_A$ is computed using the Asimov profile-likelihood formula, including a background systematic uncertainty $\sigma_b$~\cite{Cowan2011}:
\begin{equation}
Z_A =\sqrt{2\left[(S+B)\,\ln\!\left(\frac{(S+B)(B+\sigma_b^2)}{B^2+(S+B)\sigma_b^2}\right)-\frac{B^2}{\sigma_b^2}\ln\!\left(1+\frac{\sigma_b^2 S}{B(B+\sigma_b^2)}\right)\right]}
\label{eq:asimov_systematics}
\end{equation}
For the results including systematic uncertainties, we assume a relative background uncertainty of $\delta_B=10\%$, such that
$\sigma_b=\delta_B B$. In the limit $\sigma_b\rightarrow0$, this expression reduces to the standard Asimov significance obtained from the profile-likelihood ratio.

In Eq.~\eqref{eq:asimov_systematics}, $S$ and $B$ denote the expected numbers of signal and background events after all the selection criteria are applied. 
The events at a center-of-mass energy of $\sqrt{s} = 100$ TeV are analysed across varying event-classifier thresholds~($T$). We report both raw event counts and physics-weighted yields normalised to the integrated luminosity. 
Approximate process-dependent $K$-factors are applied to account for
next-to-leading-order (NLO) QCD corrections to the leading-order (LO)
cross sections. Correction factors of $1.6$, $1.7$, $1.3$, and $1.4$
are applied to the $t\bar{t}$ and $W+\mathrm{jets}$ processes,
the heavy-flavor processes ($b\bar{b}$ and $c\bar{c}$), the $tW$
process, and diboson production, respectively, consistent with standard phenomenological studies and NLO calculations~\cite{Mangano:2016jyj, FCC_CDR}.

Table~\ref{threshold table} summarises the raw and weighted event yields after applying different thresholds on the DNN classifier output, $T$. Before applying any classifier selection ($T\geq0$), the $t\bar{t}$ process constitutes the dominant contribution to the total background, resulting in a poor signal-to-background ratio.
The background suppression reflects differences in the event topologies. Because $b\bar{b}$ and $c\bar{c}$ backgrounds lack genuine boosted leptonic top quarks, their non-isolated leptons arise instead from semileptonic heavy-flavour decays. This allows the classifier to effectively suppress these backgrounds. $W+\mathrm{jets}$ events can contain genuine leptons and missing transverse momentum, mimicking some features of the signal.
However, it lacks the characteristic top-quark decay structure and the correlated event features associated with the $B^\prime\to tW$ signal, and is therefore also strongly suppressed. The $VV$ background, which can contain genuine $W$ and $Z$ bosons produced at high transverse momentum, can reproduce some of the signal's event characteristics, and a fraction of its simulated events, comparable to that of $t\bar{t}$, passes the selection.
Owing to its minor initial contribution, its weighted yield remains a negligible fraction of the total background after classifier selection. All the weighted background yields are reduced by at least two orders of magnitude by the classifier at  $T\geq0.4$ while keeping roughly $70\%$ of the signal. Raising the threshold to $T\geq0.8$ provides further background rejection, but drops the signal efficiency to about $36\%$. At this tighter selection, $VV$ and $W+\mathrm{jets}$ make very small contributions to the remaining background.

Kinematically, the $t\bar{t}$ and $tW$ backgrounds behave much closer to the signal. They contain genuine top quarks and readily populate boosted regions with high top-tagger scores. In particular, single-top $tW$ production yields the same final-state particles as the $B^\prime\to tW$ signal, making it the irreducible background. Because the presence of top-quark candidates offers little discrimination here, we use the multi-TeV kinematic scale and correlated event structure inherent to the $B^\prime\to tW$ topology; these subtle correlations are captured collectively by the classifier through global and transverse observables.
Even so, the $t\bar{t}$ and $tW$ processes remain the dominant surviving backgrounds at $T\geq0.8$ for the $m_{B^\prime}=5$~TeV benchmark. This selection yields the highest statistical significance in the statistics-only limit ($\sigma_b\rightarrow0$) of Eq.~(\ref{eq:asimov_systematics}). When background systematic uncertainties are included, the optimal classifier threshold can change and is determined separately by maximising the corresponding significance.

We note that, at the tightest classifier threshold, $T\geq0.8$, the
surviving simulated event counts for some background
processes become small. In particular, only 12 $W+\mathrm{jets}$ events survive the selection, corresponding to a weighted yield of
approximately 269 events. Although the limited Monte Carlo statistics
may lead to a relatively large uncertainty on this individual
contribution, its contribution to the total background yield remains
small compared with the dominant $t\bar{t}$ and $tW$ backgrounds.
Similarly, the $b\bar{b}$ and $c\bar{c}$ contributions are negligible
at this threshold. Therefore, the limited statistics of these
subdominant background samples are not expected to affect the total background estimate significantly.

\begin{table}[htb]
    \centering
    \scriptsize
    \renewcommand{\arraystretch}{1.2}
    \begin{tabular}{|c|c|c|c|c|c|c|c|l|}

\hline
\textbf{Threshold} &\textbf{Type} & \textbf{$B^\prime$ Signal} & $t\bar{t}$ & $tW$ & $b\bar{b}$ & $c\bar{c}$ & $VV$ & $W+jet$ \\
\hline
$T\ge0$   &  Raw & 19943 & 67543 &89065 & 5000 & 1222  & 8294 & 6838\\
         & Weighted & 4.07$\times10^4$  & 1.19$\times10^8$ & 9.6$\times10^6$ & 1.05$\times10^6$ & 6.6$\times10^4$  & 1.6$\times10^5$ & $6.87\times10^5$\\
\hline  
$T \ge0.4$ & Raw & 13868 &  11124 & 37898 & 5 & 2 & 1805 & 71\\
       & Weighted & 2.83$\times10^4$ & 6.91$\times10^4$ & 7.8$\times10^4$  & 3.80  & 0.55 & 6.10$\times10^2$ & $1.39\times10^3$\\

\hline
$T \ge0.8$ & Raw & 7184 & 4246 & 16106 & 0  & 0 & 713 & 12
\\
       & Weighted & 1.46$\times10^4$ & 9.7$\times10^3$ & 8.0$\times10^3$ & 0.0 & 0.0 & 1.56$\times10^2$ & 2.69$\times10^2$
\\
\hline
\end{tabular}

\caption{Event yields at $\sqrt{s}=100$~TeV for different classifier thresholds. Both raw event counts and physics-weighted yields ($\sigma \times \mathcal{L}$ normalisation) are shown for signal and background processes. The signal corresponds to a benchmark $B^\prime$ VLQ of mass $5$~TeV.}
\label{threshold table}
\end{table}

The event-level classifier is trained using the $m_{B^\prime}=5$~TeV
benchmark and subsequently applied to the other mass points. The resulting mass-dependent sensitivity therefore quantifies the classifier's generalisation across the range of $B^\prime$ masses considered in this work.

\begin{figure}[H]
	\begin{center}
		\includegraphics[width=12cm]{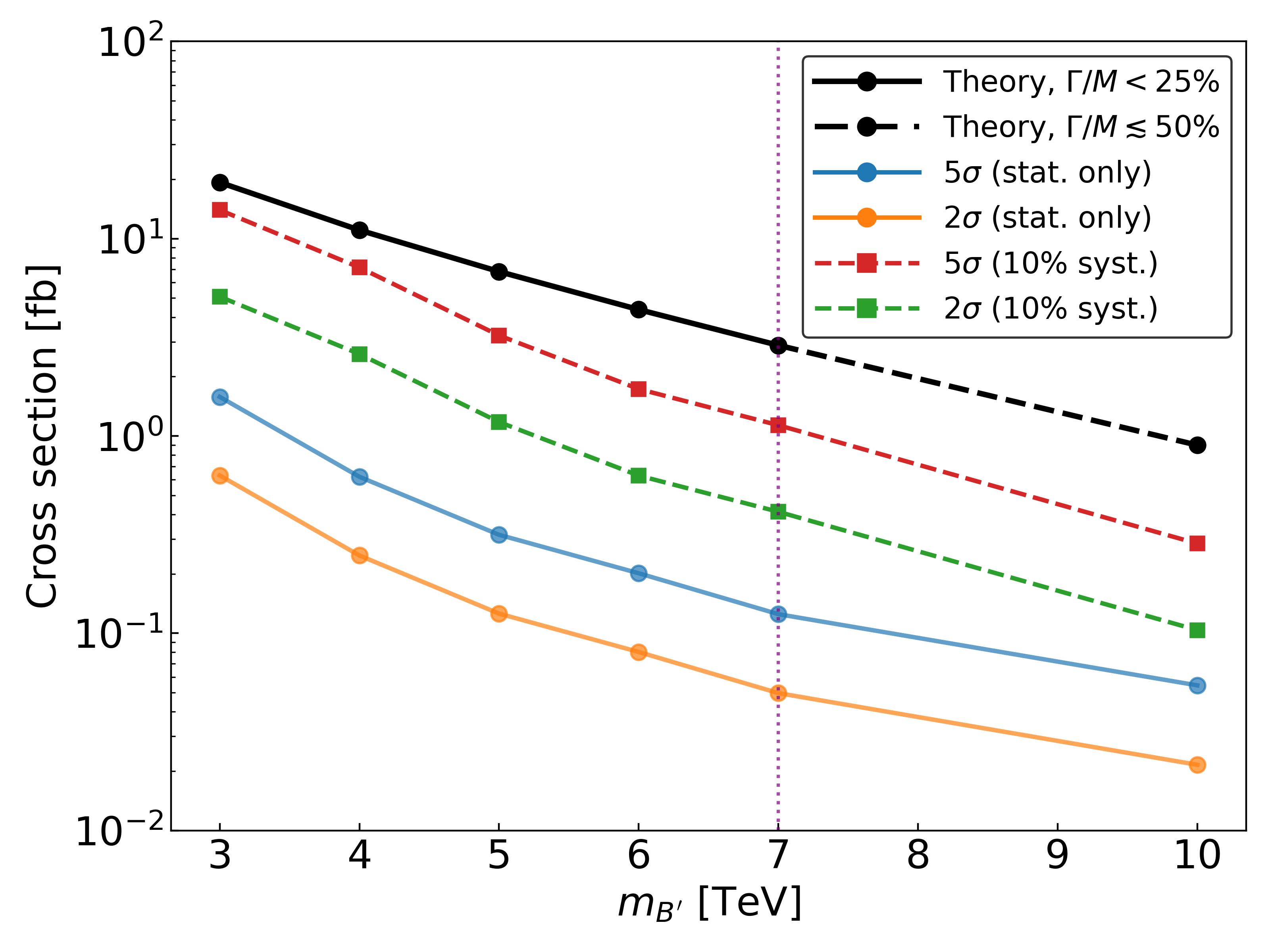}
	 \end{center}
 \caption{Minimum signal cross section required as a function of the
$B^\prime$ mass to achieve either a $5\sigma$ discovery or a $2\sigma$
sensitivity at the FCC-hh. The results are shown for the
statistics-only case and for the case including a $10\%$ background
systematic uncertainty. The predicted cross section for the benchmark
scenario $(\tilde{g}=0.07,\,R_L=1.0)$ is also shown. The theory curve
includes $\mathrm{BR}(B^{\prime}\to tW)$ as well as the branching fractions
corresponding to the leptonic decay topology considered in this
analysis. The solid part of the theory curve corresponds to the
region satisfying the imposed width criterion,
$\Gamma/m_{B^\prime}<25\%$, while the dashed segment corresponds to
$m_{B^\prime}\gtrsim7$~TeV, where the width-to-mass ratio increases up to
$\Gamma/m_{B^\prime}\lesssim50\%$.}
 
 \label{significance}
\end{figure}

The sensitivity to the $B^{\prime}$ signal is evaluated in terms of the production cross section required to achieve a given significance as a function of the $B^{\prime}$ mass. Although the signal production rate is higher at lower masses, the signal is more difficult to distinguish from the SM backgrounds in this region due to their greater kinematic overlap. At higher masses, the signal becomes more distinctive, allowing the optimised event selection to suppress a larger fraction of the backgrounds. Figure~\ref{significance} shows the cross sections corresponding to $2\sigma$ sensitivity and the $5\sigma$ discovery threshold. 

As the $B^{\prime}$ mass increases, the minimum cross section required to meet the $2\sigma$ and $5\sigma$ thresholds drops, because the event selection becomes highly efficient for heavier, more boosted resonances. Assuming $10\%$ systematic uncertainty on the background shifts these curves to higher cross sections, as expected. We also include the theoretical cross section for the $\tilde{g}=0.07$ and $R_L=1.0$ benchmark to show exactly where the $\Gamma_{B^\prime}/m_{B^\prime}<25\%$ constraint is satisfied. For this specific choice of parameters, a $5\sigma$ discovery is achievable up to a mass of roughly $7$~TeV. This remains true even when the $10\%$ systematic error is applied. If we allow for broader resonances and relax the limit to $\Gamma_{B^\prime}/m_{B^\prime}\lesssim50\%$, the mass reach extends to $10$~TeV.

At $m_{B^\prime}=10$~TeV, reaching $5\sigma$ discovery requires a minimum signal cross section of roughly $0.054$~fb in the statistics-only case, compared to $0.28$~fb when a $10\%$ background systematic uncertainty is included. Overall, the joint framework pairing the leptonic jet tagger with the event-level DNN classifier maintains strong sensitivity to heavy $B^\prime$ quarks over a broad mass range at the FCC-hh, even in the presence of a $10\%$ background systematic uncertainty.

\section{Discussion}
\label{discussion}
While our main analysis is presented in the last section, we report additional checks in this section. 

\subsection{Performance of the tagger revisited}

\begin{figure}[H]
	\begin{center}
		\includegraphics[width=12cm]
        {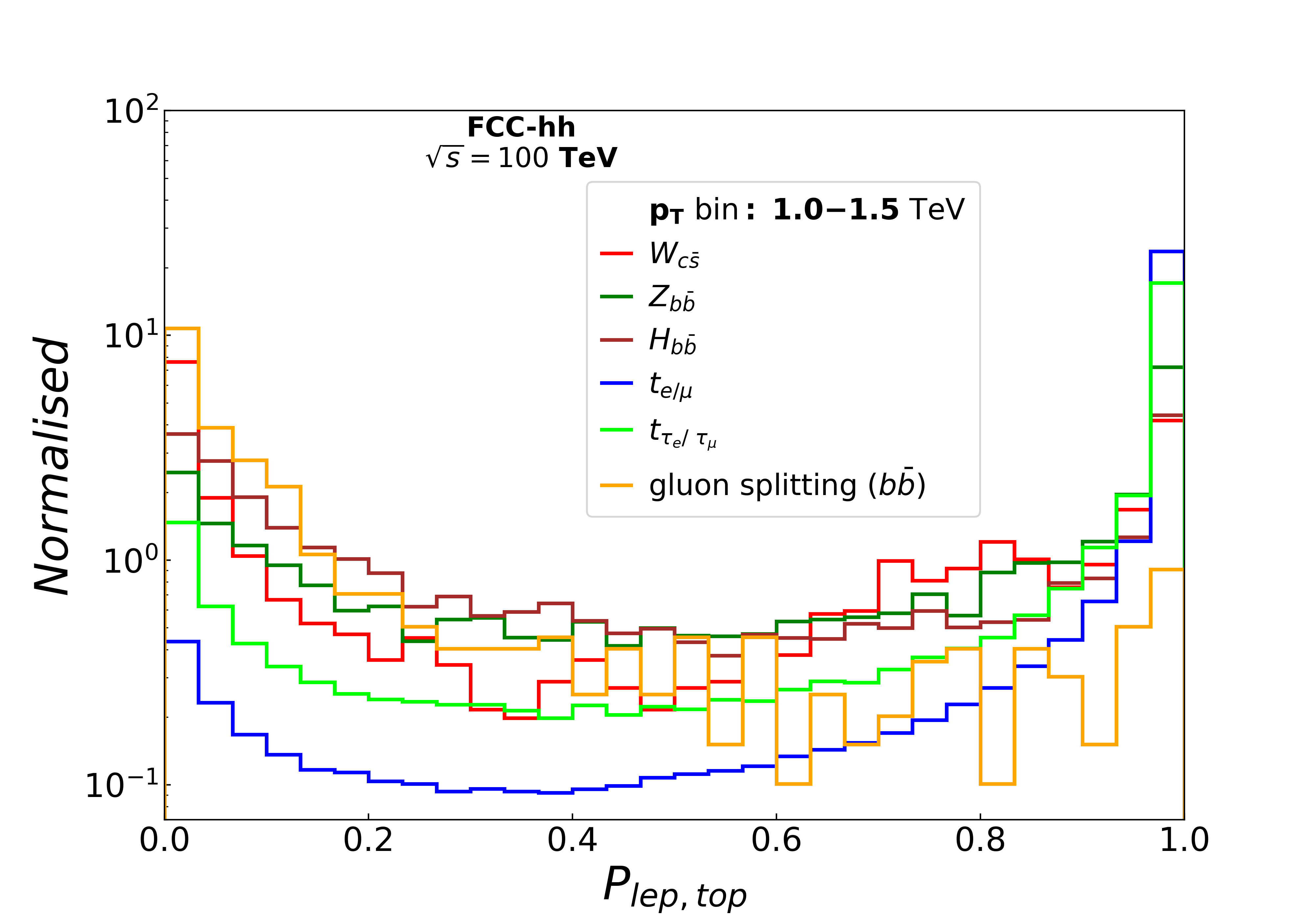}
	 \end{center}
 \caption{ DNN score distributions of the top tagger for leptonic-top jets and additional background samples in the $1.0-1.5$~TeV lepton $p_T$ interval. }
 \label{scores}
\end{figure}

While training the top tagger to identify $t_\text{lep}$, where $\text{lep} = e, \mu, \tau_e, \tau_\mu$, we considered the following background jets: $t_\text{had}$, $c$-jet and $b$-jet. Here, we report the performance of the tagger for some other background jets. 
Figure~\ref{scores} shows a distribution of jet tagger score $P_{\mathrm{lep,top}}$ for the lepton $p_T$ bin of $[1.0,\,1.5]$~TeV. 
As expected, leptonic decays of top quark ($t_{e}$, $t_{\mu}$, $t_{\tau_e}$, $t_{\tau_\mu}$) peak near $P_{\mathrm{lep,top}} \approx 1$. In contrast, the background jets $W_{c\bar{s}}$ (red), $Z_{b\bar{b}}$ (dark green), and $H_{b\bar{b}}$ (maroon)  are concentrated in the low-score region. We have also checked gluon-initiated jets where the gluon splits into $b\bar{b}$ during parton showering. This type of background jets also peaks near $P_{\mathrm{lep,top}} \approx 0$. Overall, the $P_{\mathrm{lep,top}}$ variable provides strong discrimination against the background jets that were not even used in the training process.

\subsection{Sensitivity to Soft Track Activity}

A full treatment of pile-up at FCC-hh design luminosities is beyond the scope of this work. The FCC-hh environment is expected to involve a very large number of additional proton--proton interactions per bunch crossing, and a realistic assessment of their impact requires both pile-up overlay and dedicated mitigation techniques. As discussed in Refs.~\cite{Bhattacharyya:2025kvl,Mangano}, such mitigation strategies are expected to play an important role in the FCC-hh detector environment. Since the Delphes FCC-hh card used in this work does not model a complete realistic pile-up-mitigation strategy, naively overlaying up to $\langle\mathrm{PU}\rangle\sim1000$ minimum-bias interactions would not provide a fully realistic estimate of the tagger performance.

\begin{figure}[H]
\centering
\includegraphics[width=7.5cm]{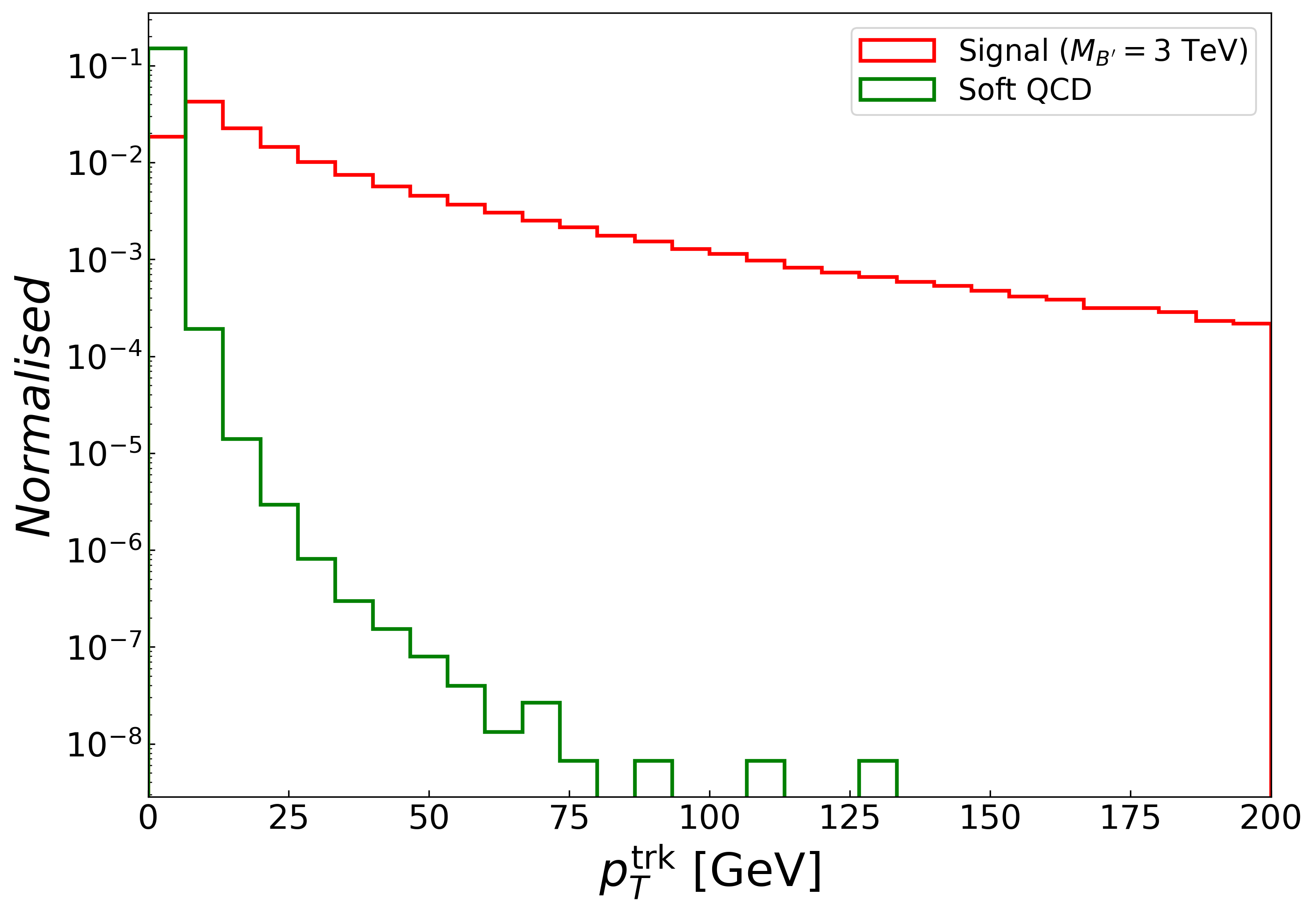}
\includegraphics[width=7.5cm]{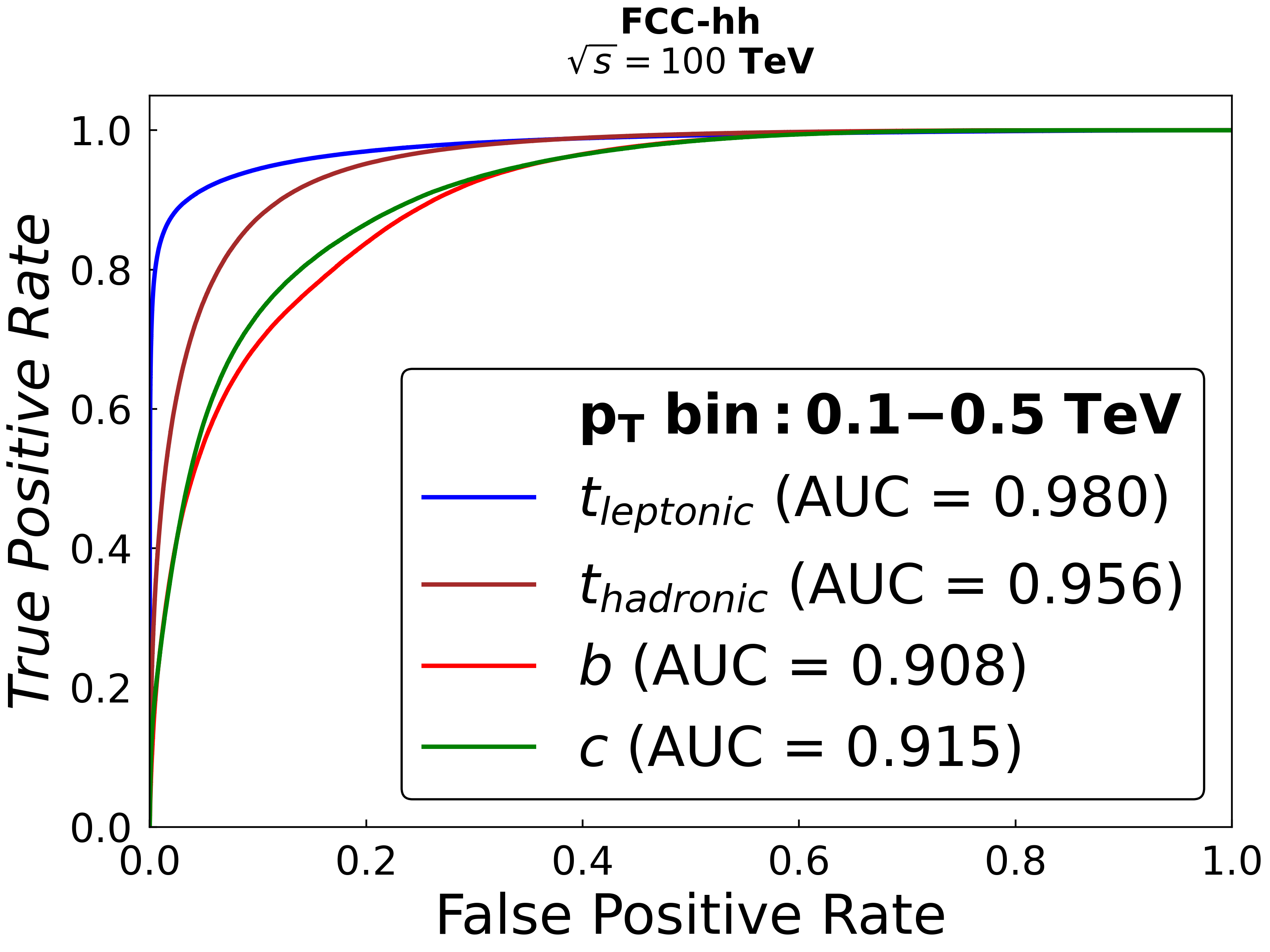}
\caption{Left: normalised transverse-momentum distributions of charged particles from soft-QCD events and of tracks within
$\Delta R(\ell,\mathrm{trk})<0.4$ of the lepton in the $m_{B'}=3$~TeV
signal sample at $\sqrt{s}=100$~TeV. Right: ROC curves for the
$0.1$--$0.5$~TeV lepton-$p_T$ bin obtained using minimum track
transverse-momentum thresholds of $p_T^{\rm trk}>20$~GeV.}

\label{softqcd}
\end{figure}

As a preliminary test of the sensitivity of the tagger to soft additional track activity, we compare the transverse-momentum spectrum of charged particles from soft-QCD events generated using \textsc{PYTHIA}~8.315 with that of tracks within $\Delta R(\ell,\mathrm{trk})<0.4$ of the lepton in the boosted signal sample with $m_{B'}=3$~TeV. The soft-QCD activity is 
dominated by low-$p_T$ particles, whereas the signal contains a
substantially harder track component. We therefore compare the DNN tagger performance obtained using the baseline track selection with
$p_T^{\rm trk}>5$~GeV and after imposing a harder threshold of
$p_T^{\rm trk}>20$~GeV. The resulting AUC values remain essentially unchanged: $0.981 \to 0.980$, $0.954 \to 0.956$, $0.905 \to 0.908$, and $0.914 \to 0.915$ for the $t_{\rm leptonic}$, $t_{\rm hadronic}$,
$b$, and $c$ classes, respectively, indicating that the tagging performance is not strongly dependent on tracks in the $5$--$20$~GeV momentum range. Calorimeter-based observables may be more affected by pile-up than track-based measurements, particularly due to additional energy deposits from neutral particles. This study does not constitute a full pile-up simulation; rather, it provides a simplified robustness test against additional soft-track activity. The results are shown in Figure~\ref{softqcd}.

\subsection{Architecture comparison and feature sensitivity}

To test the limits of our tagging framework, we compared the previously developed DNN and Particle Transformer (ParT) architectures against a Graph Neural Network (GNN). For a simple comparison, we restricted all three models to track-based observables, removing calorimeter dependencies entirely so they rely solely on tracking system data.

In the GNN configuration, tracks with $p_T^{\mathrm{trk}}>5$~GeV and lying within $\Delta R<0.4$ of the lepton are treated as graph nodes. In addition, 10 graph-level features, corresponding to the track-based observables listed in Table~\ref{object_features}, are included as global graph features and concatenated with the pooled graph embedding before the final classification layers. The GNN architecture consists of three EdgeConv layers with 128 hidden channels each, employing ReLU~\cite{Nair2010} activations, batch normalisation, and a dropout rate of 0.3. The graph connectivity is constructed using a $k$-nearest-neighbor algorithm with $k=8$. For comparison, the DNN is trained using the same track-based observables listed in Table~\ref{object_features}, while the ParT is trained exclusively on track-based constituent features. Apart from this restriction to track-based inputs, the DNN and ParT architectures and training configurations are identical to those described in Section~\ref{jet tagger}.

\begin{figure}[H]
	\centering
     \subfloat[]{\includegraphics[width=0.32\textwidth]{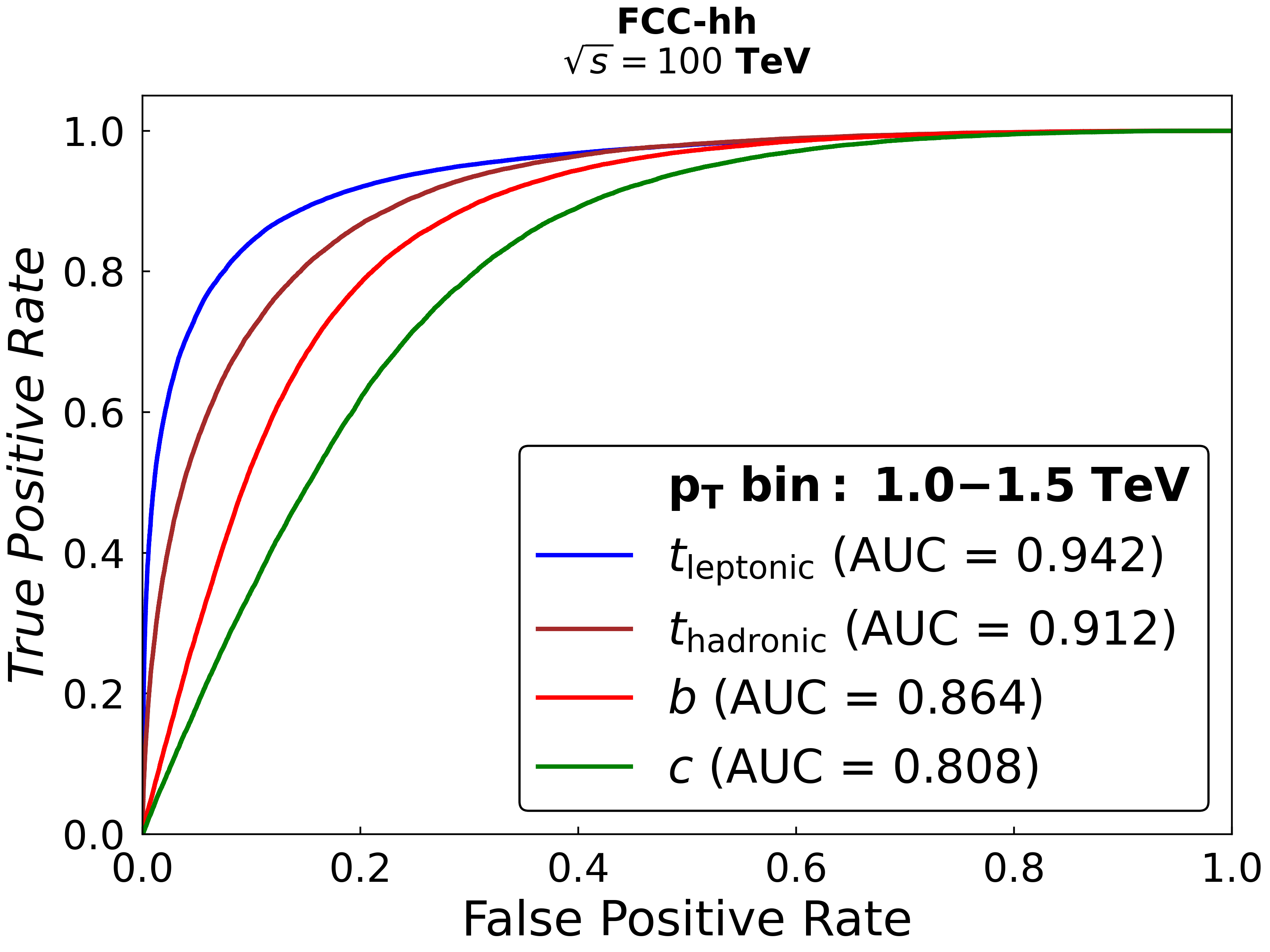}}
    \hfill
    \subfloat[]{\includegraphics[width=0.32\textwidth]{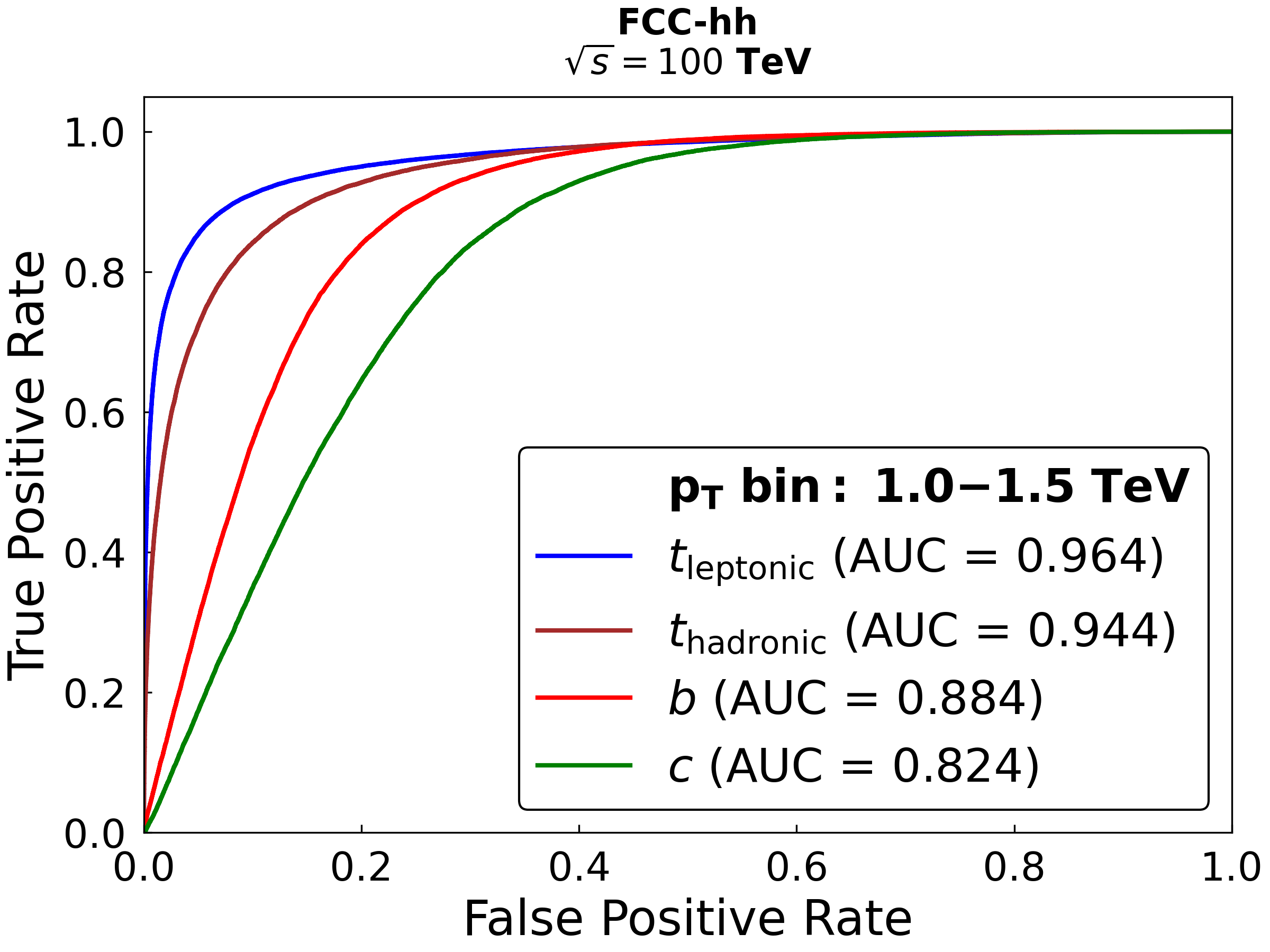}}
    \hfill
    \subfloat[]{\includegraphics[width=0.32\textwidth]{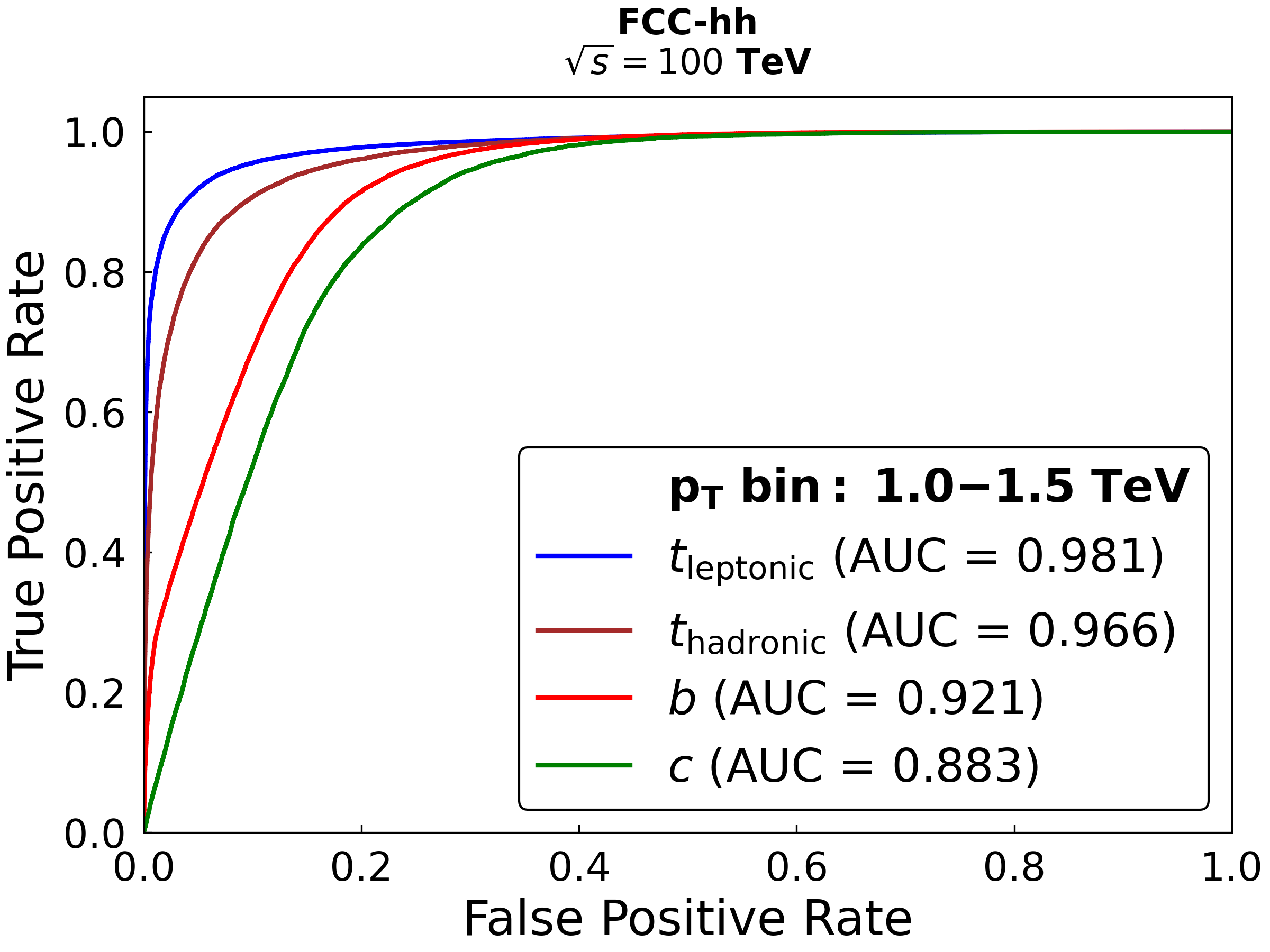}}
    \caption{
    ROC for the $1.0-1.5$ TeV lepton $p_T$ bin: (a) DNN classifier utilising track-based variables (b) GNN classifier processing tracks within $\Delta R=0.4$ as graph nodes along with global features (c)Particle Transformer classifier using lepton-centered tracks within $\Delta R=0.4$.
}
    \label{roc_track}
\end{figure}   

We trained all three models across six lepton $p_T$ intervals spanning $0.1$--$3.0$~TeV, together with an additional bin for $p_T>3$~TeV. Figure~\ref{roc_track} shows the representative ROC curves for the $1.0$--$1.5$~TeV bin. Among the three approaches, the Particle Transformer (ParT) provides the best discrimination performance for leptonic top tagging, achieving an AUC of $0.981$, compared with $0.964$ for the GNN and $0.942$ for the DNN. A similar hierarchy is observed for the other jet classes, including hadronic top, $b$, and $c$ jets.

These results indicate that, when the input is restricted to
track-level information, the ParT is particularly effective at exploiting the correlations and spatial relationships among the constituent tracks. In contrast to the DNN, which operates on a fixed set of engineered observables, both the GNN and ParT retain information about the individual tracks and their relations. The superior performance of the ParT in this comparison suggests that its attention-based architecture provides a more effective representation of the track-level structure of ultra-boosted jets.

Motivated by the strong performance of the Particle Transformer (ParT), we investigate whether its learned representations can also be used for anomaly detection. Unlike the supervised taggers, which are trained against specific background classes, an anomaly detector is trained only on the nominal signal topology and identifies jets that deviate from the learned representation of that topology. Such an approach can provide sensitivity to alternative signatures without requiring all possible backgrounds or new-physics topologies to be specified during training. This provides a complementary and less model-dependent approach to identifying unusual boosted-jet topologies at the FCC-hh.

Following the strategy proposed in Ref.~\cite{SVDD}, the pretrained ParT is used as a fixed feature extractor, with its parameters frozen during anomaly-detector training. The jet-level representation obtained from the CLS token is passed to a Deep Support Vector Data Description (DeepSVDD) network~\cite{ruff2018deepsvdd}, which maps the 128-dimensional ParT embedding to a lower-dimensional feature space through layers of dimensions $128\rightarrow64\rightarrow16$, with ReLU activations. The center $c$ of the DeepSVDD hypersphere is fixed using the mean embedding of the nominal training sample, while only the DeepSVDD network is optimised. Weight-decay regularisation is used to reduce the tendency towards hypersphere collapse.

The network is trained exclusively on the nominal leptonic-top jet sample. The hadronic-top, $b$-jet, and $c$-jet samples are not used during optimisation and are instead treated jointly as anomalous samples when evaluating the anomaly-detection performance. The DeepSVDD network maps each ParT jet representation into this lower-dimensional feature space, where the nominal leptonic-top jets are encouraged to cluster around the hypersphere center. The anomaly score is defined as the squared Euclidean distance between the resulting feature representation and the hypersphere center, so that jets more consistent with the learned nominal topology receive smaller scores, while those deviating more strongly from it receive larger scores.

For the model trained in the $1.0$--$1.5$ TeV $p_T$ bin, an AUC of 0.894 is obtained for distinguishing the nominal leptonic-top jet sample from the pooled anomalous classes. This demonstrates that the pretrained ParT representation retains useful information about the leptonic-top topology even when the alternative classes are not included during training. The corresponding anomaly-score distributions and ROC curve are shown in Appendix~\ref{DeepSVDD}.

Although the AUC is lower than that of the fully supervised taggers, the anomaly detector achieves this separation without exposure to the background classes during DeepSVDD optimisation. This approach is not fully model-independent, but it provides another way to identify jets that differ from the nominal jet topology. This illustrates the applicability of pretrained ParT representations to one-class anomaly detection in boosted-jet analyses at the FCC-hh.

\section{Conclusion}
\label{conclusion}
In this work, we have investigated the phenomenology of a VLQ and developed a search strategy for its detection at the FCC-hh, operating at a center-of-mass energy of $\sqrt{s}=100$~TeV with an integrated luminosity of $30$~ab$^{-1}$. We consider a $SU(2)$-singlet bottom-like VLQ, $B^{\prime}(3,1,-1/3)$, which couples to the first and third generations of SM quarks. Since single production generally dominates over pair production at multi-TeV mass scales, we concentrate on the channel $pp\to B^\prime j\to tWj$, where both the top quark and the $W$ boson decay to leptons. The final state contains an isolated lepton from the $W$ boson and a non-isolated, highly boosted lepton within the top-quark jet, accompanied by a large amount of missing transverse momentum. We have scanned the model parameter space $(\tilde{g},R_L)$ under the width constraint $\Gamma_{B^\prime}/m_{B^\prime}<25\%$,  and have constrained $(\tilde{g},R_L)=(0.07,1.0)$  for the benchmark selection. This bounds the resonance masses up to roughly $7$~TeV while preserving a substantial production cross section. The width-to-mass ratio is largely insensitive to $R_L$, while the production cross section is highly sensitive to it, prompting us to fix $R_L=1.0$ for the rest of the study.

A key challenge in this search is identifying highly boosted leptonic top jets in the presence of various SM backgrounds. These include a top quark jet formed by a hadronically decaying W boson, which contains genuine top-quark decay products but no prompt lepton from a leptonic top decay, as well as $ b$- and $c$-jets containing non-isolated leptons from the semileptonic decays of $B$ and $D$ hadrons. To distinguish these jet categories, we developed dedicated multi-bin tagging algorithms based on lepton kinematics, track-based variables (including displaced ones), calorimeter-based observables, and jet-substructure variables. The DNN and Particle Transformer (ParT) taggers were trained separately in seven lepton-$p_T$ bins, from $0.1$~TeV to above $3$~TeV. Throughout the full momentum range, both classifiers maintain high efficiencies for leptonic-top tagging. The DNN efficiency increases from $89\%$ in the lowest $p_T$ interval up to $96\%$ at the highest momentum region considered in this work. 

The particle transformer ParT starts at $94\%$ in the lowest bin and remains at $96$--$97\%$ across the remaining bins. The leptonic-top AUC remains above $0.98$ for the DNN and exceeds $0.99$ for ParT across all intervals. Their relative performance depends on the lepton transverse momentum; ParT performs better at low and intermediate $p_T$, but at the highest energy scale--where tagging efficiencies converge--the DNN yields slightly lower mistag rates. The high-level, physics-motivated observables provide discrimination comparable to that of a sophisticated transformer architecture under full-feature conditions. At the same time, the DNN and ParT prove equally powerful in the ultra-boosted regime, where resolving the individual constituents of highly collimated jets poses a severe experimental challenge. In the presence of high pile-up at FCC-hh, calorimeter-based observables will be more affected than track-based observables. For this reason, we also investigated the performance of track-based-only taggers using DNN, GNN, and ParT architectures. Within the 1.0--1.5~TeV lepton $p_T$ interval, ParT leads with a top AUC of 0.981, outperforming the GNN (0.964) and the track-only DNN (0.942). The results show that track-based architectures also identify correlations well and will be particularly advantageous for the FCC-hh under extreme pile-up conditions. Given ParT's strong performance, we also fed its learned representations into a DeepSVDD network trained entirely on leptonic top-jet samples to test one-class anomaly detection. Treating the hadronic top, $b$-jet, and $c$-jet samples as anomalies unseen by the DeepSVDD, the model achieved an AUC of $0.894$ in separating the baseline leptonic-top topology from alternative jet classes. In reality, pile-up vertices can also be very close to the primary vertex, and these additional soft tracks can also get included in the leptonic top jets. To suppress these pile-up tracks, one needs to apply higher-momentum cuts. We conducted a preliminary performance study of the tagger by increasing the $p_T$ threshold for tracks. This simplified test showed no major degradation in tagging performance.

Next, we built a dedicated event-level classification framework to separate the $pp\to B^\prime j$ signal from various SM backgrounds. The final state contains exactly one isolated lepton from the $W$ boson, one non-isolated lepton from the boosted top jet, and a large amount of missing transverse momentum from the neutrinos. The dominant SM backgrounds $t\bar{t}$ and $tW$ can be suppressed, and the event-level DNN achieves an AUC of $0.994$ for the $5$~TeV $B^\prime$ benchmark.

Finally, we evaluated the signal significance both in the statistics-only case and with a $10\%$ systematic uncertainty on the background yield. The corresponding $2\sigma$ sensitivity and $5\sigma$ discovery reach were determined for $B^\prime$ masses between $3$ and $10$~TeV. For our reference benchmark, the analysis secures a $5\sigma$ discovery up to masses of approximately $7$~TeV, even factoring in a $10\%$ background systematic uncertainty, while satisfying the width constraint $\Gamma_{B^\prime}/m_{B^\prime}<25\%$. If we relax that width requirement to $\Gamma_{B^\prime}/m_{B^\prime}\lesssim50\%$, the discovery reach stretches to $10$~TeV. At $m_{B^\prime}=10$~TeV, claiming a $5\sigma$ discovery takes a minimum signal cross section of $0.054$~fb (statistics-only) or $0.28$~fb (with $10\%$ systematics). For the $5$~TeV benchmark, the respective cross sections are $0.31$~fb and $3.2$~fb. In summary, the integration of dedicated leptonic-top tagging and event-level classification offers an effective and realistic avenue for discovering multi-TeV vector-like quarks within a 100~TeV hadron collider environment.

\section{Acknowledgments}
The work of SM is supported by an initiation grant ($\text{IITK}/\text{PHY}/2023282$) received from IIT Kanpur. The work of BB and SB are supported by the Core Research Grant CRG/2022/001922 of the Anusandhan National Research Foundation (erstwhile Science and Engineering Research Board (SERB) scheme), Government of India. BB and KK are grateful to the Centre for High Energy Physics, Indian Institute of Science, for the cluster facility. The authors acknowledge the use of AI-assisted tools for language editing and grammatical corrections to improve the clarity and readability of the manuscript.

\pagebreak
\appendix
\section{Appendix}
\label{appendix}

\subsection{Isolation variables}
\begin{figure}[H]
	\begin{center}
		\includegraphics[width=7.5cm, height=6cm]{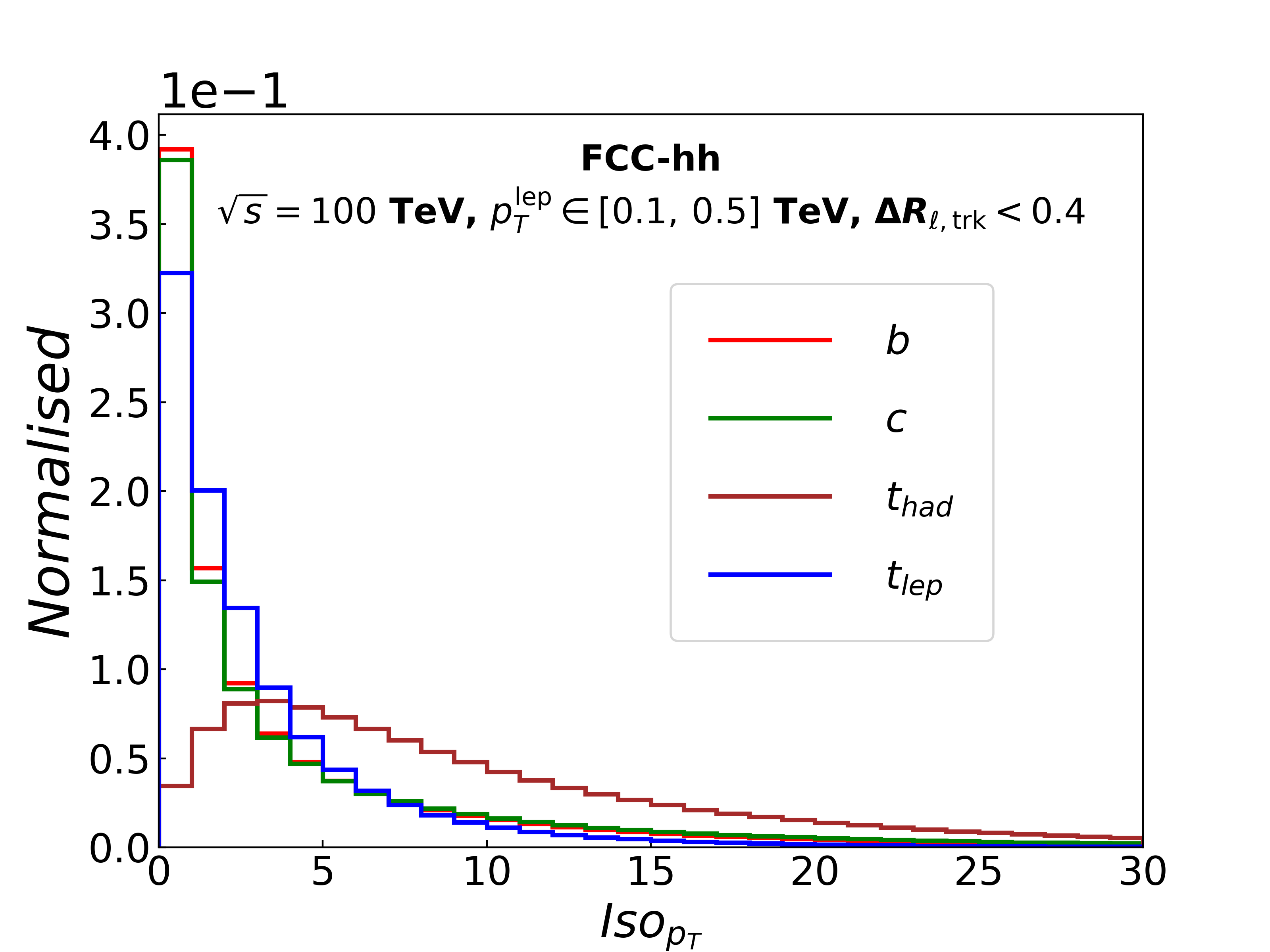}
        \includegraphics[width=7.5cm, height=6cm]
        {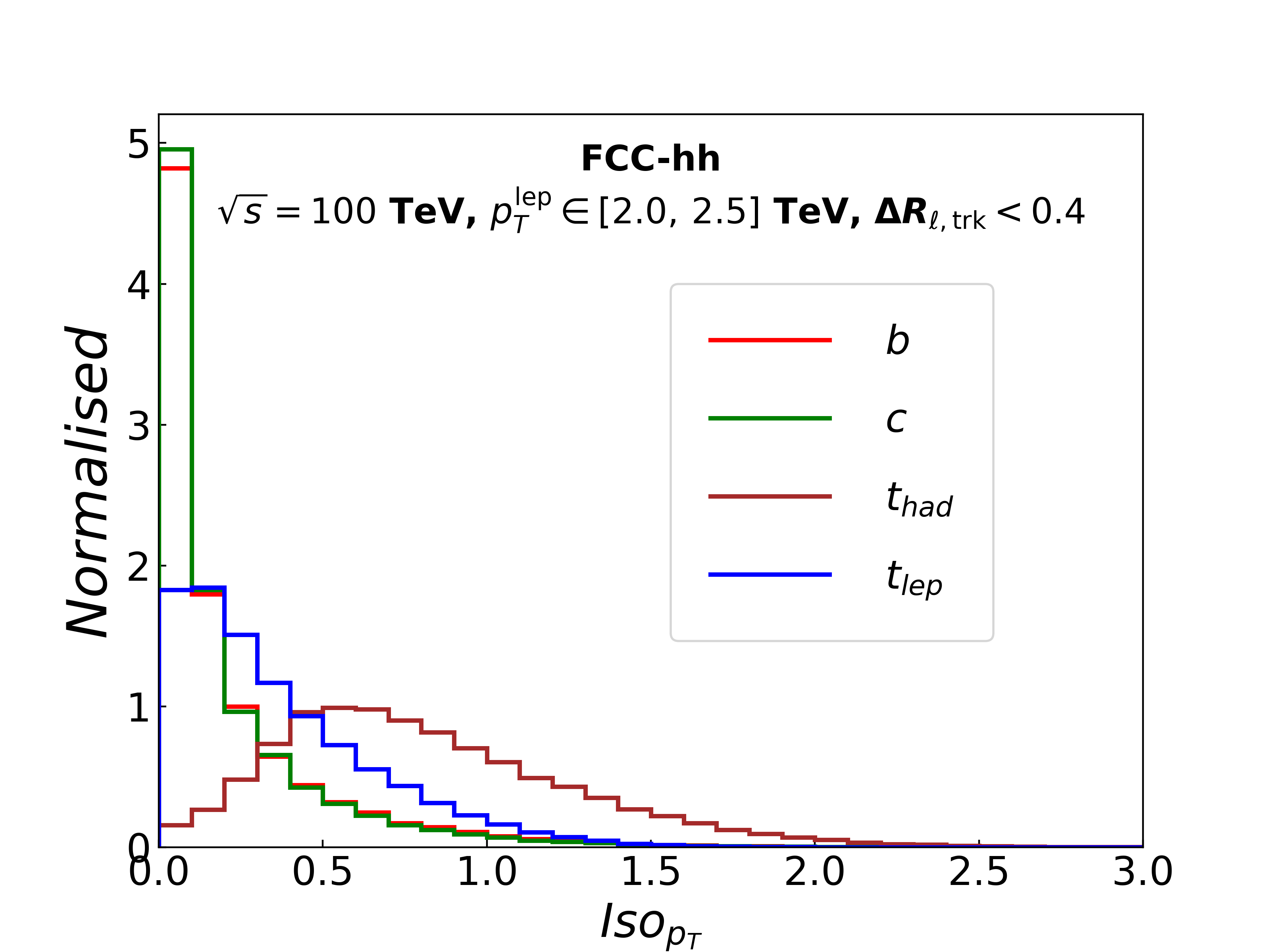}
	 \end{center}
 \caption {Track-based variable $Iso_{p_{T}}$ for 0.1 - 0.5 TeV    (left) and  2.0 - 2.5 TeV (right)  lepton pT bins }
 \label{ptr}
\end{figure}

\begin{figure}[H]
	\begin{center}
		\includegraphics[width=7.5cm, height=6cm]{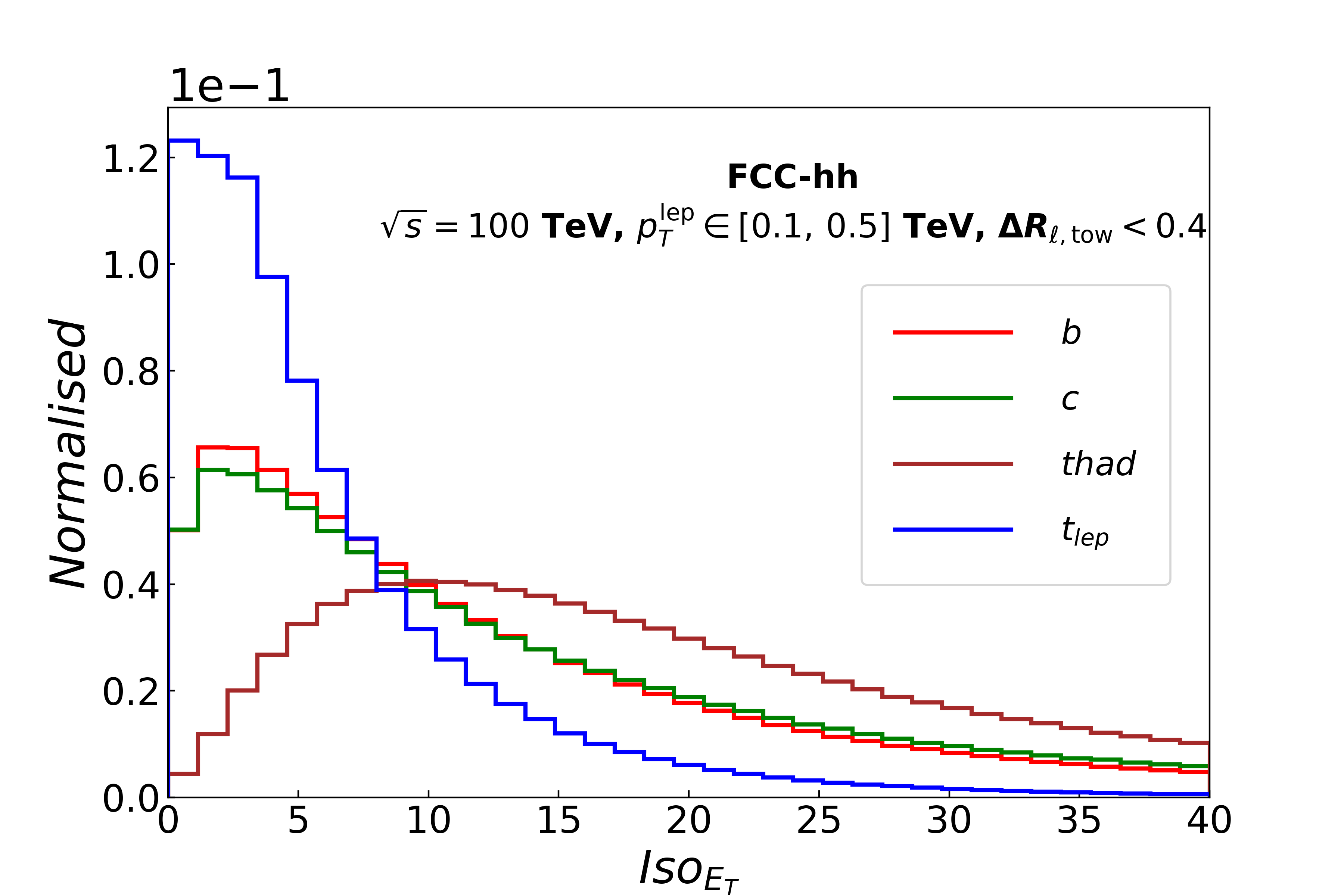}
        \includegraphics[width=7.5cm, height=6cm]{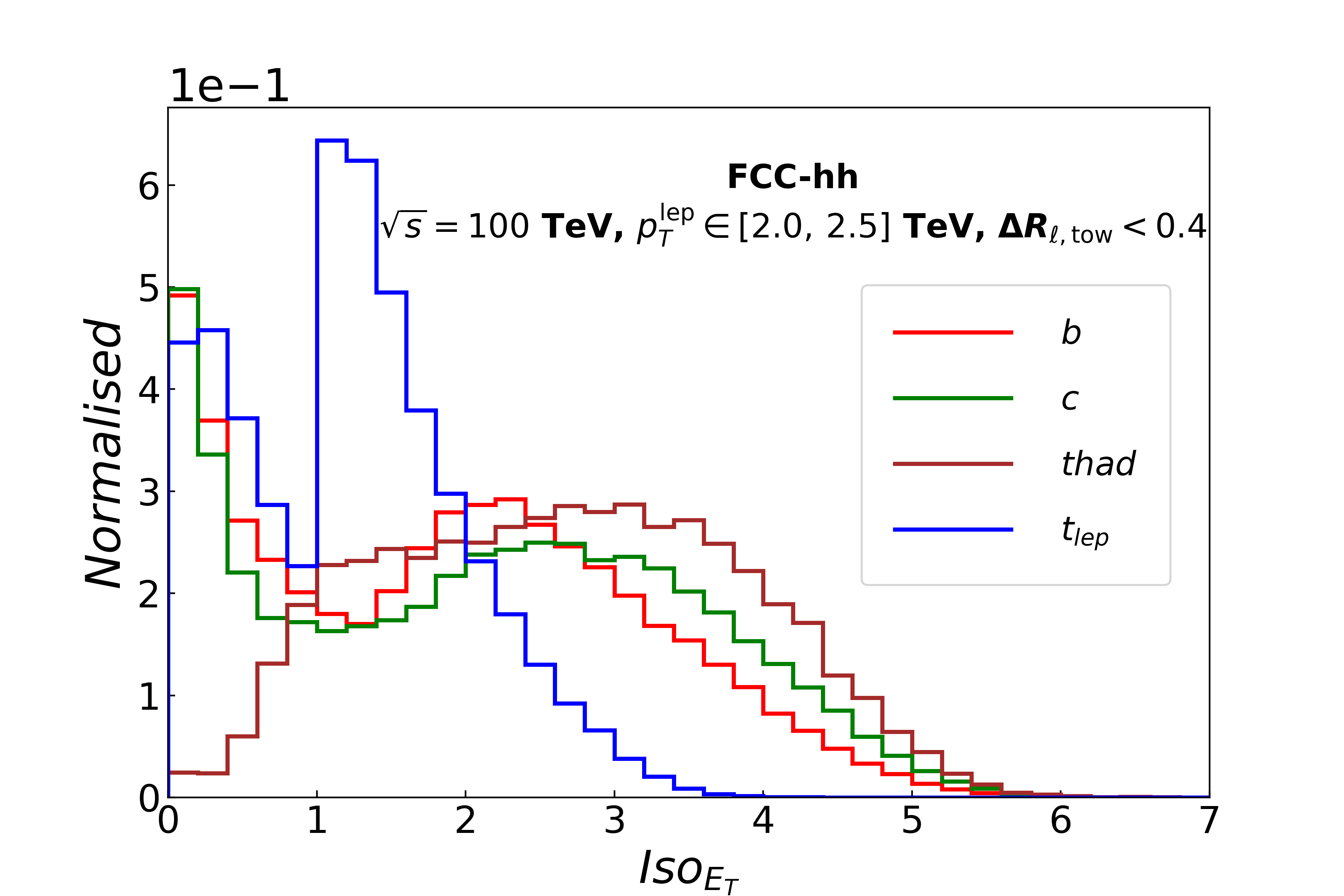}
	 \end{center}
 \caption {Tower-based isolation variable $Iso_{E_{T}}$ for 0.1 - 0.5 TeV (left side) and 2.0 - 2.5 TeV (right side) lepton pT bins}
 \label{etr}
\end{figure}

Figures~\ref{ptr} and \ref{etr} show the distributions of the isolation-related variables used to distinguish isolated from non-isolated leptons.

\subsection{Anomaly detection with DeepSVDD}
\begin{figure}[H]
	\begin{center}
		\includegraphics[width=7.5cm, height=6cm]{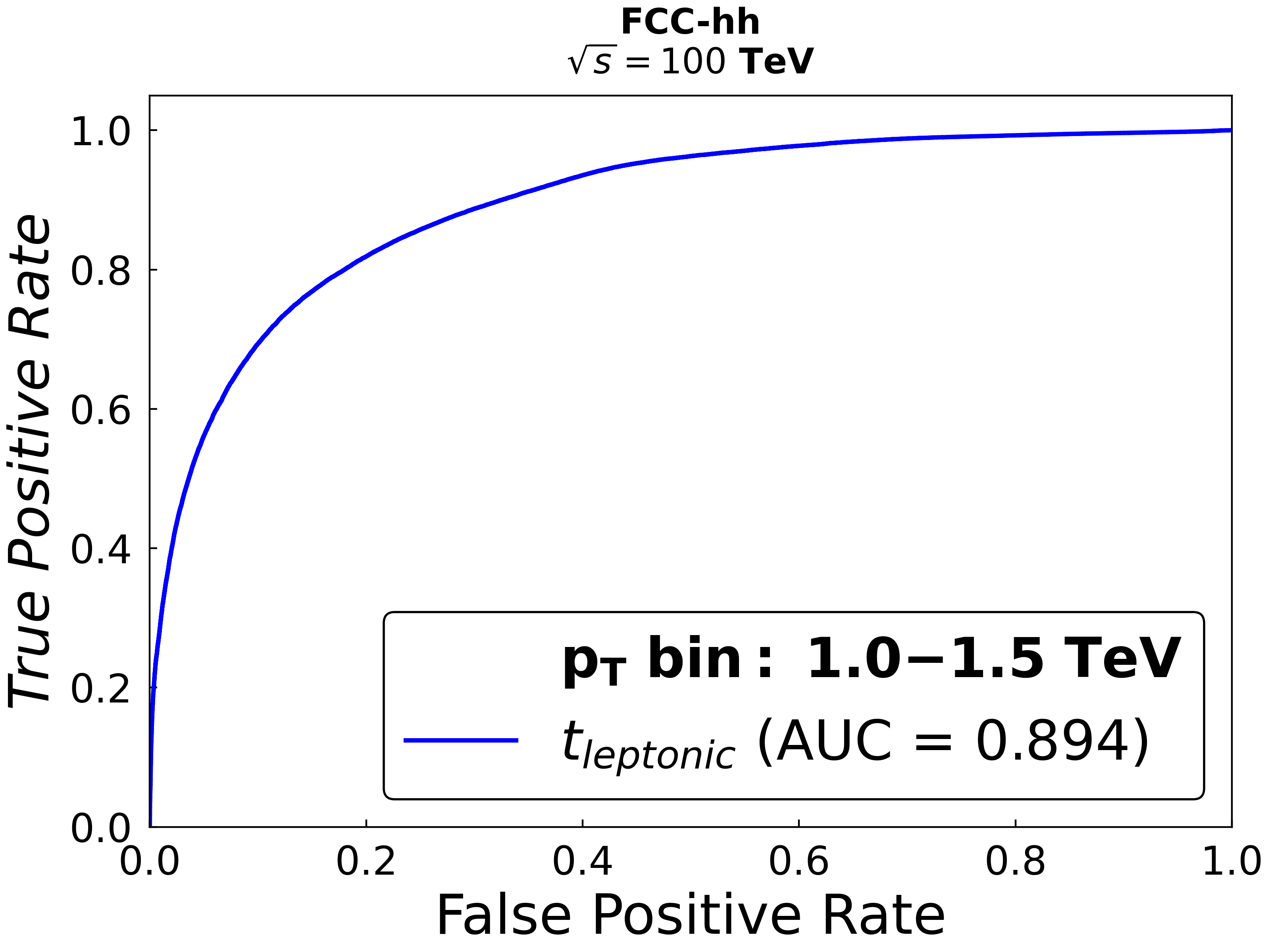}  
         \includegraphics[width=7.5cm, height=6cm]{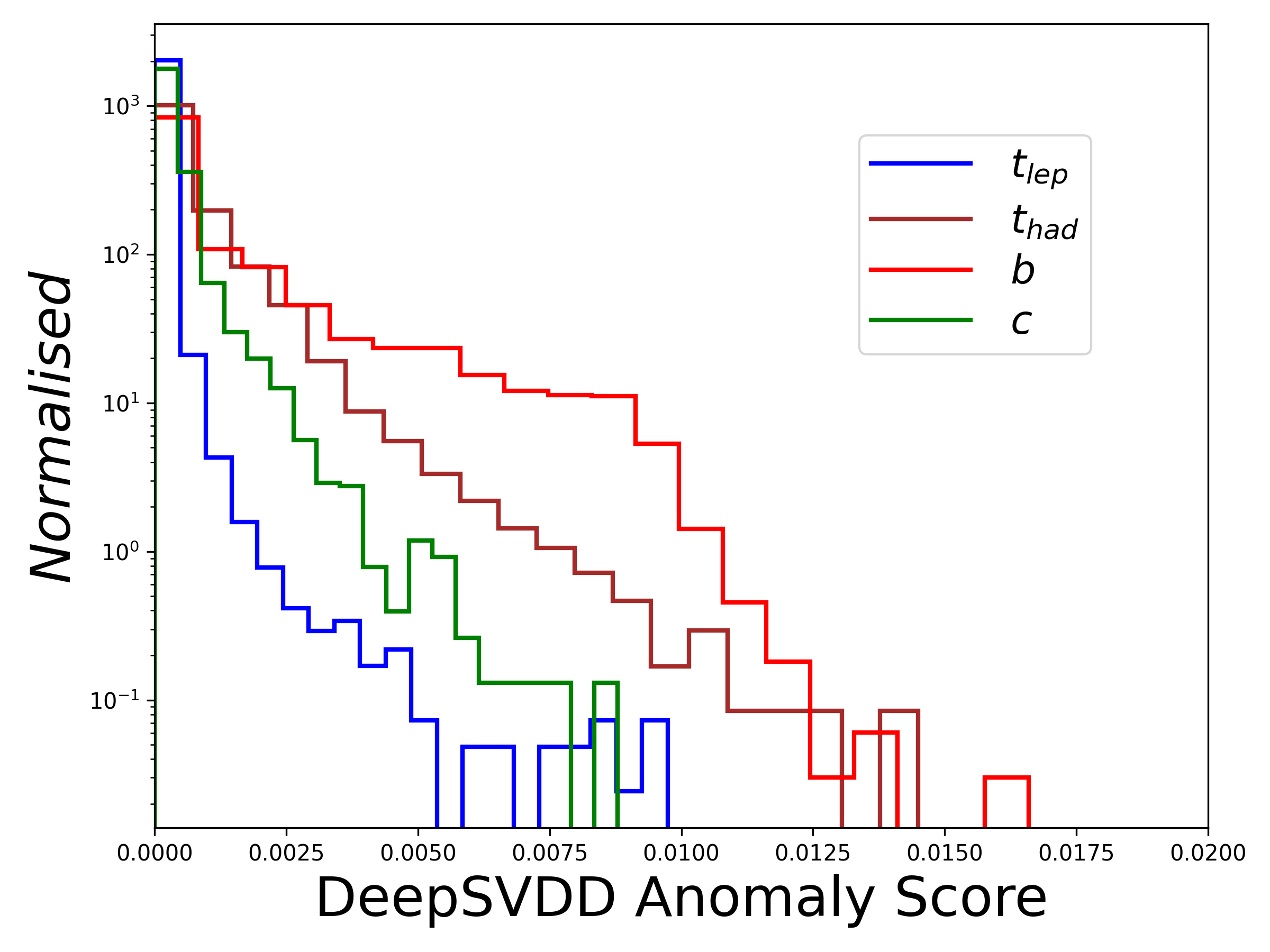}
        
	 \end{center}
 \caption { DeepSVDD model anomaly score (right) and Receiver operating characteristic (ROC) curve(left) for the DeepSVDD model in the 1.0–1.5 TeV $p_T$
 bin, with an AUC of 0.894.}
 \label{deepsvdd}
\end{figure}

The anomaly-score distribution and ROC curve for the DeepSVDD model are shown in Figure~\ref{deepsvdd}.

\label{DeepSVDD}

\subsection{Sensitivity to the isolation-cone size}
\begin{figure}[H]
	\centering
	\subfloat[]{\includegraphics[width=7.5cm, height=6cm]{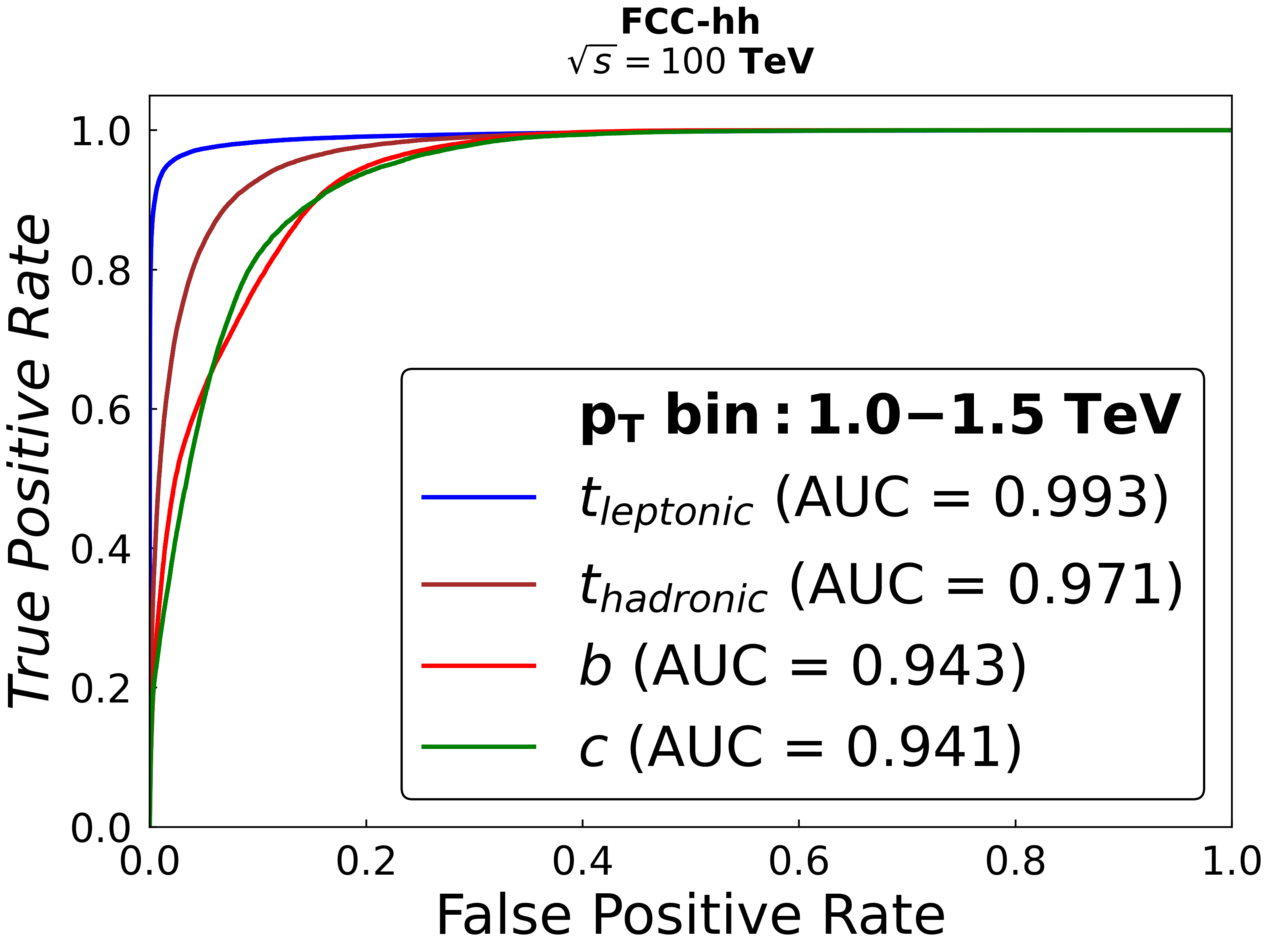}}
    \subfloat[]{\includegraphics[width=7.5cm, height=6cm]{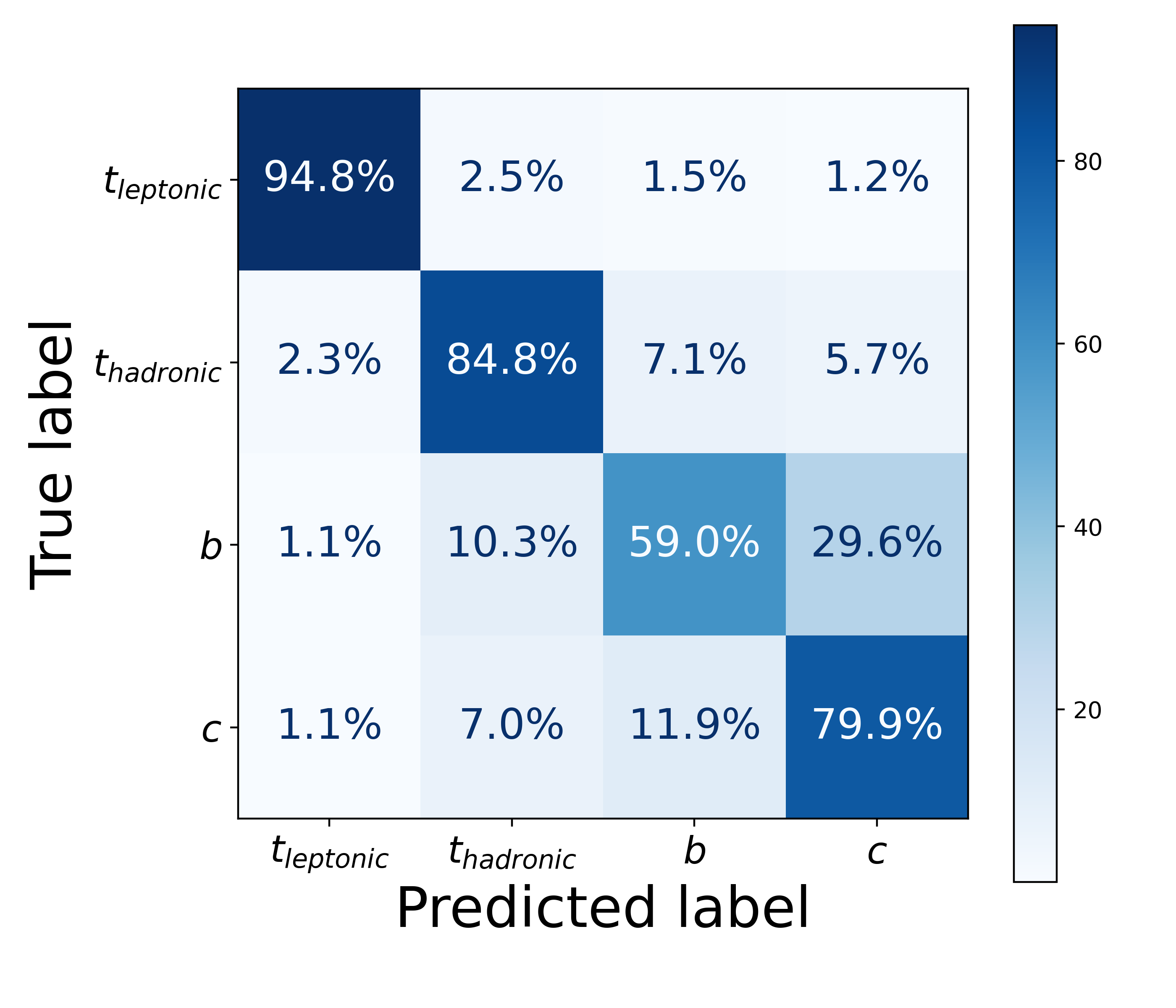}}
 \caption{Receiver operating characteristic (ROC) curve(a)  and confusion matrix(b) for 1.0 - 1.5  TeV lepton $p_T$ bins for $b$, $c$, hadronic top and leptonic top processes with $\Delta R < 0.2$ between lepton and all objects(tracks, towers, and jets).}
 \label{roc_02}
\end{figure}

The ROC curve and confusion matrix for the $\Delta R < 0.2$ cross-check are shown in Figure~\ref{roc_02}.
\label{cone_02}

\subsection{DNN and Particle Transformer architectures}
\label{dnn_part_arct}
The leptonic-top tagger is implemented using two complementary architectures.
The first is a fully connected deep neural network (DNN) operating on
physics-motivated observables, while the second is a Particle Transformer
(ParT) operating directly on the jet constituents. Their architectures and
training hyperparameters are summarized in Tables~\ref{tab:dnn_architecture}
and~\ref{tab:part_architecture}.

\begin{table}[htbp]
\centering
\small
\renewcommand{\arraystretch}{1.05}
\begin{tabular}{|l|c|}
\hline
\textbf{Parameter} & \textbf{Value} \\
\hline
Input features & 25 \\
Hidden layers & 512, 256, 128 \\
Output classes & 4 \\
Activation & GELU \\
Dropout & 0.2 \\
Loss & Cross-entropy \\
Optimiser & Adam \\
Learning rate & $10^{-3}$ \\
Batch size & 32 \\
Maximum epochs & 100 \\
Early stopping patience & 10 \\
\hline
\end{tabular}
\caption{Architecture and training hyperparameters of the DNN classifier.}
\label{tab:dnn_architecture}
\end{table}

\vspace{-4pt}

\begin{table}[htbp]
\centering
\small
\renewcommand{\arraystretch}{1.05}
\begin{tabular}{|l|c|}
\hline
\textbf{Parameter} & \textbf{Value} \\
\hline
Constituent features & 9 \\
Embedding dimension & 128 \\
Transformer blocks & 4 \\
Attention heads & 4 \\
Feed-forward network & $128\rightarrow512\rightarrow128$ \\
CLS token & Yes \\
Transformer dropout & 0.1 \\
Classifier dropout & 0.2 \\
Output classes & 4 \\
Loss & Cross-entropy \\
Optimiser & Adam \\
Learning rate & $10^{-3}$ \\
Batch size & 32 \\
Maximum epochs & 50 \\
Maximum constituents & 100 \\
Constituent ordering & Descending $p_T$ \\
Padding & Zero padding \\
Padding mask & Yes \\
\hline
\end{tabular}
\caption{Architecture and training hyperparameters of the Particle Transformer (ParT) classifier.}
\label{tab:part_architecture}
\end{table}

\clearpage

\bibliographystyle{JHEP}
\bibliography{refs}

\end{document}